\documentclass[final,5p,times,twocolumn]{elsarticle}

\usepackage{graphicx}
\usepackage{amssymb}
\usepackage{amsmath}
\usepackage{color}

\usepackage{axodraw}

\usepackage[linktocpage=true,unicode=true]{hyperref}
\usepackage{hypcap}

\newcommand{\citeeq}[1]{Eq.~(\ref{#1})}

\newcommand{\citeeqss}[2]{Eqs.~(\ref{#1})~and~(\ref{#2})}
\newcommand{\citeeqp}[1]{Eq.~\ref{#1}}
\newcommand{\citesec}[1]{Sect.~\ref{#1}}

\newcommand{\citeapp}[1]{~\ref{#1}}

\newcommand{\citetab}[1]{Tab.~\ref{#1}}
\newcommand{\citefig}[1]{Fig.~\ref{#1}}

\newcommand{\eg}{{\it e.g.}}
\newcommand{\ie}{{\it i.e.}}

\newcommand{\ben}{\begin{eqnarray}}
\newcommand{\een}{\end{eqnarray}}
\newcommand{\be}{\begin{equation}}
\newcommand{\ee}{\end{equation}}
\newcommand{\besubeq}[1]{\begin{subequations} \label{#1}}
\newcommand{\eesubeq}{\end{subequations}}
\newcommand{\nn}{\nonumber}

\newcommand{\mchi}{\mbox{$m_{\chi}$}}
\newcommand{\mchisq}{\mbox{$m_{\chi}^2$}}
\newcommand{\mchin}[1]{\mbox{$m_{\chi}^{#1}$}}

\newcommand{\sigv}{\mbox{$\langle \sigma v \rangle $}}

\newcommand{\tqcd}{\mbox{$T_{\rm QCD}$}}
\newcommand{\geff}{\mbox{$g_{\rm eff}$}}
\newcommand{\heff}{\mbox{$h_{\rm eff}$}}
\newcommand{\gstar}{\mbox{$g_{\star}$}}
\newcommand{\sqrgstar}{\mbox{$g_{\star}^{1/2}$}}

\newcommand{\spaa}{\mbox{$S\!P_{aa}$}}
\newcommand{\sphh}{\mbox{$S\!P_{hh}$}}
\newcommand{\sptot}{\mbox{$S\!P_{\rm tot}$}}

\newcommand{\gafive}{\mbox{$\gamma_{5}$}}

\newcommand{\vbar}{\mbox{$\bar{v}$}}
\newcommand{\myslash}[1]{\mbox{$\slash\!\!\!{#1}$}}

\newcommand{\sthe}{\mbox{$s_\theta$}}
\newcommand{\cthe}{\mbox{$c_\theta$}}
\newcommand{\sphi}{\mbox{$s_\varphi$}}
\newcommand{\cphi}{\mbox{$c_\varphi$}}
\newcommand{\spsi}{\mbox{$s_\psi$}}
\newcommand{\cpsi}{\mbox{$c_\psi$}}
\newcommand{\tbe}{\mbox{$\tan\beta$}}
\newcommand{\cbe}{\mbox{$c_\beta$}}
\newcommand{\sbe}{\mbox{$s_\beta$}}

\newcommand{\cchii}{\mbox{$c_{\chi i}$}}
\newcommand{\cchiit}{\mbox{$\tilde{c}_{\chi i}$}}
\newcommand{\cchij}{\mbox{$c_{\chi j}$}}
\newcommand{\cchijt}{\mbox{$\tilde{c}_{\chi j}$}}
\newcommand{\cchik}{\mbox{$c_{\chi k}$}}
\newcommand{\cchikt}{\mbox{$\tilde{c}_{\chi k}$}}
\newcommand{\cchil}{\mbox{$c_{\chi l}$}}
\newcommand{\cchilt}{\mbox{$\tilde{c}_{\chi l}$}}

\newcommand{\cchiisq}{\mbox{$c_{\chi i}^2$}}
\newcommand{\cchiitsq}{\mbox{$\tilde{c}_{\chi i}^2$}}
\newcommand{\cchijsq}{\mbox{$c_{\chi j}^2$}}
\newcommand{\cchijtsq}{\mbox{$\tilde{c}_{\chi j}^2$}}

\newcommand{\cip}{\mbox{$c_{i+}^2$}}
\newcommand{\cjp}{\mbox{$c_{j+}^2$}}

\newcommand{\cijp}{\mbox{$c_{ij+}^2$}}

\newcommand{\ccijm}{\mbox{$c_{i/j-}^2$}}

\graphicspath{{./FinalFigs/}}
\journal{Nuclear Physics B}

\begin{document}

\begin{frontmatter}



\title{Cosmic-ray antiproton constraints on light singlino-like dark matter 
  candidates}


\author[ift,dft]{David G. Cerde\~no}
\ead{davidg.cerdeno@uam.es}

\author[ift]{Timur Delahaye}
\ead{timur.delahaye@uam.es}

\author[ift,dft]{Julien Lavalle\fnref{mfellow}}
\ead{lavalle@in2p3.fr}
\fntext[mfellow]{Multidark fellow}

\address[ift]{Instituto de F\'isica Te\'orica (UAM/CSIC)\\
Universidad Aut\'onoma de Madrid, Cantoblanco,
E-28049 Madrid --- Spain
}

\address[dft]{Departamento de F\'isica Te\'orica (UAM)\\
Universidad Aut\'onoma de Madrid, Cantoblanco,
E-28049 Madrid --- Spain
}

%% Abstract
\begin{abstract}
The CoGeNT experiment, dedicated to direct detection of dark matter, has 
recently released excess events that could be interpreted as elastic collisions
of $\sim$10 GeV dark matter particles, which might simultaneously explain the 
still mysterious DAMA/LIBRA modulation signals, while in conflict with results 
from other experiments such as CDMS, XENON-100 and SIMPLE. It was shown that
5-15 GeV singlino-like dark matter candidates arising in singlet extensions of 
minimal supersymmetric scenarios can fit these data; annihilation then mostly 
proceeds into light singlet-dominated Higgs (pseudo)scalar fields. We develop 
an effective Lagrangian approach to confront these models with the existing data
on cosmic-ray antiprotons, including the latest PAMELA data. Focusing on a 
parameter space consistent with the CoGeNT region, we show that the predicted 
antiproton flux is generically in tension with the data whenever the produced 
(pseudo)scalars can decay into quarks energetic enough to produce antiprotons, 
provided the annihilation S-wave is significant at freeze out in the early 
universe. In this regime, a bound on the singlino annihilation cross section is 
obtained, $\sigv\lesssim 10^{-26}\,{\rm cm^3/s}$, assuming a 
dynamically constrained halo density profile with a local value of 
$\rho_\odot = 0.4\,{\rm GeV/cm^3}$. Finally, we provide indications on how 
PAMELA or AMS-02 could further constrain or detect those configurations 
producing antiprotons which are not yet excluded.
\end{abstract}


\end{frontmatter}

\noindent \mbox{Preprint: FTUAM-11-49, IFT-UAM/CSIC-11-41}

% \linenumbers

%% Introduction

\section{Introduction}
\label{sec:intro}

The CoGeNT~\cite{2011PhRvL.106m1301A} Collaboration has recently claimed for 
excess events in their data, while not confirmed by complementary searches like 
XENON-100 \cite{2010PhRvL.105m1302A,2011arXiv1104.2549X} or SIMPLE
\cite{2011arXiv1106.3014F}. Although CDMS-II had reported a few excess 
events~\cite{2010Sci...327.1619C}, while not statistically significant, their 
latest results are also in conflict with the CoGeNT 
data~\cite{2011PhRvL.106m1302A}. The situation has become even more confusing, 
though quite exciting, since CoGeNT announced hints for annual modulation of the
event counting rate~\cite{2011arXiv1106.0650A}. Scatterings of weakly 
interacting massive particles (WIMPs) with masses around 10 GeV off the detector
targets could explain these measurements, and interpretations in some specific 
particle physics models have been discussed in
\eg~\cite{2010PhRvD..81j7302B} and \cite{2010JCAP...02..014K} 
(see also~\cite{2010JCAP...09..022M} for less conventional models).
Intriguingly, this mass range turns out to be close to what is needed to fit 
the still mysterious DAMA/LIBRA signals
\cite{2003NCimR..26a...1B,2008EPJC..tmp..167B,1999PhRvD..59i5004B}.
Nevertheless, the authors of Ref.~\cite{2010JCAP...02..014K} have notably 
shown that some of the parameter space individually favored by the cited 
experiments may not be compatible among each other 
(see also~\cite{2011arXiv1105.3734F,2011arXiv1107.0715F,2011arXiv1107.0741M}).

These controversial data have triggered a lot of activity in the field, and 
different phenomenological directions have been pursued. While light neutralinos
arising from the minimal supersymmetric extension of the standard model (MSSM) 
are still debated (see arguments against this proposal in 
e.g.~\cite{2010PhRvD..82k5027A}, to which very detailed highlights are opposed
in \cite{2011PhRvD..83a5001F}), singlet extensions of supersymmetric scenarios
(including the next-to-MSSM --- NMSSM) have received a particular attention, 
\eg~\cite{2011PhRvD..83h3511C,2010arXiv1009.0549B,2010arXiv1009.2555G,2011PhRvL.106l1805D,2011PhLB..695..169K}. 
The NMSSM, which provides a very elegant way to solve the so-called 
$\mu$-problem occurring in the MSSM (see 
\cite{1997NuPhB.492...21E,2010PhR...496....1E} for reviews), could actually 
lead to a phenomenology consistent with the direct detection hints, though not 
generically, and some interesting regions of the parameter space are worth being
investigated more deeply
\cite{2008PhRvL.101y1301A,2010arXiv1009.2555G,2011PhRvL.106l1805D}. 
In slight contrast, more general singlet extensions uncorrelated with the 
$\mu$-problem can quite easily be tuned to accommodate the data, 
\eg~\cite{2011PhLB..695..169K}, because more flexible. In these models, the dark
matter candidate is usually a light singlino-dominated neutralino.

Interesting constraints on light WIMPs may actually come from colliders
\cite{2011PhLB..695..185G,2010PhRvD..82k6010G,2010JHEP...12..048B}, but 
most of the singlet models mentioned above seem to escape them
\cite{2010arXiv1009.2555G,2011PhRvL.106l1805D}.
From the astrophysical point of view, the annihilation of such light WIMPs may 
also generate gamma-ray fluxes at the edge of being excluded with current 
measurements~\cite{2011JCAP...01..011A}; observational limits on 
high-energy neutrinos from the Sun also give interesting constraints, as shown 
in~\eg~\cite{2011NuPhB.850..505K}. 
Another very powerful astrophysical constraint comes from measurements of local 
cosmic-ray antiprotons, which severely limits {\em direct} annihilation into 
quarks~\cite{2010PhRvD..82h1302L}.
In this paper, we aim at extending the latter analysis to singlet-like
phenomenologies in which singlino-like dark matter particles $\chi$ annihilate 
into light scalar and/or pseudo-scalar particles (generically referred to $h$ 
and $a$ in the following), which can further decay, for instance into quarks. 
For the sake of generality, we will adopt a model-independent approach by using
an effective Lagrangian relevant to any particular singlet model, and will
derive constraints on all involved masses and couplings. We note that indirect 
detection constraints come into play only when annihilation is not 
velocity-suppressed, which implies that the dominant annihilation channel is 
$\chi\,\chi\to a\,h$. This defines the mass range we are going to focus on, 
namely $2\,\mchi\geq m_a+m_h$.

The outline of the paper is the following. In \citesec{sec:efflag}, we
present our effective Lagrangian approach and discuss the annihilation 
cross section and issues related to the relic density. In \citesec{sec:direct},
we briefly review the calculation of the spin-independent WIMP-nucleus cross
section, and delineate the parameter space to be considered for interpretations 
of the direct detection hints. In \citesec{sec:pbars}, we derive the cosmic-ray 
antiproton spectra generated by decays of light scalar and pseudo-scalar Higgs 
bosons, and we determine the constraints current flux measurements put on the 
effective parameter space. We conclude in \citesec{sec:concl}. Further details 
on our calculations are given in the appendices.

%% Section 2: Effective model

\section{Effective Lagrangian approach to singlino-like dark matter scenarios}
\label{sec:efflag}
The phenomenology of light singlino-like dark matter scenarios is mostly set by 
(i) the couplings between the neutralino, a Majorana fermion, and the light 
scalar and pseudo-scalar particles of the Higgs sector, (ii) the masses of the
involved fields, and (iii) the couplings of the light scalars to standard
model particles. We address the former point here; the latter will be discussed
in~\citesec{subsec:coupl_to_ms}. Denoting $\chi$ the neutralino field and 
$\phi_i$ a generic field that contains a scalar part $h_i$ and a pseudo-scalar 
part $a_i$ such that $\phi_i\equiv h_i + i\,a_i$, we consider the following 
effective interaction Lagrangian:
\ben
{\cal L}_{\rm eff} = 
-\frac{1}{2}\sum_i\bar\chi\,{\cal C}_{\chi i}\,\chi\,\phi_i
-\sum_{i,j,k}\frac{\lambda_{ijk}}{\eta_{ijk}}\,\phi_i\,\phi_j\,\phi_k\;
+{\rm h.c.},
\label{eq:lag}
\een
where we $\lambda_{i,j,k}$ is a dimensional real coupling ($\eta_{ijk}$ is a 
combinatory factor), and where
\ben
\label{eq:coupl}
{\cal C}_{\chi i} &\equiv& \cchii + i\, \cchiit\,\gafive 
=  \bar{{\cal C}}_{\chi i}
\een
features both the couplings of the scalar and pseudo-scalar fields $\phi_i$ to 
the dark matter fermionic field $\chi$. We further define $\cchii$ (scalar 
coupling) and $\cchiit$ (pseudo-scalar coupling) as real numbers.
\subsection{Higgs sector}
\label{subsec:higgs}
We consider singlet extensions of the MSSM in which the lightest neutralino
is singlino dominated, and the light Higgs scalar and pseudo-scalar are
singlet dominated. Nevertheless, the physical Higgs fields are still mixtures 
of the Higgs doublets $H_u$ and $H_d$ and of the singlet field $S$, which 
allows us to define the mixing matrix ${\cal S}$ as
\ben
\label{eq:mixing}
{\cal S}_i^j = 
\left(
\begin{array}{ccc}
  \sthe \, \sphi & -\cthe \,\sphi & \cphi\\
  -\cthe\, \sphi - \sthe\, \cpsi\, \cphi & 
  \cthe \,\cpsi\, \cphi - \sthe \,\spsi  & 
  \cpsi\, \sphi\\
  \cthe \,\cpsi - \sthe \,\spsi \,\cphi &
  \cthe\, \spsi \,\cphi - \cthe \,\cpsi & 
  \spsi\, \sphi
\end{array}
\right )\;,
\een
where $s_k$ and $c_k$ stand for $\sin k$ and $\cos k$, respectively.
The mass eigenstates corresponding to the lightest pseudo-scalar and scalar 
fields, which we respectively denote $a$ and $h$, are conventionally given by 
the first row:
\ben
h &=& \sphi\,\left[ \sthe\,{\rm Re}(H_d) - 
\cthe\, {\rm Re}(H_u) \right] + \cphi\,{\rm Re}(S)\\
a &=& \sphi_a \,\left[ \sthe_a \,{\rm Im}(H_d) - 
\cthe_a \,{\rm Im}(H_u) \right] + \cphi_a \,{\rm Im}(S)\nn \\
 &=& \sphi_a \,\left[ A_{\rm MSSM} \right] + \cphi_a \,{\rm Im}(S)\;,\nn 
\een
where $A_{\rm MSSM}$ corresponds to the same linear combination of the imaginary 
parts of $H_u$ and $H_d$ that describes the pseudo-scalar Higgs in the MSSM.
The mixing angles can be determined in specific models from the mass terms and 
couplings featuring the superpotential and the soft potential. The Higgs mixing 
will actually play an important role not only when calculating the direct 
detection signals, as discussed in~\citesec{sec:direct}, but also to determine 
the decay channels of the light Higgs fields, which the antiproton production 
depends on (see~\citesec{subsec:pbar_prod}). It also helps us understanding the 
collider constraints, as discussed later on.

In this paper, we are assuming the lightest CP-even Higgs to be singlet-like, 
in order to avoid collider constraints, and therefore we take 
$\cphi\ge 0.9$. For the two heavy CP-even Higgs bosons, we adopt a 
MSSM-like scenario in which the lightest one is of $u$-type ($H_u$) and the 
heavy one is approximately $H_d$. Thus, we fix $\cthe =\cpsi=0.9$.

Last but not least, as it is customary in the MSSM, we define $\tan\beta$ as 
the ratio of the Higgs doublet fields 
vacuum expectation values, $\tan\beta\equiv
\langle H_u\rangle/\langle H_d\rangle$.
\subsection{Couplings to standard model fermions}
\label{subsec:coupl_to_ms}
To complement this effective setup, we also need to define the couplings between
the scalar and pseudo-scalar fields to the standard model (SM) fermions, which
will further determine the branching ratios of their decay products 
(annihilation final states) and the strength of the WIMP-nucleon interaction 
(direct detection signals). For a field $\phi_i$ having both a scalar and a 
pseudo-scalar part, the couplings to SM fermion fields differs for up-type 
quarks (or neutrinos) and down-type quarks (or charged leptons) as:
\ben
\label{eq:coupl_phi_ff}
{\cal C}_{\phi_i,{\rm up,up}} &=& 
-\frac{g\,m_u}{2\,M_W\,\sbe } \,\left\{ \sphi\,\cthe 
- i\,\gafive \,\sphi_{a} \,\cthe_{a} \right\} \\
{\cal C}_{\phi_i,{\rm down,down}} &=& 
-\frac{g\,m_d}{2\,M_W\,\cbe} \, \left\{ \sphi \,\sthe 
- i\,\gafive \,\sphi_{a} \,\sthe_{a} \right\}\;,\nn
\een
where $m_u$ ($m_d$) is the up-type (down-type) quark mass.

These couplings are useful to compute the decay branching ratios of the
light Higgs fields, which will further determine the induced antiproton 
spectrum. In practice, we use the light (pseudo-)scalar decay widths that
we derive as:
\ben
\label{eq:widths}
\Gamma_{\phi_i \to f\bar{f}} &=& 
N_c \,\frac{{\cal C}^2_{\phi_i,f,\bar{f}}}{8\,\pi}\,m_{\phi_i}\,
\left[1 - \left(\frac{2\,m_f}{m_{\phi_i}} \right)^2\right]^{3/2} \\
\Gamma_{\phi_i \to \phi_j^\star\phi_j^\star} &=& 
\frac{\lambda^2_{ijk}}{32\,\pi\,m_{\phi_i}}\, 
\sqrt{1 - \left(\frac{2\,m_{\phi_j}}{m_{\phi_i}} \right)^2} \;, \nn 
\een
where $N_c$ is a color factor (3 for quarks, 1 for leptons), and the latter
expression is relevant to the cases where the process $h\to a\,a$ adds up to 
decays into SM fermions ($m_h\geq 2\,m_a$).

In the following, we will only consider those interactions between singlinos 
and SM particles which are mediated by the scalar and pseudo-scalar Higgs 
bosons, assuming all other contributions to be negligible. In other words,
we assume that other sectors of the complete underlying theory decouple. This
is a reasonable assumption given the current mass limits on 
squarks~\cite{2011PhRvL.106m1802A,2011PhLB..698..196C} and charged 
sleptons~\cite{2004PhR...403..221T}. Moreover, as regards the Higgs sector, we 
impose that the interactions with standard matter are driven by light field 
exchanges only (see~\citesec{sec:direct}).

\subsection{Some collider constraints}
\label{subsec:colliders}

It is important to derive ranges for our effective model parameters, namely 
masses, mixing angles and couplings, which are consistent with current collider 
constraints. These constraints necessarily involve the couplings to SM 
particles, which have been made explicit in~\citeeq{eq:coupl_phi_ff}. We 
basically use the constraints provided in~\cite{2011PhLB..695..169K} for light 
scalars, and in~\cite{2009JHEP...01..061D} for light pseudo-scalars. The only 
additional feature we impose to our singlino dark matter candidate is to 
reproduce the region compatible with the WIMP interpretation of the CoGeNT 
signal, given the parameter space delineated for the scalar and pseudo-scalar 
fields below.

Limits can be derived because of the potential contributions light scalars 
and pseudo-scalars may give to rare mesons decays. From measurements of $B$ 
decays by Belle~\cite{2005PhRvD..72i2005I} and SM background calculations 
by~\cite{2002PhRvD..66c4002A}, the following limit has been derived by the 
authors of~\cite{2011PhLB..695..169K} for the light scalar mass, given its 
non-singlet content $\sphi$:
\begin{align}
\label{eq:mhlim}
{\rm Br}&(B \to h+X_s) \times {\rm Br}(h\to \mu^++\mu^-) =\\
&0.058 \left[ \frac{\sphi }{0.1} \right]^2
\left[ 1 - \frac{m_h^2}{m_b^2} \right]^2 \times {\rm Br}(h\to \mu^++\mu^-)
< 2.5 \times 10^{-6}\;.\nn
\end{align}
The decay branching ratio ${\rm Br}(h\to \mu^++\mu^-)$ is readily computed from
\citeeq{eq:coupl_phi_ff}. This roughly sets a lower limit for $m_h$ in terms
of the bottom quark mass, namely $m_h\gtrsim m_b$.

There is an additional and independent limit on $\sphi$ which can
be derived from LEP data on Higgs searches, especially from the process
$e^+\,e^-\to Z\,h\to \nu\,\bar{\nu}\,h$~\cite{1996PhLB..385..454A}. This 
yields the upper limit $\sphi \lesssim 0.08$, valid as long as 
$m_h < 2\,m_a$, \ie~as long as the decay channel $h\to a\,a$ is 
forbidden~\cite{2011PhLB..695..169K}.

Constraints on the singlet content of the pseudo-scalar can be extracted from 
\eg~the CLEO Higgs search data from the $\Upsilon(1S)$ 
decays~\cite{2008PhRvL.101o1802L}. They are expressed in terms
of the reduced coupling $X_d$ of the light pseudo-scalar $a$ to down-type quarks
and leptons, which is normalized with respect to the coupling of the CP-even 
SM Higgs boson, and a very conservative limit is given by:
\ben
X_d = \sphi_a \, \tbe < 0.5\;,
\een
as read off from Fig.~1 of Ref.~\cite{2009JHEP...01..061D}. This is 
definitely not a stringent limit for the phenomenology under scrutiny here, 
since this has no consequence on the pseudo-scalar decay products (the mixing 
angles factorize out in the decay branching ratios). This still ensures that the
singlino annihilation into SM fermions via light pseudo-scalar exchange is 
always suppressed with respect to the annihilation into light (pseudo-)scalars.

Finally, we set a lower limit of $m_a \geq 1$ GeV in the following to avoid 
relying on too large fine-tuning zones in more theoretically constrained 
parameter spaces.

\subsection{Relic density}
\label{subsec:relic}

\begin{figure*}[t!]%[htp]
 \centering
\includegraphics[width=\columnwidth]{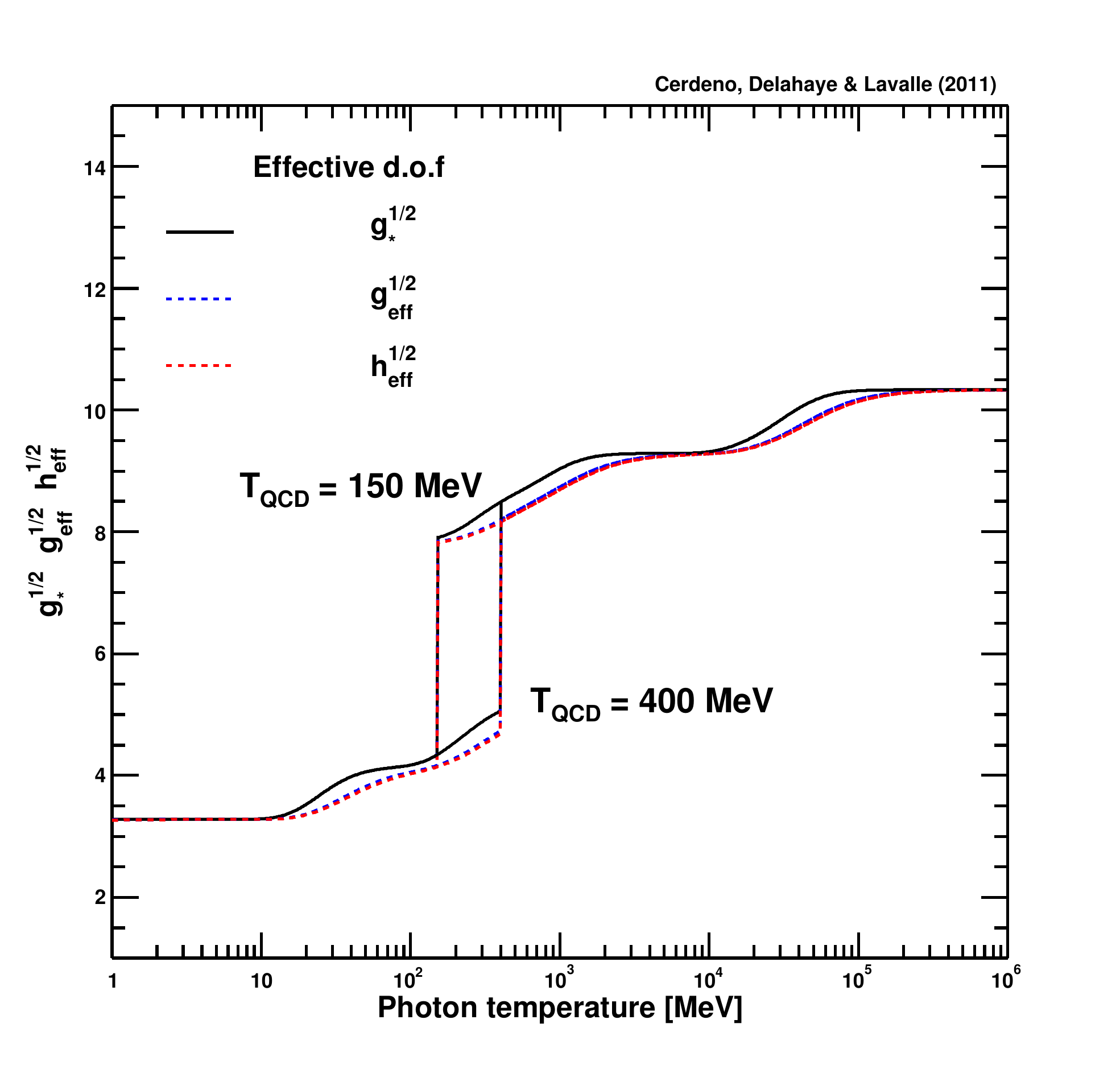}
\includegraphics[width=\columnwidth]{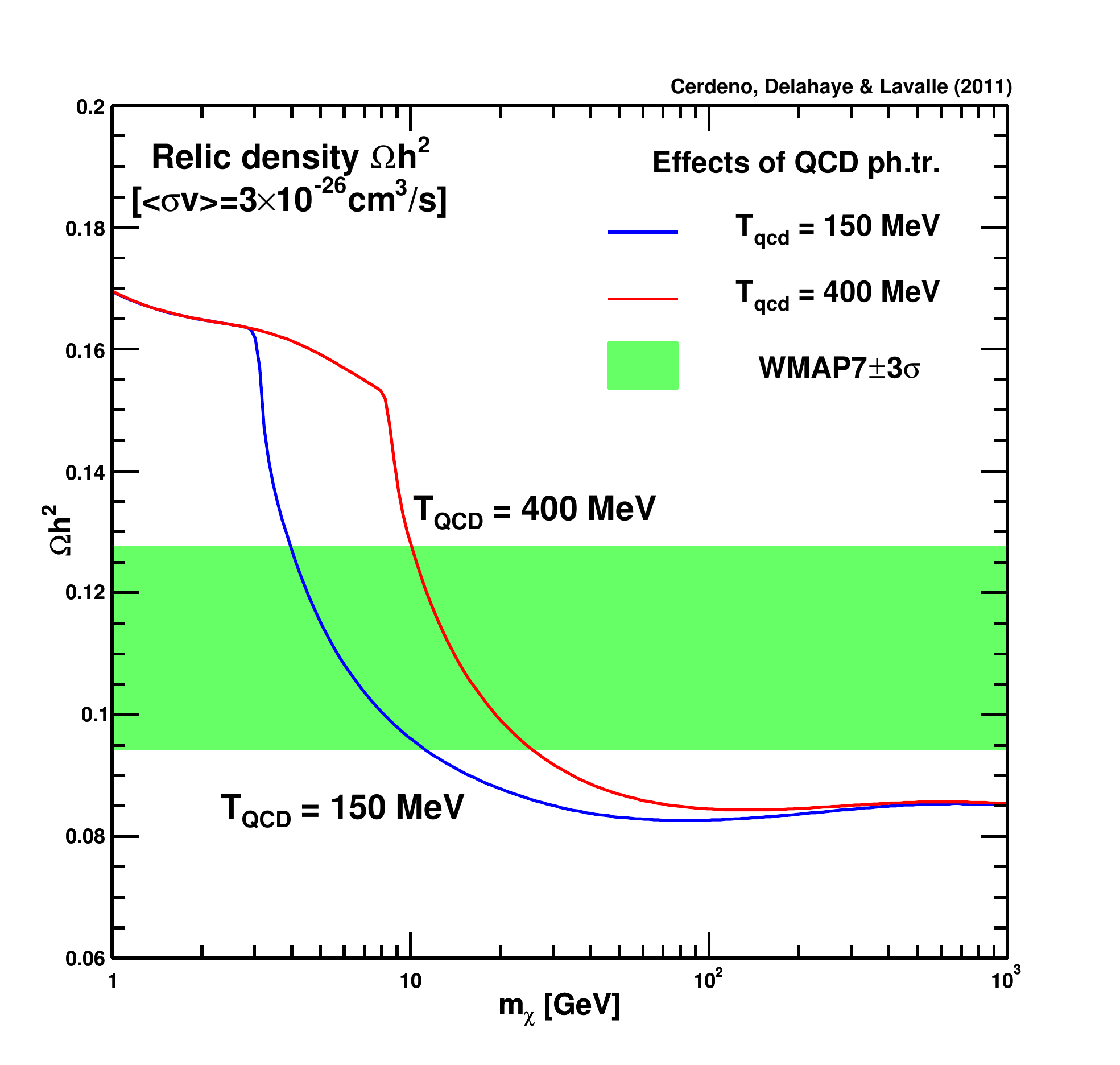}
\caption{Left: effective degrees of freedom relevant to the relic density
calculation, where a sharp first order QCD phase transition is assumed. 
Right: impact of assuming different QCD phase transition temperature on the
relic abundance of Majorana fermions --- we fix the annihilation cross-section
to $3\times 10^{-26}{\rm cm^3/s}$ for the sake of illustration.}
\label{fig:rd}
\end{figure*}

One of the most important constraints on dark matter models obviously comes
from requiring that they lead to the correct cosmological abundance, which 
usually severely reduces the available parameter space. Despite the 
uncertainties affecting our knowledge of the expansion rate of the universe 
before the big-bang nucleosynthesis (BBN) (see~\eg~
\cite{2003PhLB..571..121S}), the general formalism to derive the relic
abundance, based on solving a covariant Boltzmann equation, has been established
and tested for a long time
\cite{1980NuPhB.172..224K,1984NuPhB.237..285B,1988NuPhB.310..693S,1990eaun.book.....K,1991NuPhB.360..145G,1997PhRvD..56.1879E}. There are some public
codes like \verb!DarkSUSY!~\cite{2004JCAP...07..008G} or 
\verb!MicrOMEGAs!~\cite{2002CoPhC.149..103B}, which numerically solve the 
Boltzmann equation, originally designed for supersymmetric models. Both these 
codes essentially make use of the same numerical scheme and of the same tables 
for the effective degrees of freedom, which rely on earlier 
works~\cite{1988NuPhB.310..693S,1991NuPhB.360..145G,1997PhRvD..56.1879E}. 
Consequently, differences between them are expected to come from the 
annihilation cross section calculation only. An alternative to these numerical 
tools, but at the cost of loose accuracy, is to use approximate analytical 
expressions
\cite{1990eaun.book.....K,1991NuPhB.360..145G,2009PhRvD..80d3505D}. Here, we use
our own numerical tools, more detailed in~\citeapp{app:rd} to which we refer
the reader for the definitions of the relevant parameters. Denoting
$Y_{>f}$ the comoving dark matter density after decoupling such that 
$Y_{>f}\gg Y_{\rm eq}$, its value today $Y_0$ can be calculated from the 
following expression:
\ben
\label{eq:y0}
Y_0 = \left\{ \frac{1}{Y_{>f}} + 
\sqrt{\frac{\pi}{45}}\,\mchi \, M_p \int_{x_{>f}}dx  \, 
\frac{\sqrgstar \sigv}{x^2} 
\right\}^{-1}\;,
\een
where $x = \mchi /T$ ($T$ is the thermal plasma temperature). Note that the 
second term in the right-hand side, featuring the annihilation cross section 
$\sigv$ (see~\citesec{subsec:ann}) and the effective degrees of freedom 
$\gstar$, is usually dominant. A numerical code is basically useful to compute 
$Y_{>f}=Y(x_{>f})$ to a good accuracy from the full Boltzmann equation (see 
\citeeqp{eq:rd} in \citeapp{app:rd}), given the temperature-dependent 
annihilation cross section $\sigv$ and the effective degrees of freedom 
$\gstar$. The relic abundance then merely reads:
\ben
\Omega_\chi h^2 &=& \mchi \frac{s_0\,Y_0\,h^2}{\rho_c}
= \mchi \frac{8\, \pi \, M_p^2\, s_0\,Y_0}{3\times 100^2}\nn\\
&\simeq& 2.744\times 10^8 \,\mchi\,Y_0 \;,
\een
where we have translated the critical density $\rho_c$ in terms of the 
Hubble parameter $H$, using the relation $H^2 = (100\times h)^2 = 8\,\pi\,
\rho_c/3$. The parameter $s_0\simeq 2891.23\,{\rm cm^{-3}}$ is the entropy 
density today --- we have taken a CMB temperature of $2.72548^\circ$ K
\cite{2009ApJ...707..916F}.

One of the key issues in calculating the relic density of $\sim 10$ GeV WIMPs
is the QCD phase transition. Indeed, these WIMPs typically decouple at 
temperatures $T_f\sim \mchi/20 \sim 400$ MeV, when quarks (and gluons) go from 
asymptotic freedom to confinement into hadrons. This transition is therefore 
characterized by an important change in the relativistic degrees of freedom 
$\gstar$ appearing in~\citeeq{eq:y0}, which can be a source of systematic errors
in the predictions (as was already clear from~\eg~Figs. 5 and 6 of
\cite{1988NuPhB.310..693S}). For instance, slightly modifying the phase 
transition temperature $\tqcd$ can lead to significant changes in the 
final relic abundance, by up to 50\% (a factor of $\sim 4$ in $\gstar$), for 
those models which happen to decouple after instead of before the transition, 
after modification of $\tqcd$ (see~\citeeqp{eq:y0}). The point is that we 
do not know much about this phase transition yet despite the recent progresses
obtained in lattice QCD~\cite{2009EPJA...41..405D}, and theoretical 
uncertainties in this regime might still be sizable~\cite{2009PhRvD..79d5018H}. 
The current state of the art~\cite{2005PhRvD..71h7302H} is implemented 
in~\eg~\verb!DarkSUSY!, but modifications of the transition temperature is 
not user-free.

To try to bracket potential theoretical uncertainties coming from this phase 
transition, we have recomputed the effective degrees of freedom $\gstar$ for 
different values of $\tqcd$, considering a close to first order phase 
transition. An example is shown in the left panel~\citefig{fig:rd}, where
we display $\geff^{1/2}$, $\heff^{1/2}$ and $\sqrgstar$ for two different 
assumptions $\tqcd = 150$ or 400 MeV (see \citeeqp{eq:dof} in 
\citeapp{app:rd} for the definitions). The right panel exhibits the consequence
on the dark matter relic abundance, assuming a constant annihilation cross 
section set to its canonical value $\sigv = 3\times 10^{-26}\,{\rm cm^3s^{-1}}$, 
and only varying the WIMP mass $\mchi$. We clearly see that 
in the region of interest, $\mchi \sim 10$ GeV, the relic density can 
decrease or increase by a factor of $\sim 2$, depending on whether decoupling 
occurs just before or just after the QCD phase transition --- this can
be understood from~\citeeq{eq:y0}, where we see that $Y_0\overset{\sim}{\propto}
(\sqrgstar \,\sigv)^{-1}$. Therefore, some models in this mass range can 
easily be cosmologically excluded for one choice of $\tqcd$ but allowed for 
other choices. In the plot, as will be the case in the rest of the paper, we 
have used the WMAP-7 data for the present dark matter abundance, 
\ie~$\Omega_\chi h^2 = 0.1109 \pm 0.0056$~\cite{2011ApJS..192...16L}, which 
appears as the green band.

This illustrates why uncertainties on the quark-hadron transition and on 
$\tqcd$ should be taken into account, as often underlined in the
past, like in~\cite{2003PhRvD..68d3506B}. A detailed implementation of this 
transition is beyond the scope of this paper. Still, we consider the two 
extreme cases depicted in the left panel of~\citefig{fig:rd}, which is likely 
sufficient to encompass the full theoretical error. Finally, we stress that the 
impact on indirect detection signals amounts to the same factor of $\sim 2$. The
annihilation cross section needs to be relatively larger to get the correct 
abundance when dark matter particles decouple after the QCD phase transition, 
as clearly seen in the right panel of~\citefig{fig:rd}. 
We have checked that using other predictions for the evolution
of the relativistic degrees of freedom and their change during the QCD phase 
transition, for instance those included in the DarkSUSY package, led
to a $\lesssim 10$\% difference in the relic density predictions, or 
equivalently in the annihilation cross section\footnote{We have also made
some comparisons between our calculation and the one provided in MicrOMEGAs, and
found an agreement within $\sim 1$\% when using the MicrOMEGAS table for the
degrees of freedom (identical to the DarkSUSY table).}. Therefore, taking a 
3-$\sigma$ band for $\Omega_\chi h^2$ is well enough to encompass the 
theoretical uncertainties in the effective degrees of freedom {\em for a given
\tqcd}. Thus, we will consider any configuration as cosmologically relevant 
whenever leading to a relic density of $\Omega_\chi h^2 = 0.1109 \pm 0.0168$.

\subsection{Annihilation cross section}
\label{subsec:ann}
\begin{figure*}[!t]
\includegraphics[width=0.66\columnwidth,viewport=122 510 474 721]{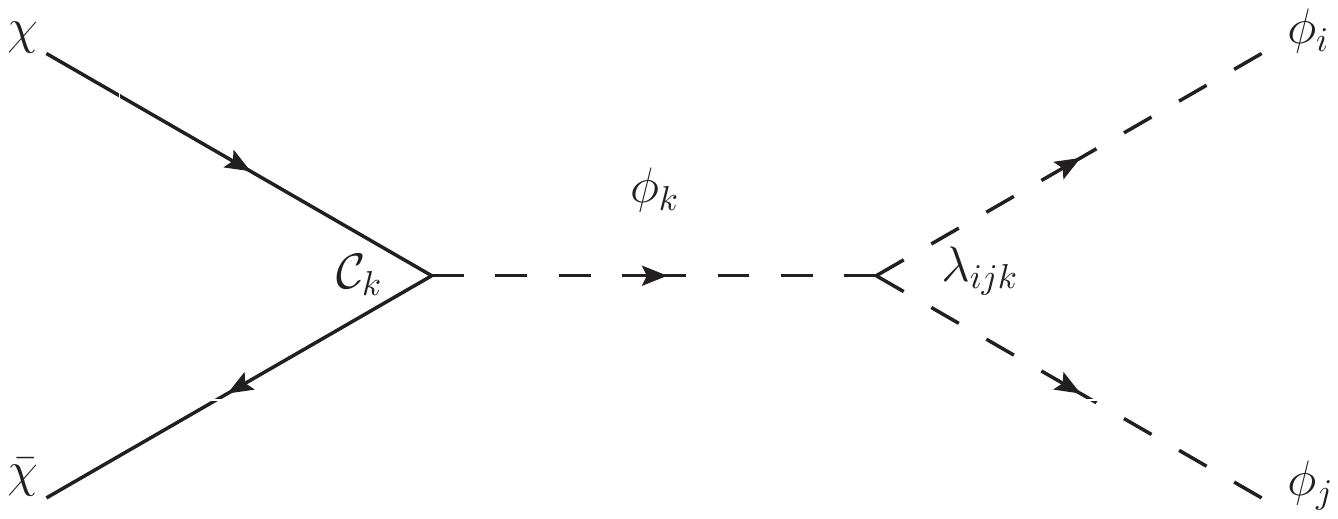}
\includegraphics[width=0.66\columnwidth,viewport=122 550 474 721]{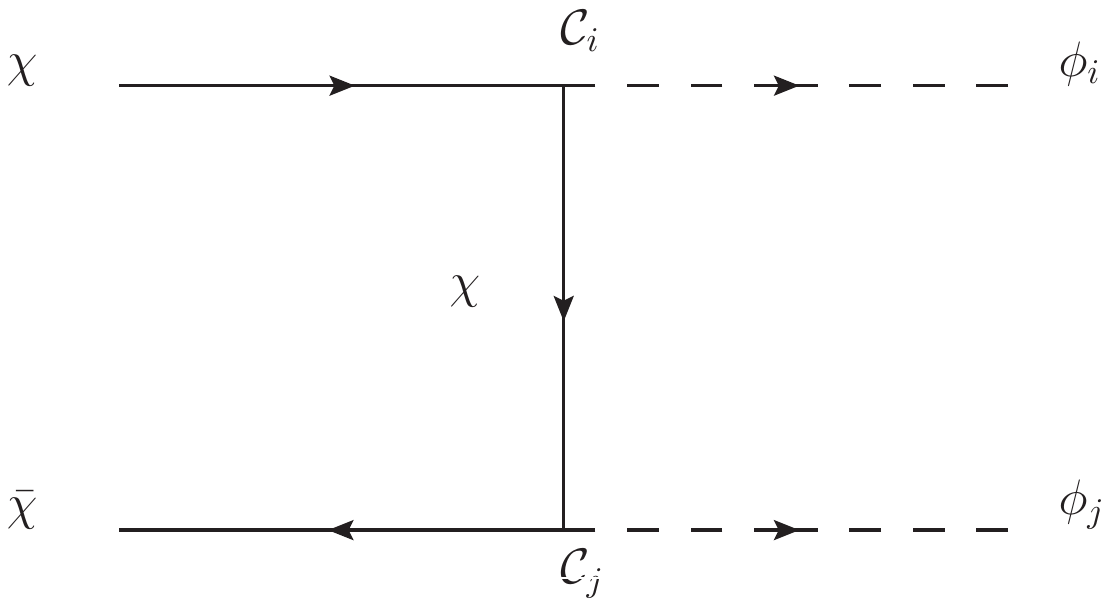}
\includegraphics[width=0.66\columnwidth,viewport=122 550 474 721]{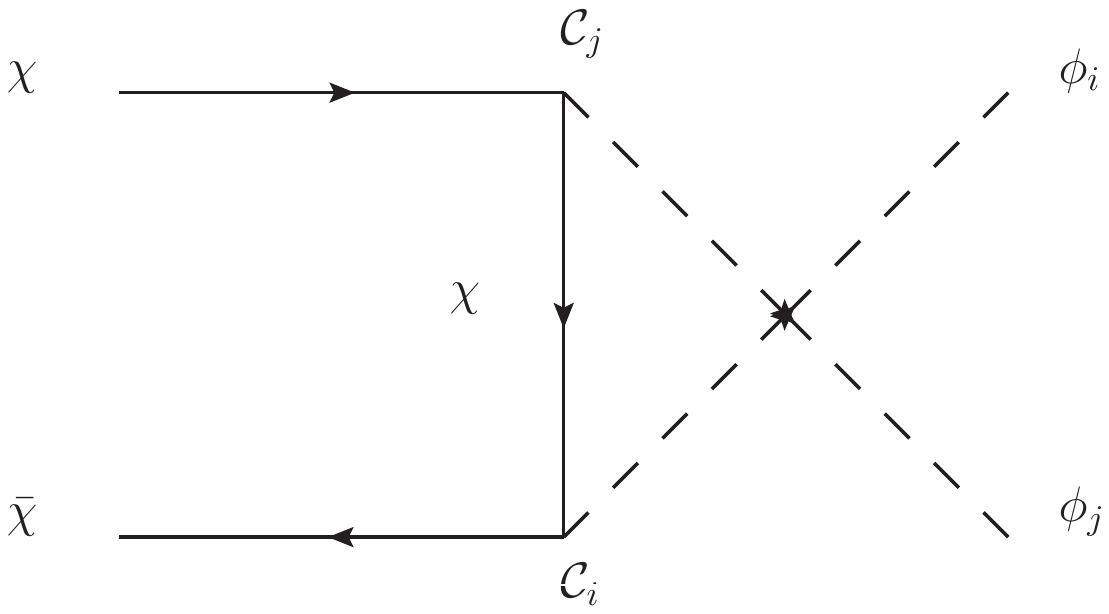}
\caption{Feynman diagrams relevant to the $\chi\chi$ annihilation process into
two (pseudo-) scalars $\phi_i\phi_j$.}
\label{fig:diagrams}
\end{figure*}
The annihilation cross section is relevant not only to indirect detection, but 
also to direct detection, since in both cases it determines the cosmological 
abundance today. In the following, we assume that the only particles 
involved in the annihilation process are the light scalar and pseudo-scalar 
Higgs bosons, $h$ and $a$ respectively. This is justified by the fact that we 
are in the very light mass and singlet/singlino dominated regime --- other Higgs
bosons, charginos and sfermions have much larger masses, while similar 
couplings. In such a regime, annihilation implies the following final states: 
$\chi\,\chi\to a\,a$, $\chi\,\chi\to h\,h$ and $\chi\,\chi\to a\,h$, which
can be summarized thanks to the Feynman diagrams represented 
in~\citefig{fig:diagrams}.

As it is well known (\eg~\cite{1996PhR...267..195J}), $\chi\,\chi\to a\,h$ is 
the only tree-level process among the three above possibilities to have a 
non-vanishing zero-velocity limit, \ie~the only process relevant to indirect 
detection. It is kinematically allowed as long as $2\,\mchi\geq m_a+m_h$, which 
defines the mass range we are dealing with in this analysis. The full 
calculation of the annihilation cross section is detailed in 
\citeapp{app:cross_section}. Here, we just summarize a few helpful results. It 
usually proves interesting to expand the thermally averaged annihilation cross 
section as a power series of the inverse of the variable 
$x\equiv \mchi/T\propto v^{-2}$ ($x_f\sim 20$ at freeze out, and 
$x_0\sim 10^7$ in the Galaxy today):
\ben
\sigv = a + b/x + {\cal O}(x^{-2})\;.
\label{eq:sigv}
\een
Both coefficients $a$ and $b$ take analytical expressions in the present study. 
The S-wave contribution is given by (we neglect the widths of the exchanged 
particles\footnote{Particle widths are included in the full numerical 
calculation.}):
\begin{align}
\label{eq:swave}
a &= a_{\chi\chi\to ah} = 
\frac{\sqrt{(1-\left[\frac{(m_a+m_h)}{2\,\mchi}\right]^2)
    (1-\left[\frac{(m_a-m_h)}{2\,\mchi}\right]^2)}}
     {16\,\pi\,m_\chi^2} \times  \\
& \frac{\tilde{c}_{\chi\chi a}^2}
     {\left(4\,m_\chi^2-m_a^2\right)^2
       \left(1-\frac{(m_a^2 + m_h^2)}{4\,m_\chi^2}\right)^2} \times \nn\\
& \Bigg[ \,c_{\chi\chi h} \, \left(4\,m_\chi^2-m_a^2\right)\, 
       \left(1 + \frac{(m_a^2-m_h^2)}{4\,m_\chi^2}\right) -\nn\\
& \lambda_{aah}\,\mchi\,\left(1-\frac{(m_h^2+m_a^2)}{4\,m_\chi^2}\right)
       \Bigg]^2\;.\nn
\end{align}
This equation fully sets the annihilation cross section for the indirect 
detection signals that we will consider later.
We immediately see that there is a competition between the $s$ and $t/u$ 
diagrams through the couplings $\lambda_{aah}$ and $c_{\chi\chi h}$. Note that 
the latter is also the one that fixes the contributions of CP-even Higgs bosons 
to the direct detection signal (more details in~\citesec{sec:direct}), which 
sketches a way to link the indirect detection constraints to the direct 
detection signal: the relation is straightforward in the case $\lambda_{aah}=0$,
corresponding not only to the suppression of the $s$-channel of 
$\chi\,\chi\to a\,h$ but also of that of process $\chi\,\chi\to a\,a$. Finally, 
for $\lambda_{aah}\sim c_{\chi\chi h}$ and close-to-degenerate masses, the $t/u$ 
diagram dominates the cross section, which again induces a significant 
correlation between the direct detection signal and the indirect detection 
constraints.

We also remark that the coupling $c_{\chi\chi a}$ does drive the $a$-term 
amplitude, which is actually quite obvious since $\chi\chi$ is bound to be a 
pseudo-scalar state at null velocity. Suppressing this coupling is a very simple
way to avoid indirect detection constraints. In that case, annihilation before 
freeze-out would proceed mostly through $\chi\,\chi\to h\,h$, but also partly 
through $\chi\,\chi\to a\,a$ via the $s$-channel, provided these final states 
are kinematically allowed.

Although the connection between the elastic interaction with nuclei and the 
annihilation into $a\,h$ may be understood from the above equation,
we still have to make sure that the $a$-term of the cross section expansion is 
the dominant one in the early universe, in order for it to be fixed 
by the relic abundance. Indeed, should the processes $\chi\chi\to aa,hh$ 
be dominant at freeze out, then that would lessen the bounding strength of 
indirect detection and also make us lose potential correlations between direct 
and indirect signals. We recall that the correlation is definitely lost when 
$\mchi \geq m_h$ ($m_a$) and $2\mchi<m_a+m_h$, for which $\chi\chi\to hh$ ($aa$)
is the only annihilation possibility left; we do not consider such cases here.

For completeness, we provide the first order terms $b_{aa}$ and $b_{hh}$ 
associated with annihilation into $a\,a$ and $h\,h$, which drive the relic 
density when $2\,\mchi<m_a+m_h$ (still neglecting the particle widths):
\ben
\label{eq:baa}
b_{aa} &=&
\frac{\sqrt{1 - m_a^2/\mchisq}}
     {64\,\pi\, (2 \,\mchisq - m_a^2)^4 } \times \\
&&  \Bigg[ 16 \tilde{c}_{\chi\chi a}^4\,\mchisq (\mchisq - m_a^2)^2 + \nn\\
&&   8\, \lambda_{aah}\,c_{\chi\chi h} \, \tilde{c}_{\chi\chi a}^2\,\mchi\, 
       (2\, \mchisq - m_a^2)^2
       \frac{( \mchisq - m_a^2)}{(4\, \mchisq - m_h^2)} \nn\\
&& + 3\,(\lambda_{aah}\,c_{\chi\chi h})^2 \,
       \frac{(2 \, \mchisq - m_a^2)^4}{(4\, \mchisq - m_h^2)^2} \Bigg]\;;\nn
\een
\ben
\label{eq:bhh}
b_{hh} &=& 
\frac{c_{\chi\chi h}^2\, \sqrt{1 - m_h^2/\mchisq }}
{64\,\pi\, (2\, \mchisq - m_h^2)^4}\times \\ 
&& \Bigg[ 
  16 \, c_{\chi\chi h}^2 \, \mchisq (9\, \mchin{4} + 2\, m_h^4 - 
  8 \,m_h^2\, \mchisq) - \nn\\ 
&& 8\, \mchi \, \lambda_{hhh} \, c_{\chi\chi h}\,  (2\, \mchisq - m_h^2)^2 
\frac{(5\, \mchisq - 2 \, m_h^2)}{(4\, \mchisq - m_h^2)} + \nn\\
&&  3 \, \lambda_{hhh}^2 \frac{(2\, \mchisq - m_h^2)^4}{(4 \,\mchisq - m_h^2)^2}
\Bigg] \;.\nn
\een
These terms are the leading terms of the annihilation cross section when 
$2\,\mchi<m_a+m_h$. We emphasize that even in the case $2\,\mchi\geq m_a+m_h$,
\ie~when $a_{ah}$ is not suppressed, the velocity-dependent terms can not only 
partly contribute to the full annihilation cross section at decoupling in the 
early universe (up to $\sim$10-20\% when couplings and masses are taken similar,
and considering also $b_{ah}$, not given here), but can also dominate in some 
cases, \eg~when the coupling $\tilde{c}_{\chi\chi a}$ is suppressed. This means 
that the full annihilation cross section needs to be considered for relic 
density calculations, which we have done in this study. In fact, we did not use 
this Taylor expansion approach to compute the thermal average of $\sigma v$ in 
the relic density calculation, but the exact numerical method presented 
in~\cite{1991NuPhB.360..145G,1997PhRvD..56.1879E}. Still, the velocity 
independent term $a_{ah}$ given in~\citeeq{eq:swave} fully characterizes the 
annihilation cross section relevant to indirect detection. Moreover, 
\citeeqss{eq:baa}{eq:bhh} will prove very useful to interpret the antiproton 
constraints (see~\citesec{sec:pbars}).

% Section 3: direct detection
\section{Direct detection of singlino-like particles}
\label{sec:direct}

In this paper, we aim at confronting the light singlino parameter space
compatible with the CoGeNT region to the antiproton constraints. It is
therefore important to provide some details on how the CoGeNT region can
be covered in the frame of our effective approach.

The singlino couplings to the Higgs mass eigenstates will be denoted as 
$\cchij$, in consistency with the Lagrangian introduced in~\citeeq{eq:lag}.

\begin{figure}[!t]
\includegraphics[width=0.66\columnwidth,viewport=122 580 350 721]{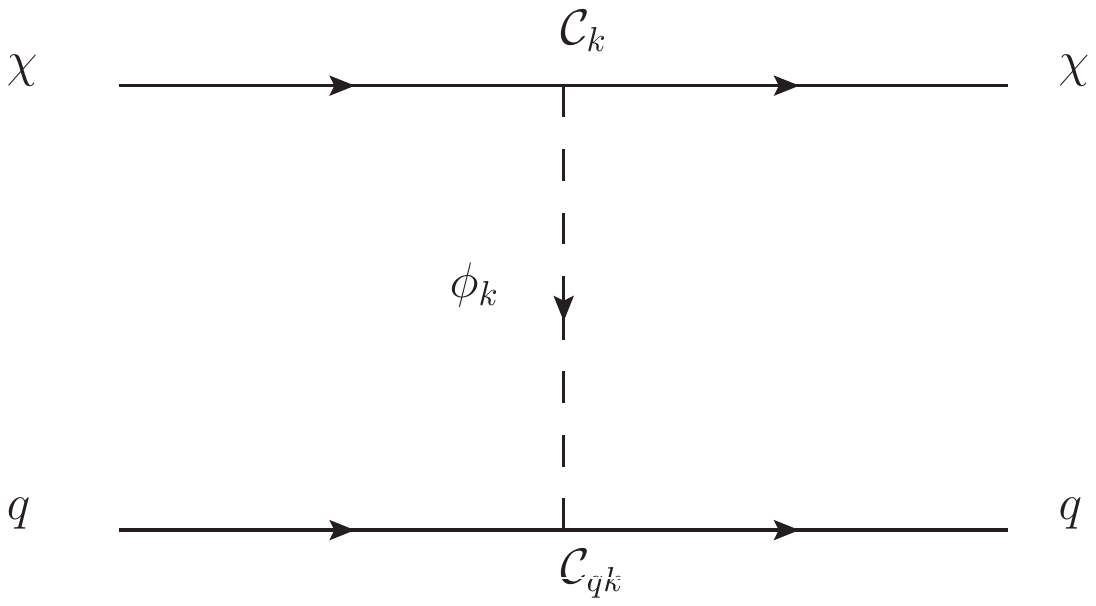}
\caption{Diagram describing the elastic interaction of a singlino
  with a quark.}
\label{fig:cross-diag}
\end{figure}

Elastic collision of a singlino off a target nucleus would take place through 
the exchange of a CP-even Higgs along a $t$-channel, as depicted 
in~\citefig{fig:cross-diag}. This scattering can be described in terms of an 
effective Lagrangian~\cite{1996PhR...267..195J},
\ben
{\cal L}_{eff}\supset \alpha_{q_i} \bar \chi\,\chi
\bar q_i q_i \;,
\een
where the coupling $\alpha_{q_i}$ results from adding the contributions of 
the exchanges of the three CP-even Higgs bosons and reads,
\ben
\alpha_{q_i}\equiv\sum_{j=1}^3\frac{g m_{q_i}}{2 M_W B_i}
\frac{\cchij }{m_{H_j^o}^2}{\cal S}^{3-i}_j\; ,
\label{eq:alphaq}
\een
where $i=1,2$ for up(down)-type quarks, $B_1=\sin\beta$ and $B_2=\cos\beta$.
The mixing matrix ${\cal S}$ has been defined in~\citeeq{eq:mixing}.
Note that the term of the above sum corresponding to our light Higgs 
boson $h$ can be straightforwardly rewritten as $\alpha_{q_i}^h = \cchij \, 
{\cal C}_{hq\bar{q}}/m_h^2$, where the coupling ${\cal C}_{hq\bar{q}}$ has been 
defined in~\citeeq{eq:coupl_phi_ff}.

The total spin-independent WIMP-proton scattering cross section yields
\ben
\sigma_{\chi-p}^{\rm SI} = \frac4\pi
\frac{m_p^2\, m_\chi^2}{(m_p+m_\chi)^2}\,f_p^2\,,
\een
where $m_p$ is the proton mass and 
\ben
\frac{f_p}{m_p}=
\sum_{q_i=u,d,s}f_{Tq_i}^p\frac{\alpha_{q_i}}{m_{q_i}}+ 
\frac{2}{27}\ f_{TG}^p\sum_{q_i=c,b,t}\frac{\alpha_{q_i}}{m_{q_i}}\ .
\een
The quantities $f_{Tq_i}^p$ and $f_{TG}^p$ are the hadronic matrix elements which
parameterize the quark content of the proton (the second term is due to the 
interaction of the WIMP with the gluon scalar density in the nucleon). We will 
use $f^p_u=0.023$, $f^p_d=0.033$ and $f^p_s=0.26$ which follows from the latest 
results on the pion-nucleon sigma-term \cite{2001hep.ph...11066P}, which 
affects the determination of the strange content of the quark.

The leading term in the scattering cross section is usually due to the coupling 
to the $s$ quark (the contribution from quarks $u$ and $d$ being almost 
negligible and that from $f_{TG}^p$  being at least a factor 5 smaller in the 
determination of $f^p$). Under this assumption it is possible to derive from 
\citeeq{eq:alphaq} a condition for the lightest Higgs contribution to be 
the dominant one in terms of the Higgs masses and the various couplings. In 
particular, we will demand
\ben
\left| \frac{c_{\chi1} }{m_{H_1^o}^2}\,{\cal S}^{1}_1 \right| > 
10\,\left| \frac{c_{\chi2}}{m_{H_2^o}^2}\,{\cal S}^{1}_2 \right|\;,
\een
and
\ben
\left| \frac{c_{\chi1} }{m_{H_1^o}^2}\,{\cal S}^{1}_1\right| > 
10\,\left| \frac{c_{\chi3}}{m_{H_3^o}^2}\,{\cal S}^{1}_3 \right| \;.
\een
Notice that since we are considering very small masses for the lightest Higgs 
boson, these conditions are normally easy to fulfil. In our case they will imply
a constraint on the Higgs doublet contribution and on the neutralino-Higgs 
coupling. For example, for a typical configuration with $m_{H_1^o}=10$~GeV and 
$m_{H_2^o}=120$~GeV the above condition implies 
$c_{\chi1} \,{\cal S}^{1}_1>7\times 10^{-2}c_{\chi2} \,{\cal S}^{1}_2 $. Under 
these conditions the spin-independent cross-section can be approximately 
expressed as
\ben
\sigma_{\chi-p}^{\rm SI}\approx 
\frac{m_p^4\, m_\chi^2}{\pi(m_p+m_\chi)^2}\,
\left(\frac{g\,f^p_s}{M_W}\right)^2
\left(\frac{c_{\chi1} S^{1}_1 }{m_{H_1^o}^2 \cbe}\right)^2
\label{eq:approxsigma}
\een

In order to explore the region consistent with the latest hints for very light 
WIMPs in direct detection 
experiments~\cite{2008EPJC..tmp..167B,2011PhRvL.106m1301A,2011arXiv1106.0650A}, 
we have constrained the input parameters to produce a singlino-proton 
spin-independent cross section between $3\times10^{-5}$~pb and 
$3\times10^{-4}$~pb, and singlino masses in the range 
$6-15$~GeV. This window in the $(\sigma_{\chi-p}^{SI},m_\chi)$ plane contains the
region compatible with the first observation by the CoGeNT collaboration 
\cite{2011PhRvL.106m1301A} and is extended in order to account for variations 
of this region when astrophysical uncertainties are taken into account 
\cite{2010JCAP...02..014K,2010arXiv1011.5432S} and to include the area 
consistent with DAMA/LIBRA result if interpreted in terms of WIMP recoils on 
NaI (assuming negligible channeling). Of course, the region consistent with the 
latest data release by CoGeNT~\cite{2011arXiv1106.0650A} is also contained 
in the whole area.

To try to delineate a parameter space as complete as possible, we have performed
different random scans over the full parameter space defined by the involved 
free masses and couplings, taking different values of $\tbe$ for each.
Since we require a given range of spin-independent cross-section and, as we said
above, we consider only those cases where the lightest Higgs contribution to 
$\sigma_{\chi-p}^{\rm SI}$ is dominant, this implies a constraint on the 
lightest Higgs composition and its coupling to singlinos. In particular, for 
$m_\chi =6-15$~GeV, \citeeq{eq:approxsigma} leads to
\ben
3\times10^{-4}\,{\rm GeV}^{-2}
\lesssim \left| \frac{c_{\chi1} \,{\cal S}^{1}_1 }{m_{H_1^o}^2 \cos\beta}\right|
\lesssim
9\times10^{-4}\,{\rm GeV}^{-2}\ .
\label{eq:ddcondition}
\een
We note that this condition implies that for large values of $\tbe$ the 
necessary value of the coupling $c_{\chi1}$ in order to reproduce the CoGeNT 
data can be smaller.

In practice, the various input parameters have been randomly drawn in the 
following ranges:
\ben
\label{eq:rand}
m_{\chi}&\subset&[6,15]\,{\rm GeV}\\
|c_{\chi1},\,c_{\chi2},\,c_{\chi3}| &\subset&[5\times10^{-6},5\times10^{-1}]\,
\nn\\
|\lambda_{ijk}| &\subset&[2\,{\cal C}_{\phi,b,\bar{b}},5\times10^{-1}]\,{\rm GeV}\nn\\
\tbe &=& 10\;{\rm and} \; 50\;,\nn
\een
where the scan in the couplings is performed in the logarithmic scale. 
Note that the lower bound on the couplings $\lambda$ is not a fixed
number, but instead twice the coupling of any light boson to $b$ quarks
(see~\citeeqp{eq:coupl_phi_ff}). This is to ensure that (the s-channel) 
annihilation dominantly proceeds into 2-body combinations of bosons $a$ and $h$,
the pair production of SM fermions being then suppressed and irrelevant in the 
computation of the relic density as well as of the indirect detection signals.
This will induce a correlation between couplings $\lambda$ and 
${\cal C}_{\chi h}$ through the range imposed on the spin-independent cross
section [see~Eqs.~(\ref{eq:coupl_phi_ff}), (\ref{eq:alphaq}), and 
(\ref{eq:approxsigma})].
The lightest CP-even and CP-odd Higgs masses are assigned a random value, 
imposing the constraint $m_h+m_a\le2\,m_{\chi}$ in order to guarantee that 
annihilation into a pair $h,a$ is possible. We have imposed the experimental 
constraints on the Higgs masses and couplings as detailed in 
\citesec{subsec:colliders} and the condition that the singlino-proton 
spin-independent cross section falls within the window 
$\sigma_{\chi-p}^{\rm SI}\subset[3\times10^{-5}, 3\times10^{-4}]\,{\rm pb}$. 
We have thus kept those configurations which fall in a region that contains the 
area compatible with the CoGeNT excess. As regards $\tbe$, after having tried 
some random scans, we have selected two extreme values only, which are 
sufficient to fully illustrate the impact of this parameter.
In \citefig{fig:sigsi}, we show our scan results in terms of $\sigma^{\rm SI}$ 
as a function of $\mchi$ for both values of $\tbe$, and for two different 
QCD phase transition, with $\tqcd = 150$ and 400 MeV. The color index refers to 
relic density and antiproton constraints, and will be specified and discussed 
in~\citesec{subsec:results}. All samples we consider in the following are made 
of $\sim 10^{5}$ random models, such as those reported in~\citefig{fig:sigsi}.

\begin{figure*}[t!]%[htp]
 \centering
\includegraphics[width=\columnwidth]{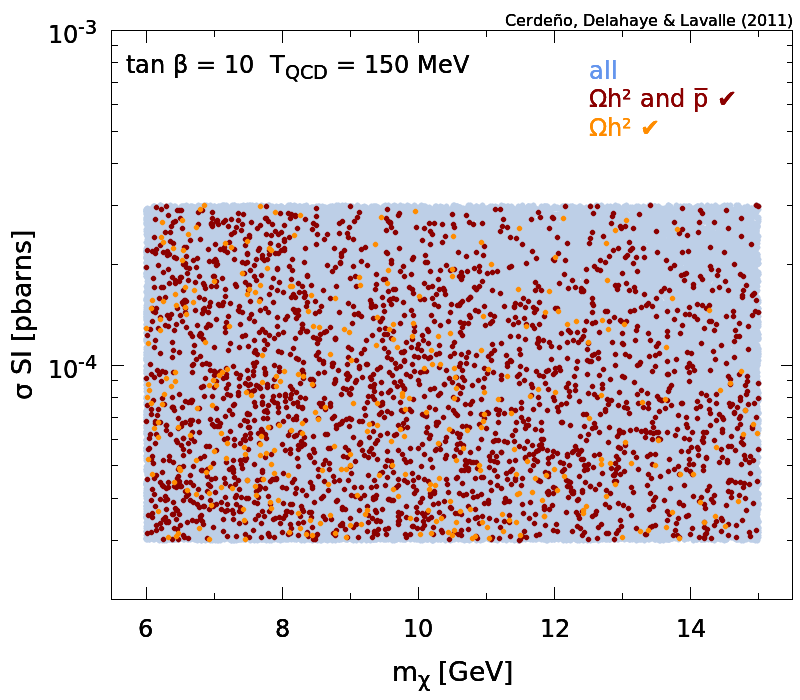}
\includegraphics[width=\columnwidth]{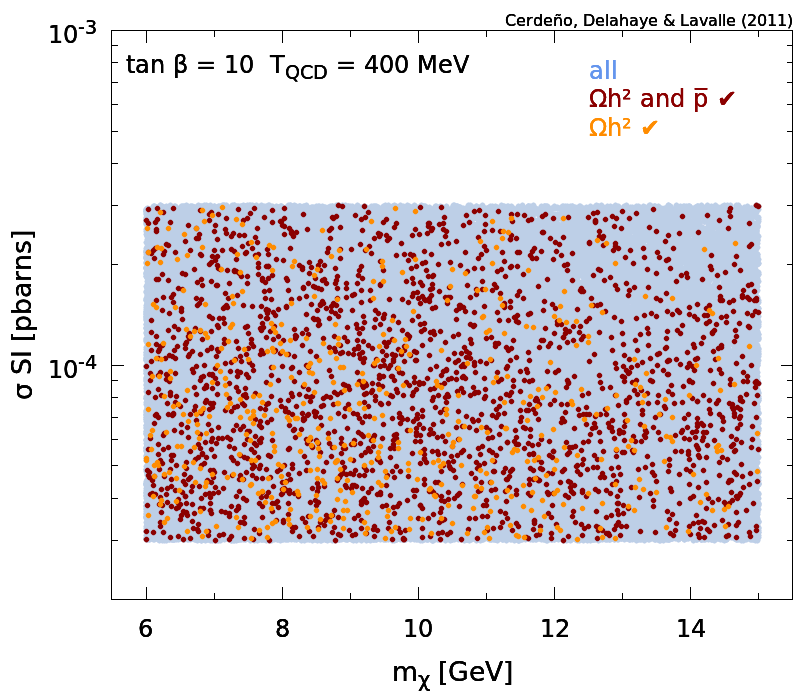}
\includegraphics[width=\columnwidth]{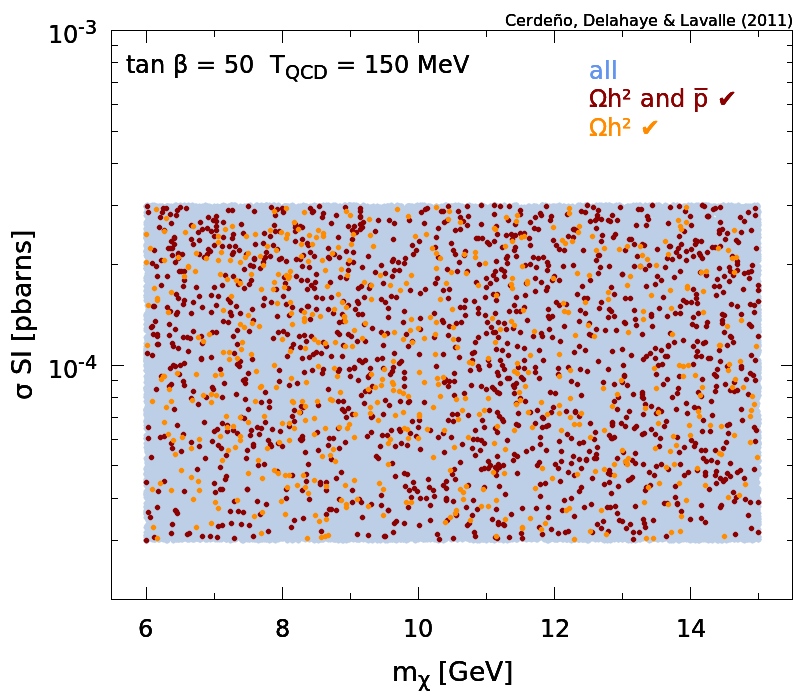}
\includegraphics[width=\columnwidth]{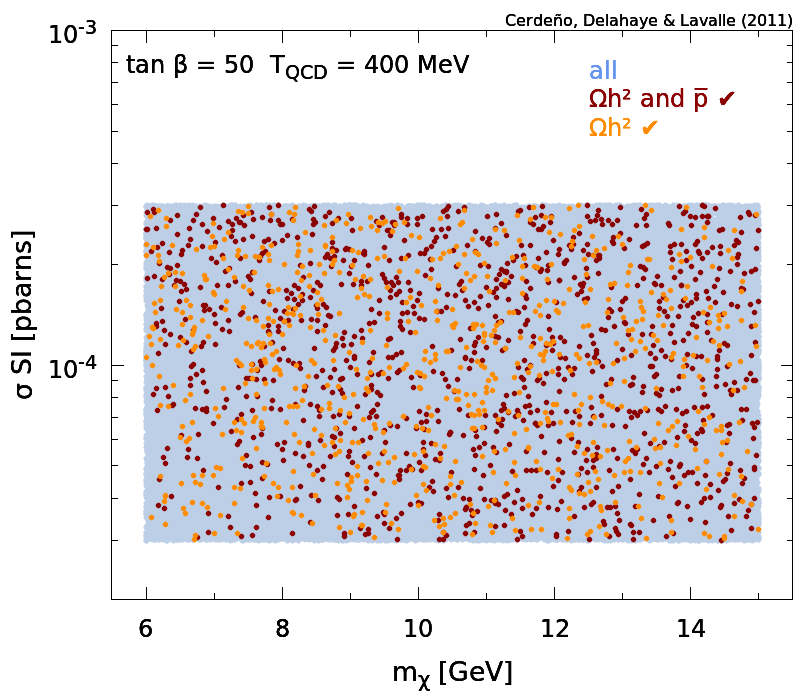}
\caption{Spin-independent cross section as a function of the singlino
mass, for different values of $\tbe$ --- 10 (left panel) and 50 (right panel).
Each panel shows $\sim 10^5$ models randomly drawn according 
to~\citeeq{eq:rand} and selected such that $\sigma^{\rm SI}$ falls in the
CoGeNT region. The color index is specified in~\citesec{subsubsec:scanres}.}
\label{fig:sigsi}
\end{figure*}

To conclude this section, we must emphasize that some uncertainties in 
the hadronic matrix elements may affect the predictions for the direct detection
rate of singlinos. The largest effect is due to the indetermination in the 
strange quark content of nucleons. This propagates into the theoretical 
predictions for the spin-independent cross section and can be responsible for a 
variation of about an order of magnitude~\cite{2000APh....13..215B,2002APh....18..205B,2005PhRvD..71i5007E,2008PhRvD..77f5026E}. 
This induces an uncertainty in the values of the couplings for which the CoGeNT 
region can be reproduced and therefore enhances the range of these with respect 
to those plotted in~\citefig{fig:sigsi}. A recent illustration has been given in
\eg~Fig.~5 of Ref.~\cite{2011arXiv1106.4667B}. We will further discuss this 
issue in~\citesec{subsubsec:unc}.

%% Section 4: pbar constraints

\section{Constraints from local cosmic-ray antiprotons}
\label{sec:pbars}

Cosmic-ray antiprotons were proposed as a potential tracers of dark matter
annihilation in~\cite{1984PhRvL..53..624S}. Dark matter-induced antiprotons
have been studied in the context of supersymmetry
\cite{1988ApJ...325...16R,1998PhRvD..58l3503B,1999ApJ...526..215B,2004PhRvD..69f3501D,2005JCAP...09..010L,2008PhRvD..78j3526L} and 
extra-dimensions~\cite{2005PhRvD..72f3507B,2008PhRvD..78j3526L}, and related
signals are likely very difficult to observe. Current observations by~\eg~
the PAMELA~\cite{2010PhRvL.105l1101A} or 
BESS~\cite{2000PhRvL..84.1078O,2001APh....16..121M,2002PhRvL..88e1101A,2005ICRC....3...13H}
experiments are in agreement with the predictions of secondary antiprotons
\cite{2001ApJ...563..172D,2007PhRvD..75h3006B,2010PhRvL.105l1101A},
\ie~those created from nuclear interactions of standard astrophysical cosmic 
rays with the interstellar gas, although some models still fail to saturate 
the data, leaving a small room for exotic primary contributions 
\cite{2002ApJ...565..280M,2011ApJ...729..106T}. The case for light 
supersymmetric neutralinos was studied in~\cite{2005PhRvD..72h3518B}.

Based on recent studies on cosmic-ray transport 
parameters~\cite{2010A&A...516A..66P} and on the local dark matter 
density~\cite{2010JCAP...08..004C,2010A&A...523A..83S}, it was shown that 
$\sim 10$ GeV light WIMPs annihilating into quark pairs were in serious tension 
with the low energy cosmic-ray antiproton data~\cite{2010PhRvD..82h1302L}. In 
this paper, we are focusing on a quite different phenomenology where WIMPs 
annihilate into light scalar $h$ and pseudo-scalar $a$ Higgs bosons, which may 
further decay into quarks. These light bosons do not have necessarily the same 
masses, so one can for instance be more Lorentz boosted than the other. 
Therefore, the energy distribution of the potentially created antiprotons is 
expected to differ from those arising from direct annihilation into quark pairs,
and the results obtained in~\cite{2010PhRvD..82h1302L} can hardly be 
extrapolated. In this section, we first determine the antiproton energy spectrum
associated with the decay of light Higgs bosons into quark pairs before 
calculating the dark matter-induced flux at the Earth.

\subsection{Antiproton production}
\label{subsec:pbar_prod}

Before estimating the antiproton flux at the Earth, one needs to know the 
antiproton spectrum before propagation,~\ie~the one arising from the 
decay/hadronization of the dark matter annihilation products, here the scalar 
$h$ and pseudo-scalar $a$. The only decay final states relevant to antiproton
production are actually quark pairs energetic enough to produce a pair
of proton-antiproton. This means that antiprotons can be created each time a 
light (pseudo-)scalar boson is heavier than twice the proton mass, roughly, 
which sets a production threshold to $m_\phi\gtrsim 2$ GeV.

The first step to get the antiproton spectrum is easy: before the decays of 
our light (pseudo-)scalars, we have a mere two-in-two-body 
annihilation so all the 4-momenta are set by kinematics. Assuming that dark 
matter particles annihilate at rest in the Galactic halo frame, the energy of 
resulting particle 1 is
\ben
E_1 = \frac{4 m_\chi^2 + m_1^2 - m_2^2}{4 m_\chi}
\een
and the norm of its momentum is
\ben
k_1 = \frac{\sqrt{\lambda\left(4 m_\chi^2 , m_1^2 , m_2^2\right)}}{4 m_\chi}\,,
\een
where $\lambda(a^2,b^2,c^2) = (a^2-(|b|+|c|)^2)(a^2-(|b|-|c|)^2)$, 
which are enough to change from the halo frame to the rest frame of particle 1: 
($\gamma,\beta$)=($E_1/m_1,k_1/E_1$). In the rest frame of particle 1, the 
quarks it decays into have energy $E_q^\ast = m_1/2$ and momentum 
$|k_q^\ast| = \sqrt{m_1^2/4 - m_q^2}$. One finally gets the energy of the quarks
in the halo frame from a mere Lorentz transform:
\ben
E_q &=& \frac{E_1}{2} - \cos(\theta) 
\frac{\sqrt{\lambda\left(4 m_\chi^2 , m_1^2 , m_2^2\right)}}{8 m_\chi} 
\sqrt{1 - \frac{4 m_q^2}{m_1}} \nn\\
    &=& \frac{E_1}{2} - \cos(\theta)\, {\cal E}.
\een
Thus, the energies of the quarks and anti-quarks coming from the decay of 
particle 1 are evenly distributed between $\frac{E_1}{2} - {\cal E}$ and 
$\frac{E_1}{2} + {\cal E}$. One finally gets the probability of having an 
antiproton of energy $E_{\bar{p}}$ from a quark of energy $E_q$ 
$f(E_q,E_{\bar{p}})$ thanks to the \verb!PYTHIA!\footnote{For this work we 
made use of version 6.4.24 with CDF tune A.} package~\cite{2006JHEP...05..026S}.
Eventually, the antiproton spectrum obtained after decay of the annihilation 
products is:
\ben
%\displaystyle
{\cal F}(E_{\bar{p}})  = \!\!
\displaystyle \sum_{i=1,2}\!\! \displaystyle \sum_{q} {\rm Br}_{i,q} 
\displaystyle \int_{\frac{E_1}{2} - {\cal E}}^{\frac{E_1}{2} + 
{\cal E}}\!\! 
\frac{f(E_q,E_{\bar{p}})+f(E_{\bar{q}},E_{\bar{p}})}{2 \,{\cal E}} 
dE_q\,, \nn\\
\een
where the first sum is done over the annihilation products (1,2)=($a,h$), and 
the second sum is over all those quark flavors for which 
$2 \, m_q \leqslant m_i$ is satisfied. The branching ratios ${\rm Br}_{i,q}$ can
be derived from the couplings to SM fermions given in~\citeeq{eq:coupl_phi_ff} 
and the decay widths given in~\citeeq{eq:widths}. For the quark
masses, we use the pole masses reported in~\eg~\cite{2010JPhG...37g5021N}.

\begin{figure*}[!t]
\centering
\includegraphics[width=\columnwidth]{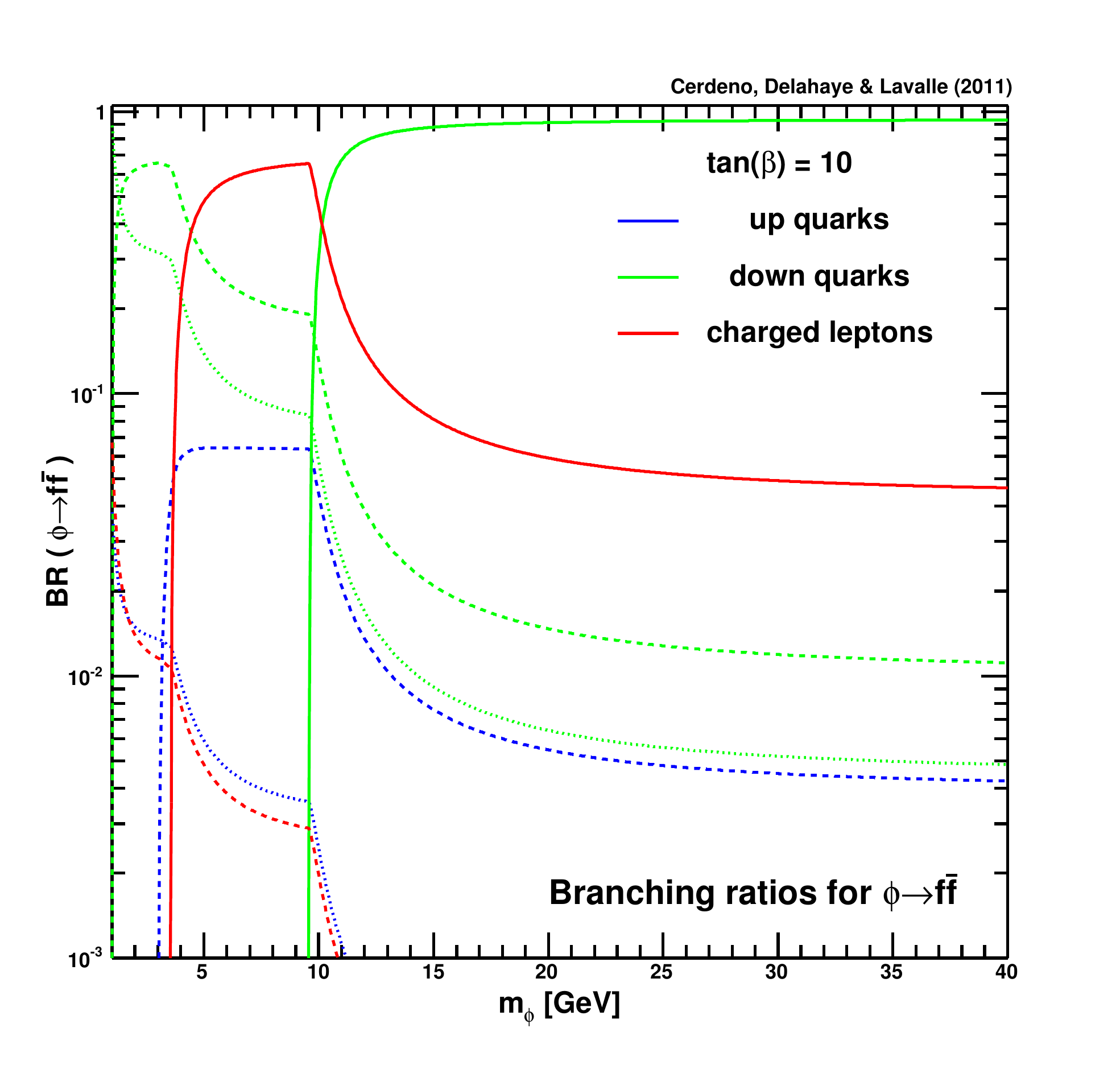}
\includegraphics[width=\columnwidth]{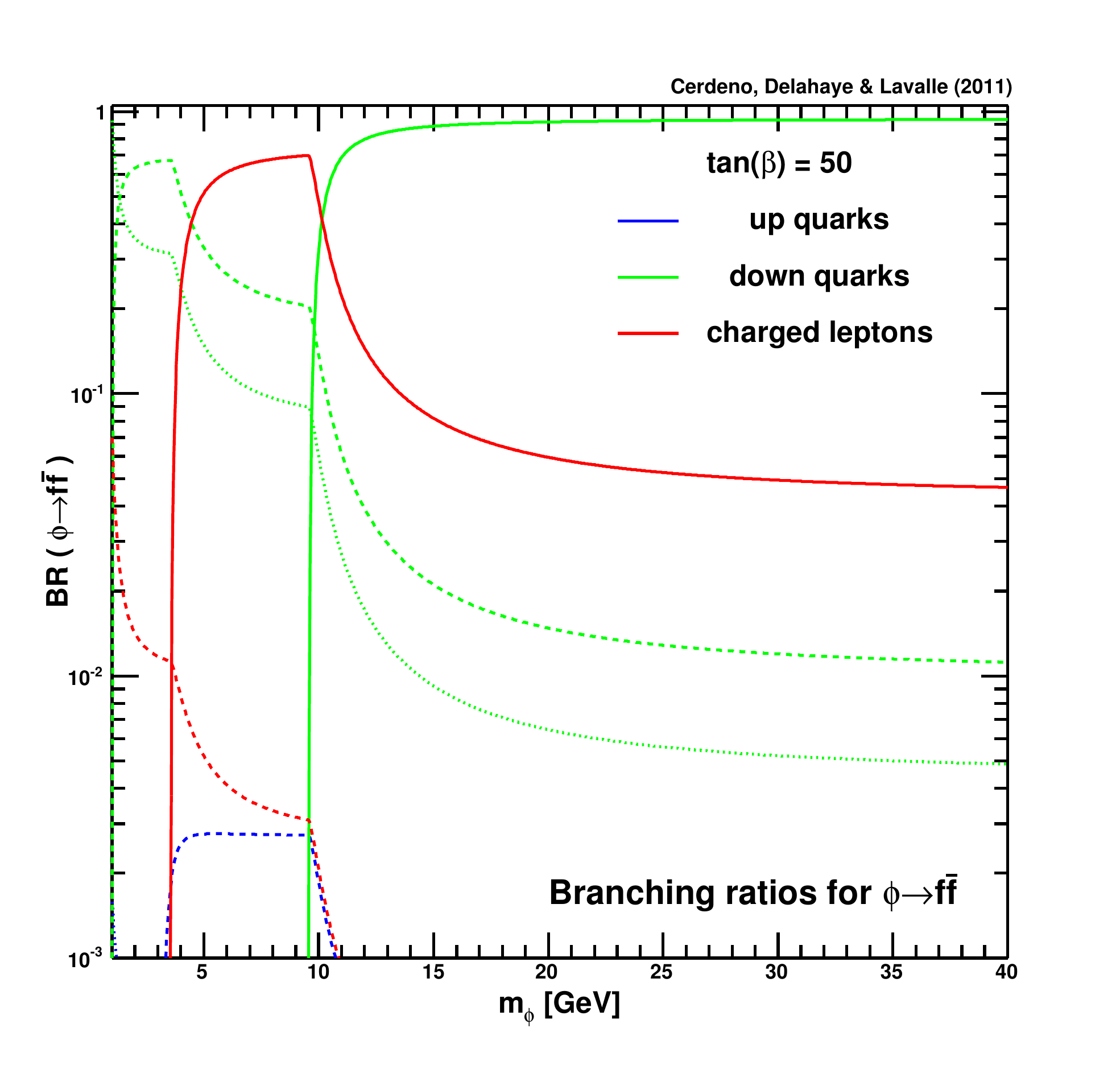}
\caption{Branching ratios for the (pseudo-)scalar $\phi$ decay into SM fermions
as a function of its mass $m_\phi$.
Contributions of the third, second and first generation fermions are reported 
with solid, dashed, dotted lines, respectively.}
\label{fig:br}
\end{figure*}

It is clear that the decay into the down-type $b$ quark-antiquark pairs will 
dominate whenever permitted. When $m_i<2\, m_b$, the dominant decay channel 
becomes $\phi_i\to \tau^+\,\tau^-$ because of the more favorable dependence into
$\tbe$ with respect to the up-type quark $c$. Therefore, the antiproton
production is reduced in the range $2\,m_\tau \lesssim m_i \lesssim 2\,m_b$, 
but not completely suppressed, since $d$ and $s$ quark contributions remain
significant. We have calculated the decay branching ratios 
explicitly in~\citefig{fig:br} assuming $\cphi=\cthe=0.9$ and $\tbe = 10$ (50) 
on the left (right) panel. Those associated with the third, second and first 
generation fermions are reported with solid, dashed, dotted lines, respectively.
These curves are relevant to $h$ and $a$, but we recall that we have only
constrained the mixing of $h$ so far. For simplicity, we take the same mixing
angles for $a$, which can be considered as conservative for the antiproton
production --- for example, a slight increase of the up Higgs doublet content
of $a$ would reduce the production of $\tau$ leptons.

Finally, we also consider the possibility that $h$ decays into 2 $a$, a
channel that is open when $m_h\geq 2\,m_a$. This channel is relevant when 
the coupling $\lambda_{aah}/m_h \gtrsim {\cal C}_{\phi q\bar{q}}$ 
(see~\citeeqp{eq:widths}). Given the range we have considered for $\lambda$ 
(see~\citeeqp{eq:rand}), this barely happens in our scan. Still, we note that if
this channel is open and dominant, then the upper limit on the mixing parameter 
$\sphi$ can be relaxed (see~\citesec{subsec:colliders}); for simplicity, we 
still do not relax it.

An illustration of the spectrum determination is presented 
in~\citefig{fig:spectra}, where we have used different combinations of the 
masses $\mchi$, $m_a$ and $m_h$ and applied the Lorentz boost procedure 
described above --- we did not include $h\to a\,a$. We have considered five 
different cases, with $\mchi = 12$ GeV for each. We first see that each time 
$m_a,m_h> 2$ GeV, the antiproton production is rather efficient and peaks with 
almost the same amplitude, within a factor of $\sim 3$, with a sharp decrease. 
The brown curve (third from top at peak energy) shows the case $\mchi=m_a=m_h$, 
which means that $a$ and $h$ both mostly decay at rest into $b$ quarks. 
Nevertheless, these quarks are created 
with a relatively small momentum ($E_b=m_{a,h}/2=6$ GeV), which reduces the 
proton and antiproton production efficiency. The violet (upper) curve shows a 
more favorable situation in which $m_a=12$ GeV and $m_h=4$ GeV,~\ie~both $a$ and
$h$ can decay into quarks with momenta large enough to fragment into antiprotons
efficiently --- the center of mass energy is $2\,\mchi$, and the production of 
$\tau$ is still not strongly dominant for the $h$ decay because very close to 
threshold (see~\citeeqp{eq:widths} and~\citefig{fig:br}). The pink curve
(second from top) characterizes an unexpected situation where, although $h$ and 
$a$ decays dominantly into $\tau$ leptons, the remaining $\sim 30\%$ going into
$s$ and $d$ quarks are very efficient at producing antiprotons thanks to their
large momenta. The orange curve (fourth from top) shows a less favorable case 
for antiproton production in which $a$ is not massive enough, and the $b$ 
quarks coming from the $h$ decay have low momenta, which translates into a 
factor of $\sim 2$ suppression with respect to the previous case. Finally, the 
blue (lower) curve corresponds to an even worse case, too small $m_a$ and $m_h =
10$ GeV, where $b$ quarks are inefficient at producing antiprotons in addition 
to threshold effects (the decay into $\tau$ comes into play --- 
see~\citefig{fig:br}).

These spectral differences that depend on the mass combination demonstrate
that there are non-trivial ways to suppress the antiproton production,
\eg~producing $b$ quarks at threshold with low momenta. Nevertheless, we will 
further show that each time $a$ or $h$ can decay into antiprotons efficiently, 
the antiproton flux is always in tension with the current data, provided the 
s-wave contribution to the annihilation cross section given in \citeeq{eq:swave}
is dominant at freeze out.

\begin{figure}[!t]
\centering
\includegraphics[width=\columnwidth]{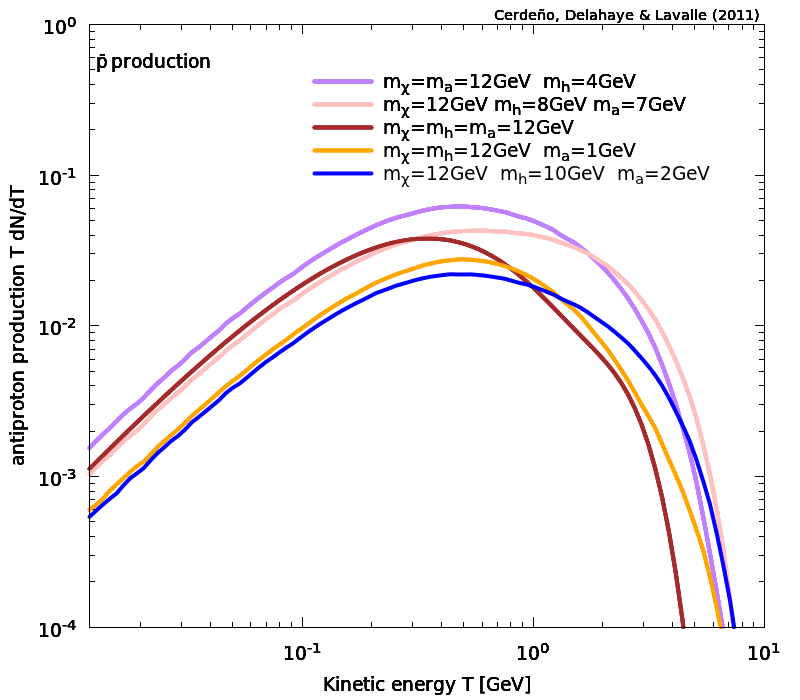}
\caption{Antiproton energy distributions for different mass combinations.}
\label{fig:spectra}
\end{figure}

\subsection{Antiproton flux at the Earth}
\label{subsec:pbar_flux}

\subsubsection{Dark matter halo}
\label{subsubsec:halo}

To predict the antiproton flux at the Earth, we need to specify the dark matter
distribution in the Galaxy. Many conventional dark matter halo functions
and related parameters have been widely used in the literature so far, but
we aim at employing a distribution as constrained as possible, both 
theoretically and observationally. From the theoretical point of view, there
seems to be a consensus in using an
Navarro-Frenk-White (NFW) profile~\cite{1997ApJ...490..493N,1996MNRAS.278..488Z}
or an Einasto profile~\cite{1968PTarO..36..357E,2006AJ....132.2685M}, which
are found to provide good fits to the dark matter distributions in 
cosmological N-body simulations (see~\eg~\cite{2010MNRAS.402...21N}). The
former is mildly cuspy in the center, behaving like $r^{-1}$, while the
latter asymptotically flattens towards the center.
Nevertheless, we stress that the very central parts of the Galaxy could
differ from what most of dark matter-only N-body simulations have predicted 
because baryons dominate the gravitational potential there, which might modify
the central behavior of the distribution~\cite{2010Natur.463..203G}. 
From the observational point of view, observations of low surface brightness 
galaxies also seem to indicate cores instead of cusps in these 
objects~\cite{2010AdAst2010E...5D}, showing that profiles are likely not
universal in the centers. Still, in contrast to gamma-rays, the antiproton 
signal is well known to have almost no dependence in the distribution in the 
Galactic center, provided it is less cuspy than $r^{-3/2}$ 
\cite{2004PhRvD..69f3501D,2007PhRvD..75h3006B,2010PhRvD..82h1302L}: this is due 
to diffusion effects. Much more important is the local dark matter density, 
which roughly sets the antiproton flux. It is therefore of stronger relevance 
to use a model dynamically constrained on kinematic data.

Recently, complementary efforts were undertaken to constrain the local density 
of dark matter $\rho_\odot$
\cite{2010JCAP...08..004C,2010A&A...523A..83S,2011MNRAS.tmp..553M},
the results of which all point to $\rho_\odot\simeq 0.4$ GeV/cm$^3$.
Here, we will use the spherically symmetric NFW profile as constrained in 
\cite{2011MNRAS.tmp..553M}, which we recall hereafter:
\ben
\rho(r) = \rho_\odot\left[\frac{r_\odot}{r}\right]
\left[\frac{1+r_\odot/r_s}{1+r/r_s}\right]^2\;,
\een
the variable $r$ denoting the galactocentric radius, and setting the 
parameters to $r_\odot = 8.28 \pm 0.16 $ kpc, $\rho_\odot = 0.40 \pm 0.04$ 
GeV/cm$^3$ and $r_s = 18\pm 4.3$ kpc. We emphasize that the antiproton flux will
be sensitive mostly to variations in $\rho_\odot$, not in the other parameters.

In principle, consistency imposes to consider dark matter 
substructures~\cite{1993ApJ...411..439S,2009NJPh...11j5027B}, which can
increase the antiproton flux at low 
energy~\cite{2008A&A...479..427L,2011PhRvD..83b3518P}. While the amplification
is expected to be small~\cite{2011PhRvD..83b3518P}, we stress that even
a factor of 2 could have important consequences in delineating the 
antiproton constraints, as will be shown later. Here, we will not consider
substructures, which can therefore be considered as a conservative approach.

\subsubsection{Antiproton transport and flux at the Earth}
\label{subsubsec:prop}

Once antiprotons are injected in the Galaxy, they diffuse on magnetic 
inhomogeneities and experience different additional processes. At low
energy, say below a few GeV, they are sensitive to Galactic winds which
translate into convection upwards and downwards the Galactic disk and into
adiabatic energy losses; they are also sensitive to nuclear interactions with 
the interstellar gas, which may destroy them and/or decrease their energy; 
finally, they can be reaccelerated because of the proper motion of the magnetic 
scatterers. Higher energy antiprotons are roughly sensitive to spatial diffusion
only. The general steady state equation that drives their evolution in 
phase-space reads:
\ben
\label{eq:diff}
\underbrace{{\cal Q} (\vec{x},E,t)}_{\rm source} =&
-\vec{\nabla}
\underbrace{\left\{ \left( K_{xx}(E)\vec{\nabla} -\vec{V}_c \right) 
{\cal N}\right\}}_{\text{spatial current}}
\\ 
&+
\partial_p 
\underbrace{\left\{ 
\left( \dot{p} - \frac{p}{3}\vec{\nabla}\cdot \vec{V}_c -  
p^2\,K_{pp}(E)\partial_p \frac{1}{p^2}\right)
{\cal N}\right\}}_{\text{current in momentum space}}
\nn\\
&+
\underbrace{\Gamma_s\,{\cal N}}_{\text{spallation}} \nn
\een
where ${\cal N} = {\cal N}(\vec{x},p) = dn/dp = \beta\, dn/dE$ is their 
differential density, $K_{xx}$ and $K_{pp}$ are respectively the diffusion 
coefficients in space and momentum, $V_c$ is the convection velocity and 
$\Gamma_s=\Gamma_s(\vec{x},p)$ the spallation rate. This equation can
be solved numerically~\cite{1998ApJ...509..212S} or semi-analytically 
\cite{berezinsky_book_90,1997A&A...321..434P,2001ApJ...555..585M}. In this 
study, we have used the semi-analytical method sketched 
in~\eg~\cite{2001ApJ...555..585M,2004PhRvD..69f3501D,2007PhRvD..75h3006B,2008A&A...479..427L}, to which we refer the reader for more details.
For the transport parameters, we have utilized the best-fit model found
in~\cite{2001ApJ...555..585M} and still in excellent agreement with more
recent analyses (\eg~\cite{2010A&A...516A..66P}). This transport model was 
dubbed {\em med} (for {\em median}) configuration 
in~\cite{2004PhRvD..69f3501D}. It is a slab transport model, with a cylindrical
diffusion volume of radius $R = 20$ kpc, half-height $L = 4$ kpc, in which the 
diffusion coefficient obeys 
$K_{xx}(E)=\beta\,K_0({\cal R}/ 1\, {\rm GV})^\delta$ (${\cal R}=p/q$ is the 
cosmic-ray rigidity). The main parameters are reported in~\citetab{tab:prop}.

\begin{table}
\centering
\begin{tabular}{ccccc}
\hline
\hline
$K_0$ & $\delta$ & $L$ & $V_c$ & $V_a$\\
${\rm kpc^2/Myr}$ & & kpc & km/s & km/s \\
\hline
0.0112 &0.70 &4 &12 &52.9\\
\hline
\end{tabular}
\caption{Transport parameters associated with the best-fit model 
of~\cite{2001ApJ...555..585M}.}
\label{tab:prop}
\end{table}

The source term is characterized by the specific dark matter model 
configuration:
\ben
{\cal Q}(\vec{x},E) = {\cal Q}(r,E)
= \frac{\sigv}{2} \left\{ \frac{\rho(r)}{\mchi}\right\}^2\frac{dN_{\rm dec}}{dE}
\;,
\label{eq:src}
\een
where $\sigv = a_{ah}$ is the annihilation cross section in the Galaxy given
in~\citeeq{eq:swave}, and $dN_{\rm dec}/dE$ is the antiproton spectrum arising
from the (pseudo-)scalar decay and determined in~\citesec{subsec:pbar_prod}.

Assuming a Green function ${\cal G}(\vec{x},p\leftarrow \vec{x}_s,p_s)$ solution
to~\citeeq{eq:diff}, the antiproton flux at the Earth is given by the 
following expression:
\ben
\phi_\odot(E) &= \frac{\beta\,c}{4\,\pi}
\int dp_s \int_{\rm slab} d^3\vec{x}_s \,
{\cal G}(\vec{x},p\leftarrow \vec{x}_s,p_s) \, {\cal Q}(\vec{x}_s,E_s)\,.\nn\\
& \label{eq:flux}
\een
In practice, we use the Bessel series method presented 
in~\cite{2004PhRvD..69f3501D,2007PhRvD..75h3006B}, but we have checked with a 
couple of configurations that results obtained from Green function 
methods~\cite{2007PhRvD..75h3006B,2008A&A...479..427L} are equivalent.

\subsection{Results and discussion}
\label{subsec:results}

\begin{figure*}[!t]
\includegraphics[width=\columnwidth]{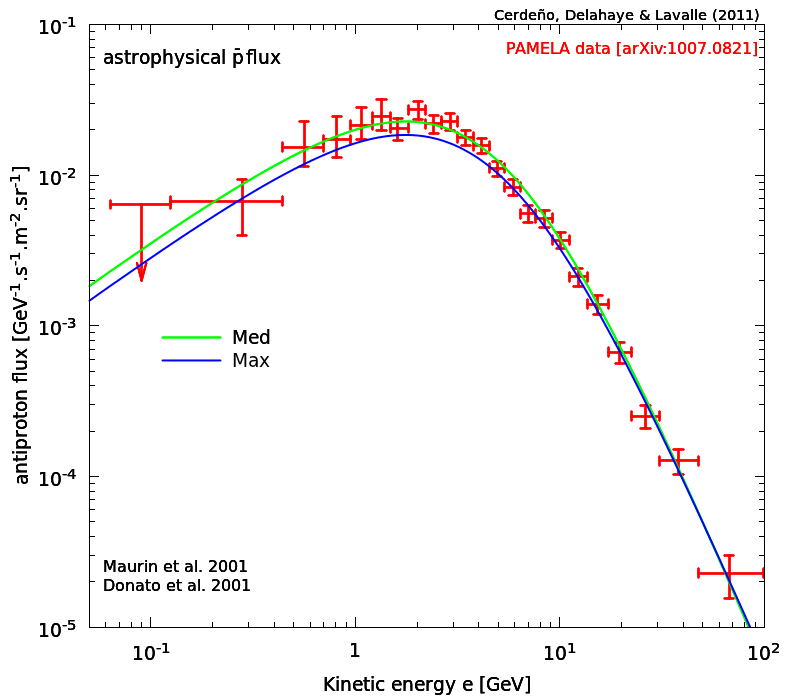}
\includegraphics[width=\columnwidth]{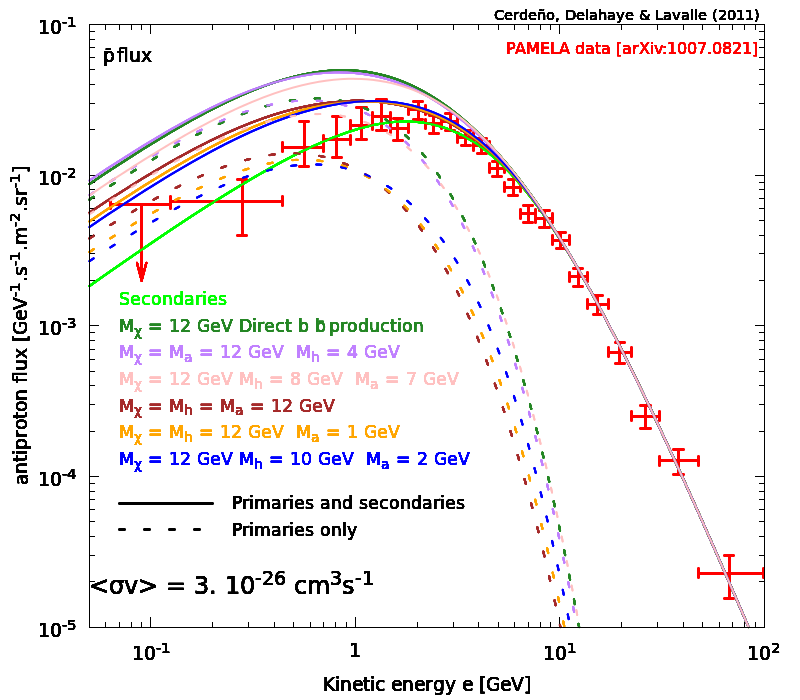}
\caption{Left: secondary antiproton flux (background) and rough theoretical 
uncertainties. Right: examples of primary fluxes with the same scalar
and singlino mass combinations as in~\citefig{fig:spectra}; they are compared
to a $b\bar{b}$ spectrum similar to the one used in~\cite{2010PhRvD..82h1302L}.}
\label{fig:fluxes}
\end{figure*}

\begin{figure*}[!t]
\includegraphics[width=0.48\textwidth]{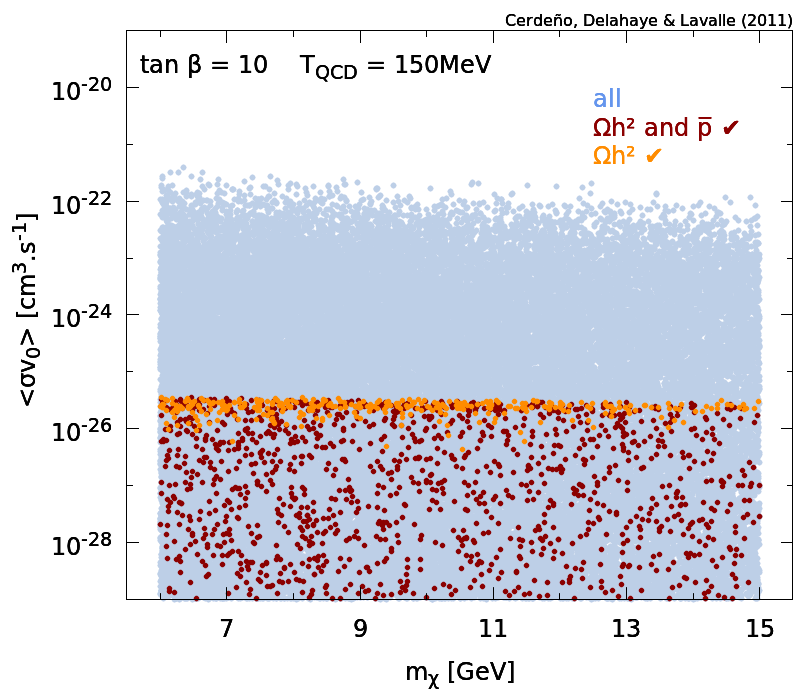}
\includegraphics[width=0.48\textwidth]{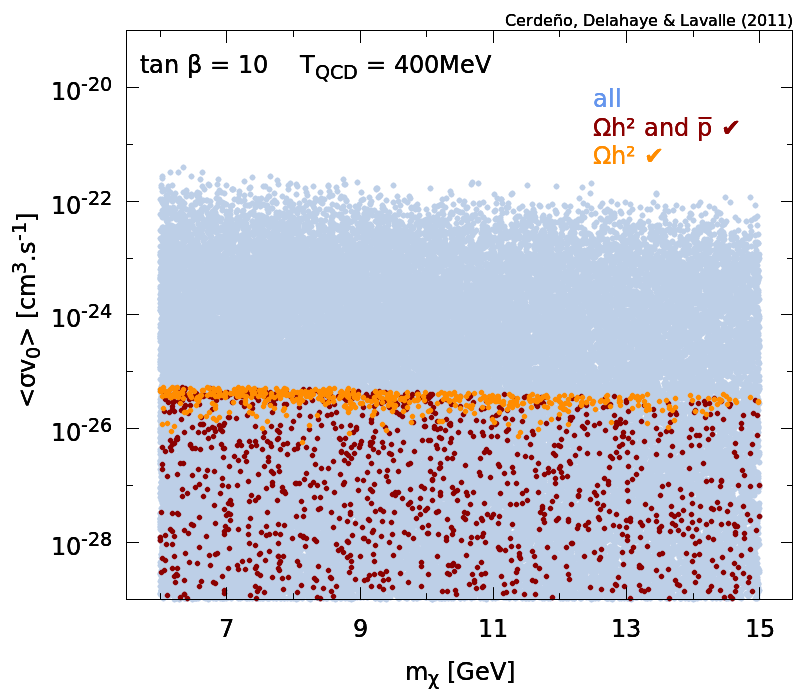}\\
\includegraphics[width=0.48\textwidth]{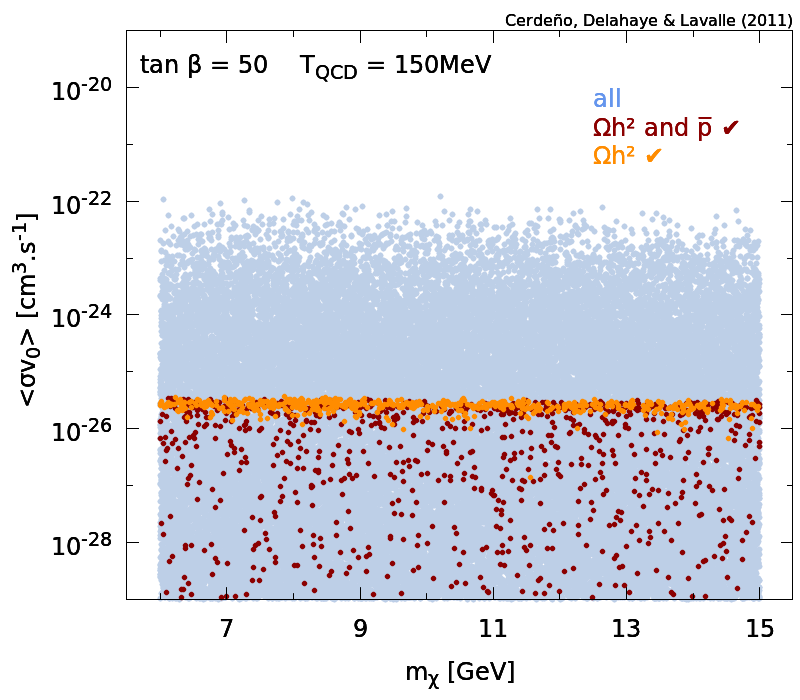}
\includegraphics[width=0.48\textwidth]{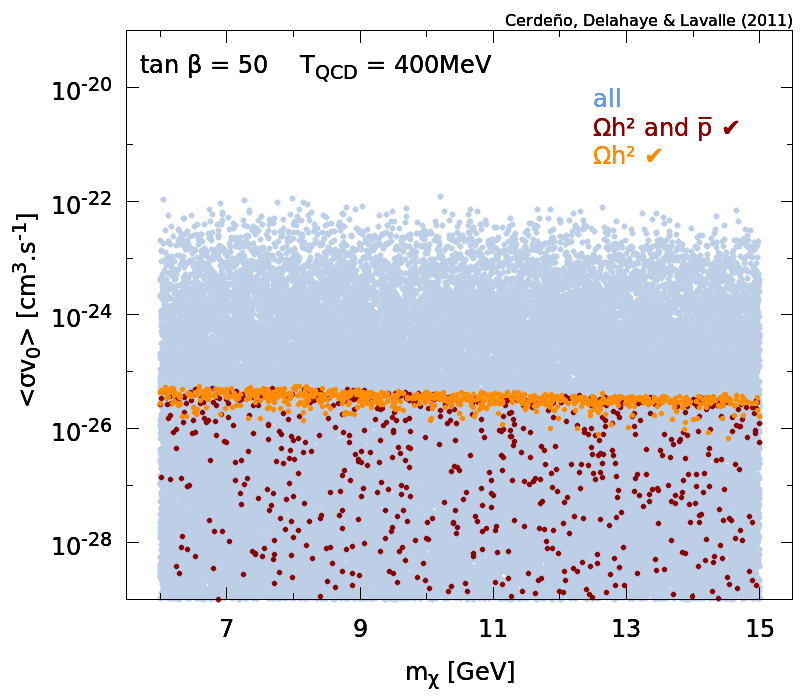}
\caption{CoGeNT points in the plane $\sigv$ (S-wave only) versus $\mchi$. Top 
panels: $\tbe = 10$. Bottom panels: $\tbe = 50$. Left panels: $\tqcd = 150$ MeV.
Right panels: $\tqcd = 400$ MeV. Dark red points are cosmologically allowed,
light orange points are cosmologically allowed but lead to an antiproton 
excess.}
\label{fig:s0m}
\end{figure*}

\begin{figure*}[!t]
\includegraphics[width=\columnwidth]{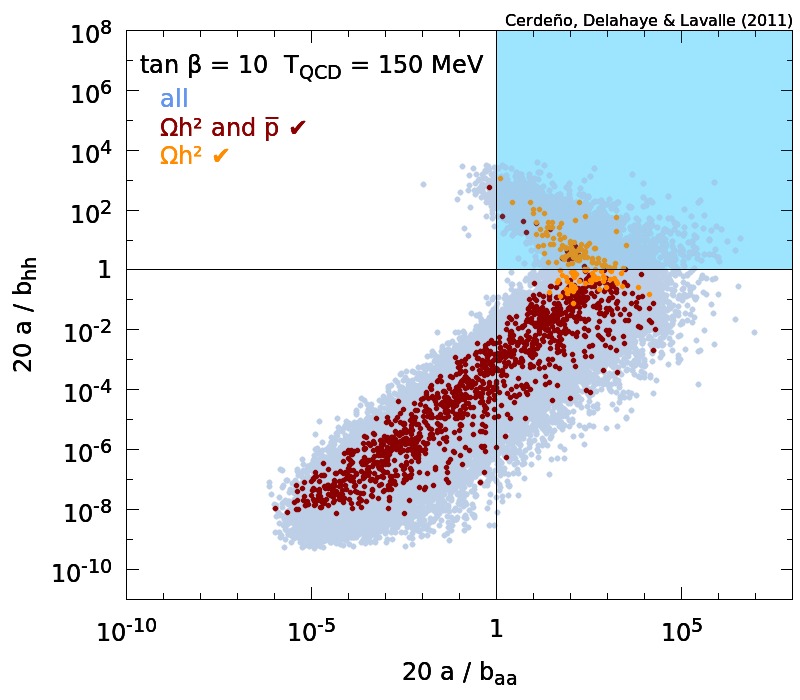}
\includegraphics[width=\columnwidth]{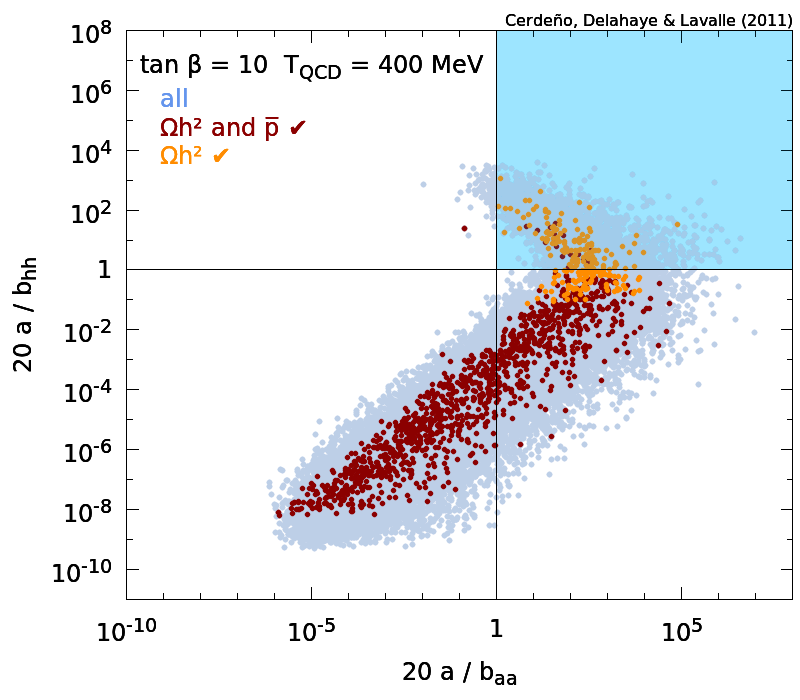}\\
\includegraphics[width=\columnwidth]{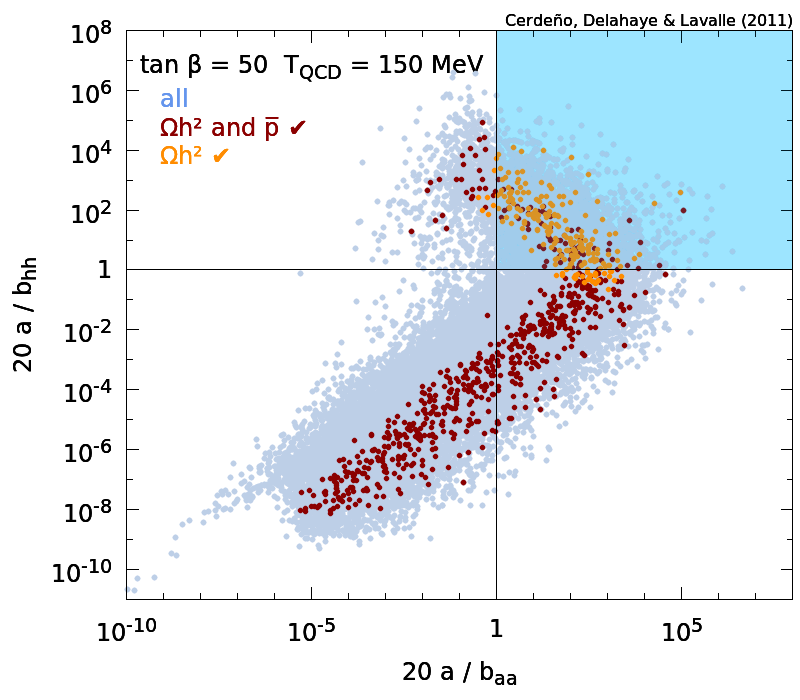}
\includegraphics[width=\columnwidth]{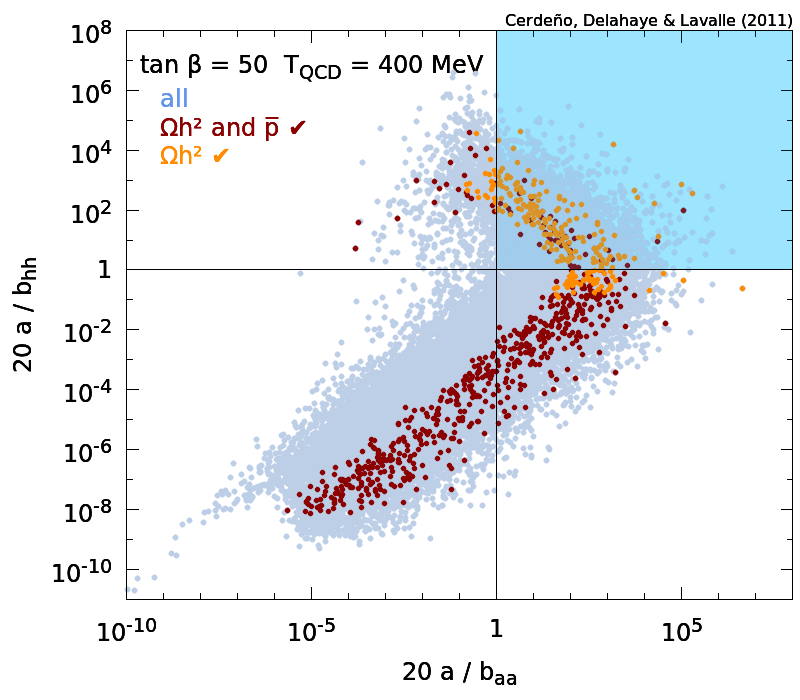}
\caption{S-wave to P-wave ratio, $a_{ah}/b_{aa}$ versus $a_{ah}/b_{hh}$, as
calculated from Eqs. (\ref{eq:swave},\ref{eq:baa},\ref{eq:bhh}).}
\label{fig:ab}
\end{figure*}

\begin{figure*}[!t]
\includegraphics[width=\columnwidth]{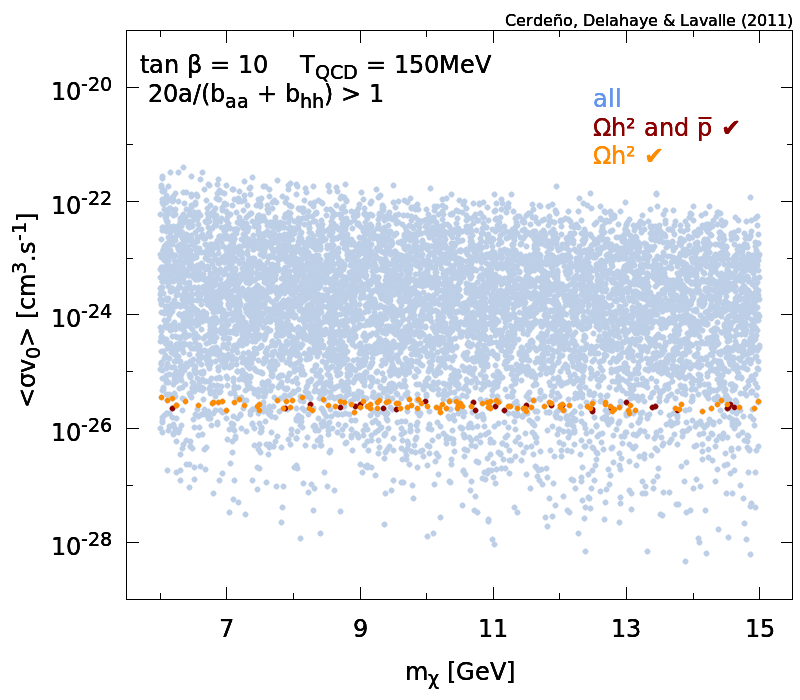}
\includegraphics[width=\columnwidth]{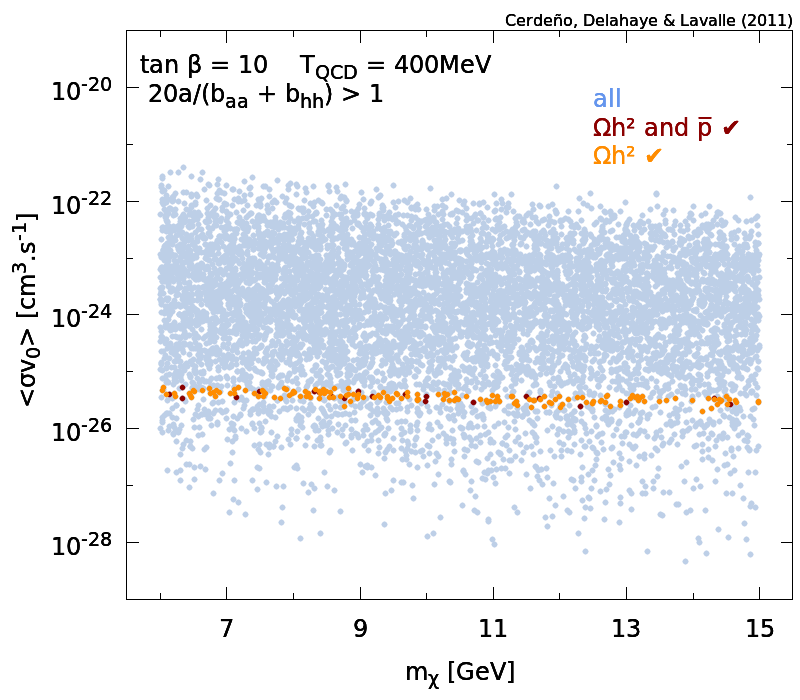}\\
\includegraphics[width=\columnwidth]{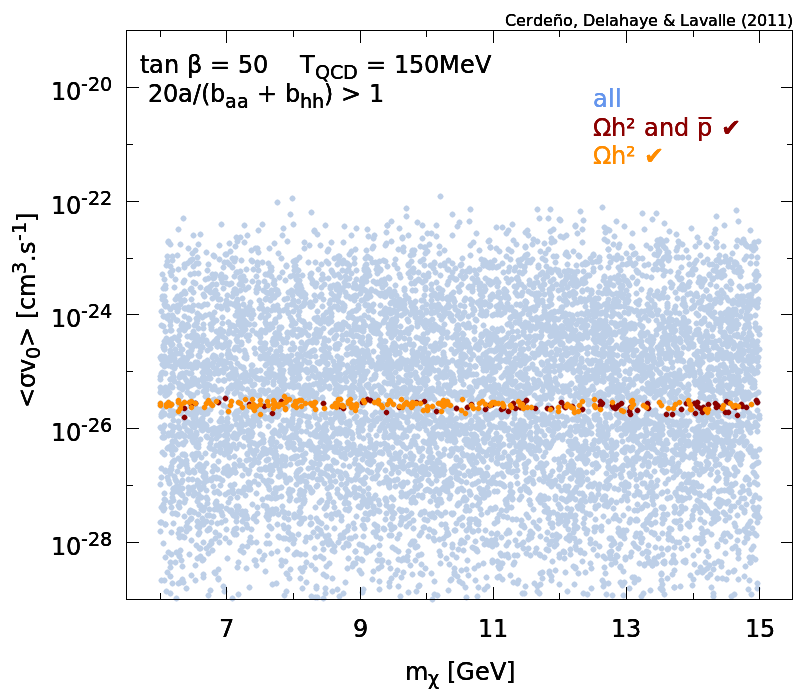}
\includegraphics[width=\columnwidth]{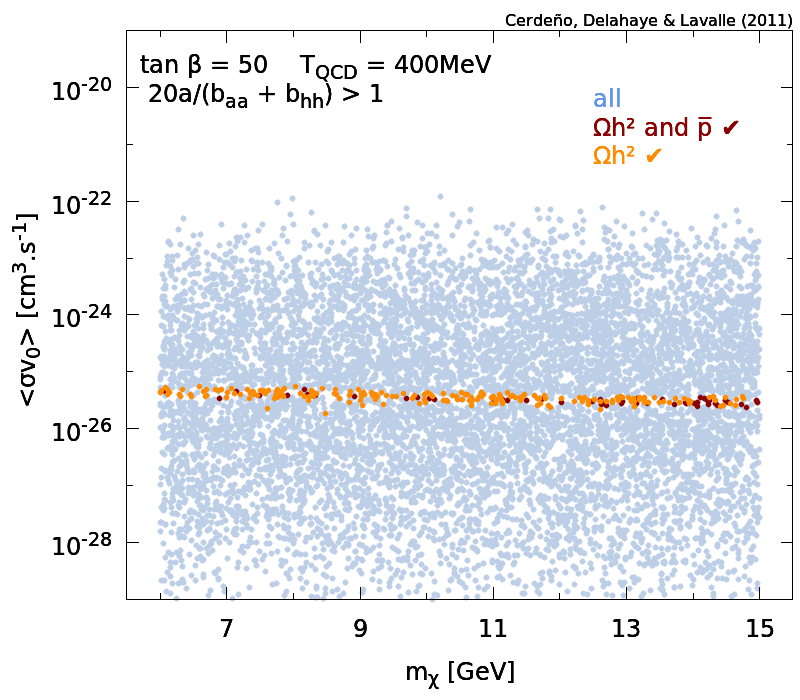}
\caption{Same as~\citefig{fig:s0m} after applying the cut $SP_{\rm tot}>1$ 
(see~\citeeqp{eq:spratio}).}
\label{fig:s0ma}
\end{figure*}

\begin{figure*}[!t]
\includegraphics[width=0.5\columnwidth]{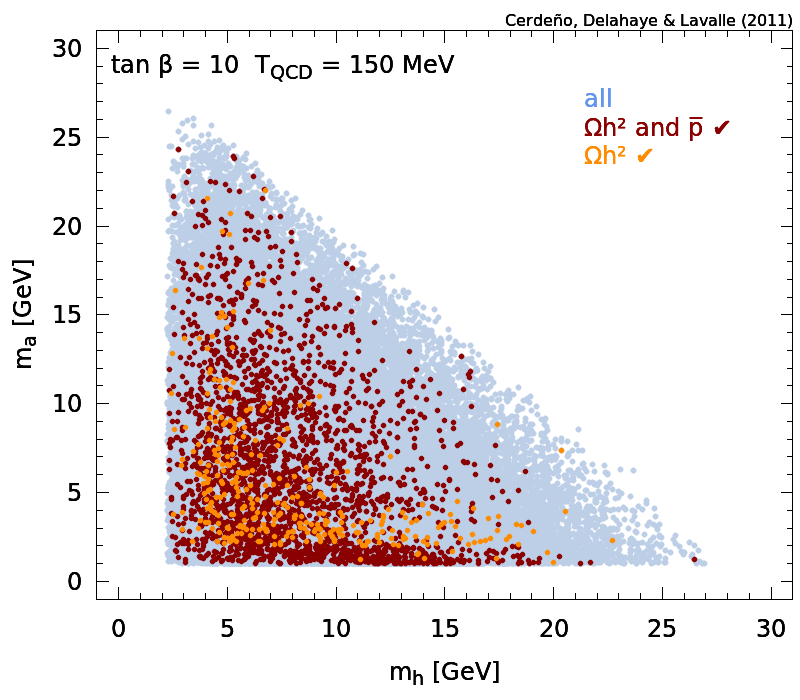}
\includegraphics[width=0.5\columnwidth]{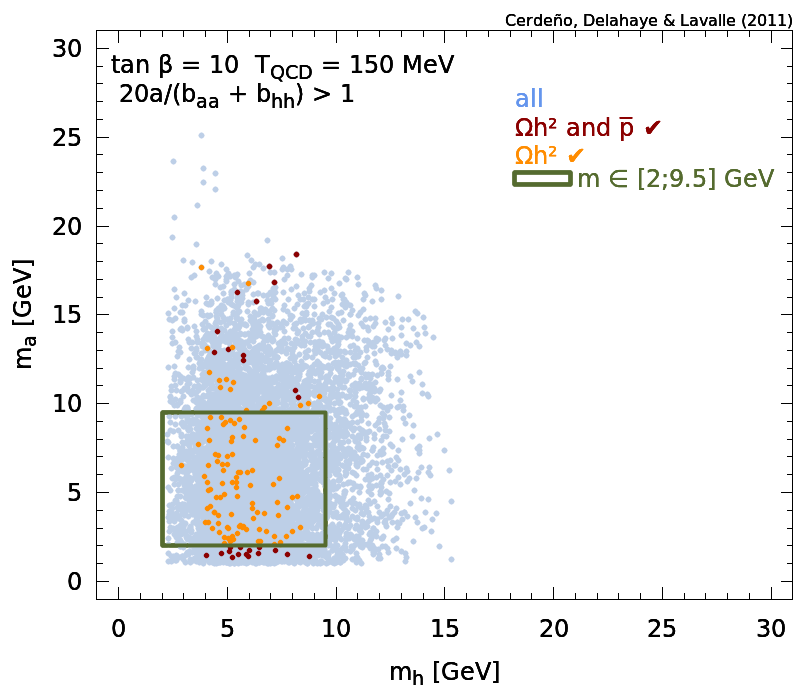}
\includegraphics[width=0.5\columnwidth]{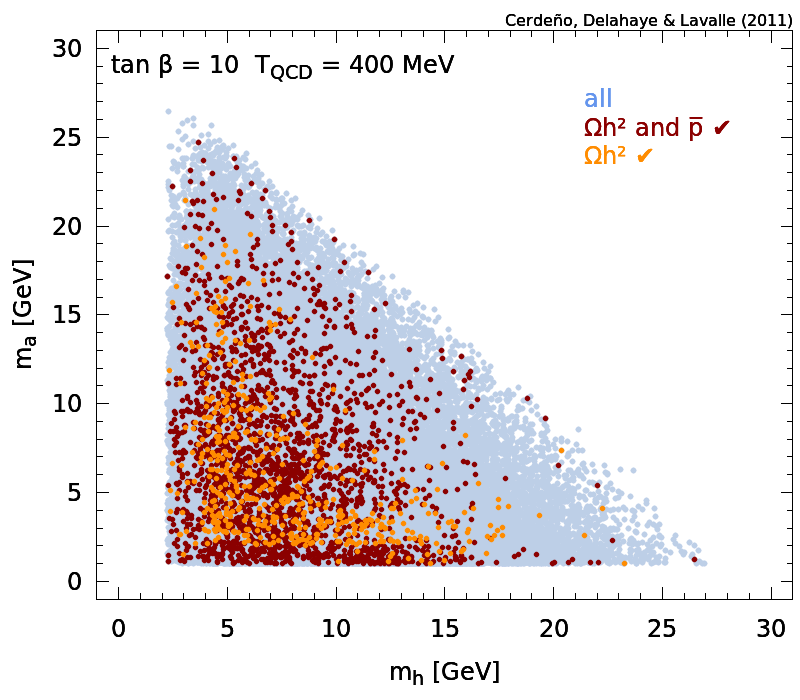}
\includegraphics[width=0.5\columnwidth]{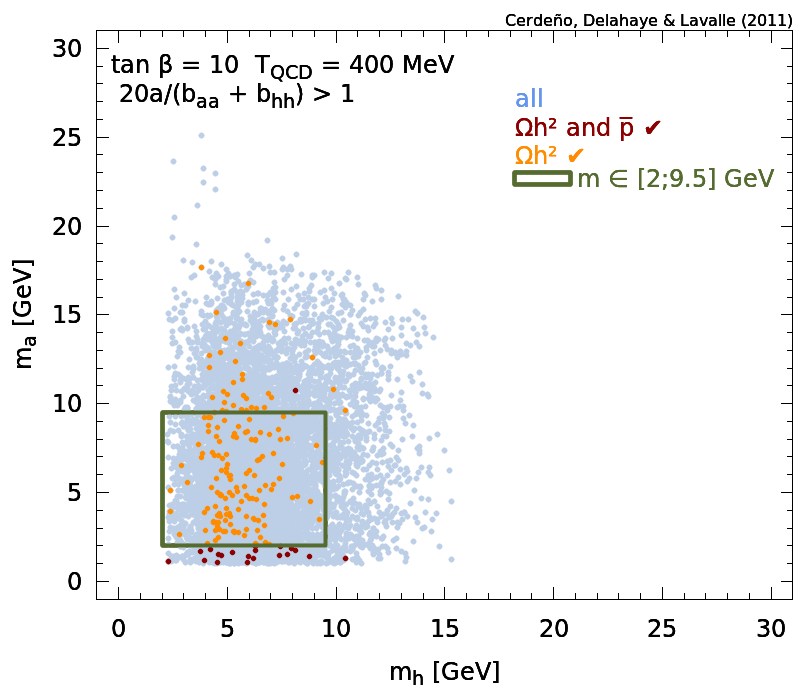}\\
\includegraphics[width=0.5\columnwidth]{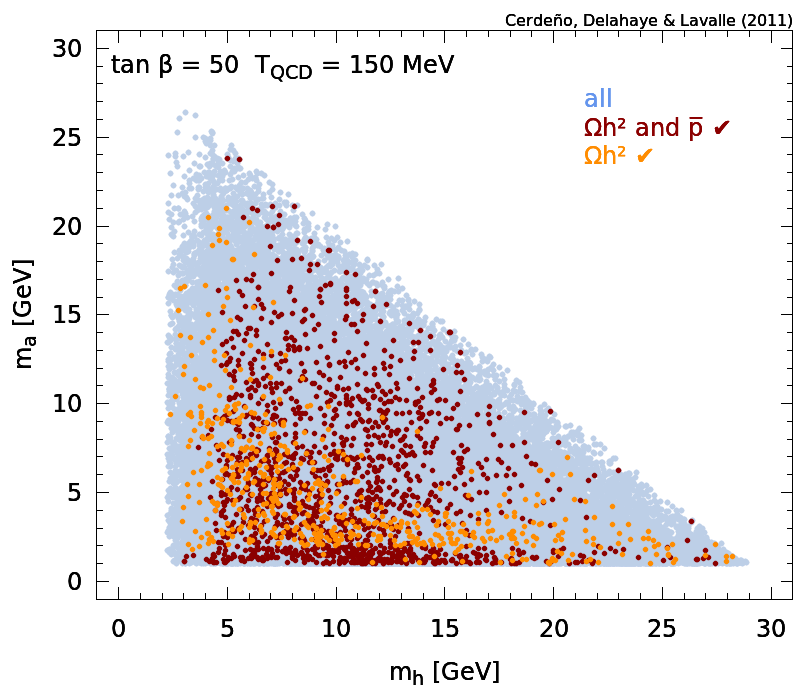}
\includegraphics[width=0.5\columnwidth]{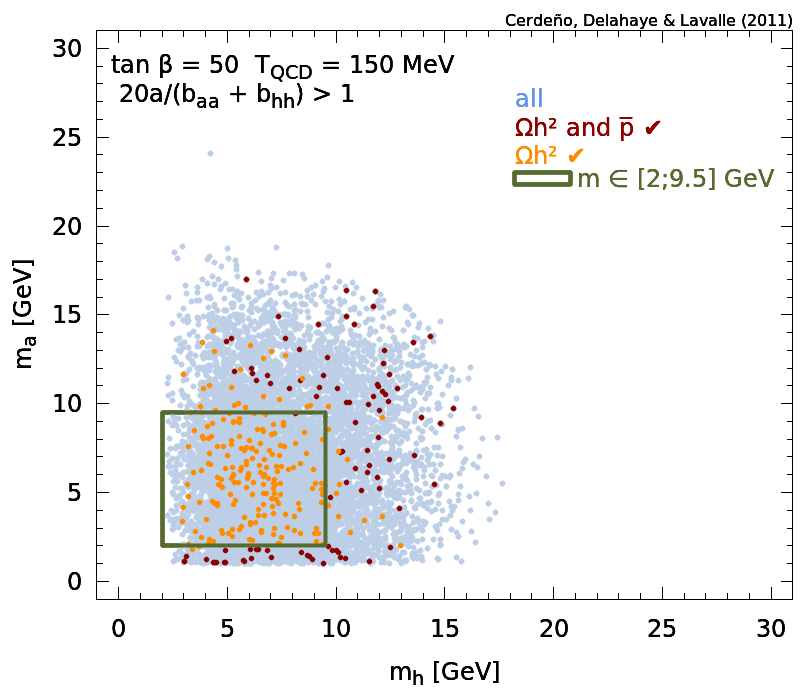}
\includegraphics[width=0.5\columnwidth]{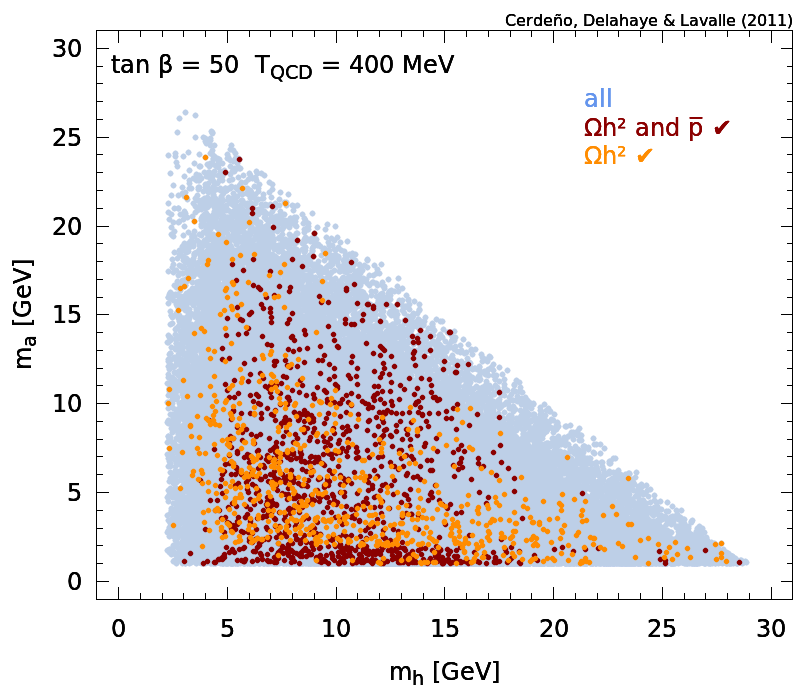}
\includegraphics[width=0.5\columnwidth]{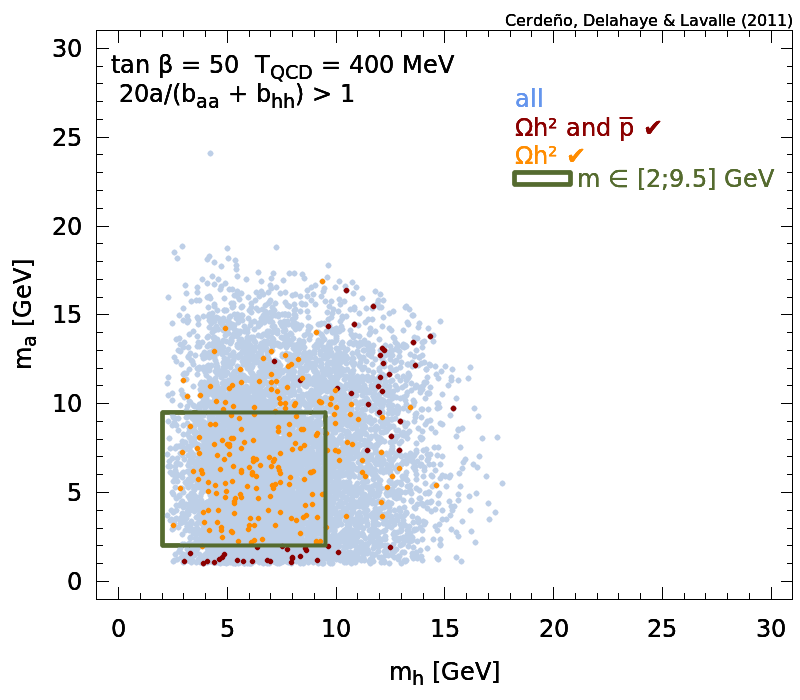}
\caption{Scan results in the plane $m_h$-$m_a$ for $\tbe=10/50$ on 
top/bottom panels, and $\tqcd=150/400$ MeV on left/right 2 columns. A cut 
$SP_{\rm tot}>1$ (see~\citeeqp{eq:spratio}) was applied in the even column 
panels.}
\label{fig:mhma}
\end{figure*}

We have performed a scan over the masses and couplings associated with
\citeeq{eq:lag} such that we keep only those configurations falling in the
so-called CoGeNT region (see~\citesec{sec:direct}). For these configurations,
we have computed the annihilation cross section summing over the three
diagrams of~\citefig{fig:diagrams} for all possible final states,~\ie~$a\,a$,
$h\,h$ and $a\,h$ as soon as kinematically allowed --- we recall that we have
imposed $2\,\mchi\geq m_a+m_h$. Then, we have derived the relic density assuming
different QCD phase transition temperatures, and computed the antiproton fluxes 
for models leading to the correct relic 
abundance,~\ie~$0.0941 \leq \Omega_\chi h^2\leq 0.1277$ in this 
study (see~\citesec{subsec:relic}). The flux is computed assuming the transport
parameters of~\citetab{tab:prop}, and a solar modulation is subsequently 
accounted for in the force-field 
approximation~\cite{1968ApJ...154.1011G,1971JGR....76..221F}, with a Fisk 
potential of $\Phi = 600$ MV, consistent with the average solar activity 
during the PAMELA data taking~\cite{2010PhRvL.105l1101A}.
For the secondary antiproton flux, we use the predictions obtained in
\cite{2001ApJ...563..172D} with the same transport parameters. It is represented
in the left panel of~\citefig{fig:fluxes} together with a prediction using 
another parameter set characterized by a larger diffusion halo with $L=15$ kpc. 
Both predictions are consistent with the current PAMELA data
\cite{2010PhRvL.105l1101A}, while the latter transport model would lead to
a significantly larger primary antiproton flux because integrating over much 
more dark matter annihilation products. It turns out that large diffusion halo 
models,~\ie~with $L\gtrsim 5$ kpc, are favored by several recent studies on 
Galactic cosmic-ray transport~\cite{2010A&A...516A..66P,2011ApJ...729..106T}, 
while small halo models with $L\lesssim 2$ are strongly disfavored from 
complementary observations~\cite{berezinsky_book_90,2007ARNPS..57..285S,2010PhRvD..82h1302L} and are to be considered as extreme cases --- we complete this
discussion in~\citesec{subsubsec:unc}. Therefore, we will 
stick to our median transport model which can be thought of as rather 
conservative.

We finally confront the predicted antiproton fluxes with the PAMELA 
data\footnotemark. An error box is defined as having the energy bin 
width and the 1-$\sigma$ error bar half-height. 
\footnotetext{Other data are not considered because data taken from space are 
supposed to have smaller systematic errors than balloon data for a given 
statistics; moreover, PAMELA operated during low solar activity periods, 
which is optimal for low energy cosmic ray studies.}
A model is considered in tension with the data when the predicted flux (adding 
up the primary and secondary contributions) is above an error box without 
crossing it\footnotemark. A few examples are shown in the right panel of
\citefig{fig:fluxes}, where the same injected spectra as 
in~\citefig{fig:spectra} have been used an compared to the case of an 
annihilation into $b\bar{b}$. A conventional annihilation cross section of
$\sigv = 3\times 10^{-26}\,{\rm cm^3/s}$ was assumed, for which all models
appear in excess with respect to the data. This clearly illustrates the 
rejection power of the antiproton analysis. In the following we present the
results we obtained from scans that cover the CoGeNT region, as detailed
in~\citesec{sec:direct}.

\footnotetext{We have dismissed those data points which systematically lie
below the secondary flux prediction by more than a standard deviation.}

\subsubsection{CoGeNT region scan analysis}
\label{subsubsec:scanres}

%%% large tbe => smaller c_chichih/mh2 required for cogent 
%%% => larger c_chichia/mchi to get correct relic abundance => chichi->ah
%%% important.
%%% => larger mh.

\citefig{fig:sigsi} shows the results of two scans we performed to cover the
CoGeNT region in the plane $\mchi-\sigma^{\rm SI}$. Dark red points are those
which fulfil the relic density constraint, and the light orange points are
those in excess with respect to the antiproton data. Both populations can 
roughly cover the whole region, for small (upper panels) as well as for large 
values (lower panels) of $\tbe$, independently of $\tqcd$. Still, taking small 
values of $\tbe$ results in more cosmologically relevant configurations in 
average, as it can be observed. Indeed, $\tbe=10$ implies a coupling 
$c_{\chi h}$ a factor of $\sim 10$ larger than in the case of 
$\tbe=50$, for a given spin-independent cross section 
(see~\citeeqp{eq:approxsigma}), which provides more room to get the correct 
relic density only from the annihilation into $h\,h$. Modifying the QCD phase
transition temperature leaves these statements valid (see left/right panels). 
More generally, \citefig{fig:sigsi} demonstrates that we cannot draw a generic 
antiproton exclusion contour in this plane, since many configurations pass 
both the cosmological and the antiproton tests. Nevertheless, it shows that 
the antiproton constraints are relevant {\em everywhere} in this plane, which is
by itself an important result. In the following, we provide details about the 
nature of those points which are (or not) in excess with respect to the 
antiproton data.

In \citefig{fig:s0m}, we replot the scan results of~\citefig{fig:sigsi} in the 
plane $\sigv_0$-$\mchi$. Note that $\sigv_0$ refers to the annihilation cross 
section relevant to indirect detection, \ie~the velocity-independent 
contribution given in~\citeeq{eq:swave}, while the full velocity-dependent 
annihilation cross section was used in the relic density calculation. We 
remark that the relic density constraint obviously translates into an upper 
bound $\sigv_0\lesssim 5(8)\times 10^{-26}\,{\rm cm^3/s}$ for $\tqcd=150(400)$ 
MeV, but that the cosmologically viable points spread down to arbitrarily small 
values of $\sigv_0$ that correspond to cases where the P-wave is dominant, 
\ie~where annihilation proceeds mostly into $h\,h$ or $a\,a$ at 
freeze out. This demonstrates that the kinematic condition $2\,\mchi\geq m_a+
m_h$ does not generically ensure that annihilation into $a\,h$ dominates. 
Indeed, suppressing the coupling between the singlino and the pseudo-scalar $a$ 
is enough to suppress the S-wave, as shown in~\citeeq{eq:swave}. We will come 
back to this later. The population located close to the upper cosmological bound
is featured by those configurations for which annihilation is saturated by the 
S-wave,~\ie~annihilation into $a\,h$. This is precisely where the antiproton 
constraint is the strongest. There is still a significant population that 
remains unaffected by the antiproton constraints in this region. This actually 
corresponds to cases where the decays of $a$ and $h$ do not produce antiprotons 
efficiently (see discussion in \citesec{subsec:pbar_prod}). Eventually, the 
impact of changing $\tqcd$ is a bit harder to figure out from our scan results. 
Typically, considering large values of $\tqcd$ implies a larger cross section at
freeze out over a broader WIMP mass range, \ie~when $\mchi/x_f \lesssim \tqcd$ 
(see~\citefig{fig:rd}). This means that going from $\tqcd = 150$ to 400 MeV, we 
increase the required annihilation cross section by a factor of $\sim 2$ in the 
singlino mass range $\sim$4-12 GeV (see~\citefig{fig:rd}).

A way to identify the impact of the S-wave content of the annihilation cross
section on the antiproton constraint strength consists in determining the
S-wave to P-wave ratio, as shown in~\citefig{fig:ab}. Each panel illustrates the
ratio $a_{ah}/b_{hh}$ versus the ratio $a_{ah}/b_{aa}$ for different combinations
of $\tbe$ (10/50 on top/bottom panels) and $\tqcd$ (150/400 MeV on left/right 
panels). The additional factor of 20 applied to the ratios comes from that we 
want to estimate their values close to the freeze out, which occurs around
$x_f=\mchi/T\sim 20$ (see \citeeqp{eq:sigv}). Let us denote
\ben
\label{eq:spratio}
\spaa &=& 20\,a_{ah}/b_{aa}\\
\sphh &=& 20\,a_{ah}/b_{hh}\nn\\
\sptot &=& 20\,a_{ah}/(b_{aa}+b_{hh})\;,\nn
\een
where $a_{ah}$, $b_{aa}$, and $b_{hh}$ are defined in 
Eqs. (\ref{eq:swave},\ref{eq:baa},\ref{eq:bhh}), respectively.
In each panel of~\citefig{fig:ab}, going from bottom to top means decreasing 
the relative contribution of the annihilation into $h\,h$, while going from left
to right means decreasing the relative contribution of the annihilation into 
$a\,a$. 
The top left part of each panel, where annihilation into $a\,a$ is 
always dominant with respect to other channels, is barely populated. In 
contrast, there is a dense wake of points going diagonally from the bottom right
part to the bottom left part, which is characteristic of an annihilation driven 
by $h\,h$ production; indeed, as long as $\sphh<\spaa<1$, this channel overtops 
the others\footnote{Note that the sharp trend is reminiscent from the 
correlation arising between couplings $\lambda$ and ${\cal C}_{\chi h}$ (see 
\citeeqp{eq:rand} and discussion below); the wake would have been more spread 
without this correlation.}. This implies that annihilation into $a\,a$ is 
always subdominant in our sample.
This is partly a consequence of requiring the CoGeNT 
region to be fulfilled, which constrains the coupling $c_{\chi h}$ to be large, 
while $c_{\chi a}$ is left free; besides, we see that going to larger values
of $\tbe$ slightly alleviates this effect, as expected 
(see~\citeeq{eq:ddcondition}). 
Another cause is less trivial: increasing $c_{\chi a}$ does not only increase 
annihilation into $a\,a$, but also that into $a\,h$ (see~\citeeqp{eq:swave}); 
the latter is always kinematically allowed here, but not necessarily the former.

\citefig{fig:ab} further lets us distinguish a particular area, made shaded in
each panel, that corresponds to $SP_{aa}>1$ and $SP_{hh}>1$. This area is the
region where the S-wave saturates the full annihilation cross section at freeze 
out, and is therefore fully fixed by the relic density (optimal case for 
indirect detection). Naturally, the largest fraction of the models inducing an 
excess in the antiproton flux (light orange points) lies in this region. 
Nevertheless, there are still some configurations unaffected by the antiproton 
constraints there, which are actually those for which $a$ and $h$ are too light
to decay into antiprotons (through quark fragmentation), or with masses 
$\gtrsim$10 GeV for which the decay into $b$ quarks occurs at threshold and is 
is then inefficient at producing antiprotons ($\sim$[9.5,12] GeV, 
see~\citesec{subsec:pbar_prod}). In the latter case, $\mchi$ usually takes 
rather large values because of our kinematical condition, which also decreases
the antiproton flux (scaling like $1/\mchisq$). We note that the density of 
unconstrained configurations increases with $\tbe$, which comes from three 
effects. First, the up/down asymmetry in the branching ratio is strongly 
dependent in $\tbe$, and taking a large $\tbe$ increases the relative decay into
$\tau$ leptons compared to that into $c$ and $s$ quarks 
(see \citeeqp{eq:widths} and~\citefig{fig:br}). Second, a larger value of 
$\tbe$ induces an increase in $m_h$ for a given spin-independent
cross section (see~\citeeqp{eq:ddcondition}), and either the density of points 
in the mass range $m_h\sim$10-12 GeV increases accordingly and $\mchi$ is pushed
towards larger values. Third, a larger value of $\tbe$ may also translate into 
a smaller $\chi\,\chi\,h$ coupling, leaving more room to the $h\,a\,a$ coupling 
to set the relic density from the $s$-channel (see~\citeeqp{eq:swave} and the 
left diagram of~\citefig{fig:diagrams}). The consequence is that $h\to a\,a$ 
opens up when masses permit, which may further reduce the antiproton production 
efficiency.

In any case, the plots of~\citefig{fig:ab} show that the antiproton constraints 
can be generically very strong whenever $SP_{aa}\gtrsim 1$ and 
$SP_{hh}\gtrsim 1$, as expected --- one can safely use 
Eqs. (\ref{eq:swave},\ref{eq:baa},\ref{eq:bhh}) to account for this constraint.

In~\citefig{fig:s0ma}, which depicts the indirect detection annihilation cross 
section $\sigv_0$ as a function of the singlino mass $\mchi$, we report the same
samples as in~\citefig{fig:s0m} after applying the cut $SP_{\rm tot}>1$ 
(see~\citeeqp{eq:spratio}). We clearly see that this cut selects all points 
for which the relic density completely sets $\sigv_0$, a large fraction of which
turns out to lead to an antiproton excess. The remaining unconstrained cases are
mostly characterized either by $m_a\lesssim 2$ GeV, for which $a$ cannot decay 
into antiprotons, or the condition $\mchi \lesssim 10$ GeV, mostly due to the
antiproton flux dependence on $1/\mchisq$ (see~\citeeqp{eq:src}). This appears
more clearly in the plane $m_h$-$m_a$ shown in~\citefig{fig:mhma}, where our 
results are presented with and without the cut $SP_{\rm tot}>1$ --- 
a rectangle is drawn in each panel to emphasize the most constrained 
mass range. We see that
once the cut is applied, all models exhibit an excess with respect to the 
antiproton data but those with $m_a\lesssim 2$ GeV, and $(m_a+m_h) \lesssim 20$ 
GeV --- the latter condition is equivalent to requiring $\mchi \lesssim 10$ GeV.

The dependence on $\tbe$ and $\tqcd$ arises in 
all~Figs.~\ref{fig:s0m},~\ref{fig:ab},~\ref{fig:s0ma} and~\ref{fig:mhma}. Small 
$\tbe$ values are 
generically more constrained by the antiproton data because they lead to 
relatively small values of $m_h$ for a given spin-independent cross section
(see~\citeeqp{eq:approxsigma}), thus favoring smaller values of $\mchi$ which 
increases the antiproton flux. In this regime, the $a$/$h$ decay into 
$\tau$ leptons is also slightly less significant than in the large $\tbe$
regime. As to the impact of $\tqcd$, things are much simpler. The cross section
needs to be relatively larger for $\mchi \lesssim 20\,\tqcd$, which makes
the indirect detection constraints more stringent for large values of $\tqcd$.

Coming back to the full results shown in~\citefig{fig:s0m} in the light of the
previous discussion, we may summarize our results with a generic upper limit on 
the annihilation cross section relevant to the singlino-like phenomenology as
\ben
\label{eq:uplim}
\sigv_0 & \lesssim & 10^{-26}\,{\rm cm^3/s} 
\,\left[\frac{\rho_\odot/0.4\,{GeV/cm^3}}{\mchi/10 \, {\rm GeV}}\right]^2\\
{\rm for} &&
\begin{cases}
\; m_a,m_h \in [2,9.5] \, {\rm GeV}\\
SP_{\rm tot}\gtrsim 0.3
\end{cases}
\;.\nn
\een
This is not an accurate limit, but rather a way to avoid regions of the 
parameter space that could lead to an antiproton excess. This is illustrated
in~\citefig{fig:optcut}, where we used the worst situation for the antiproton
constraints, $\tbe = 50$ and $\tqcd = 150$ MeV. We see that a looser cut of
$SP_{\rm tot}>0.3$ still leads to an excess in the specified mass range.

Finally, we show in~\citefig{fig:am} how the antiproton constraint may translate
into a constraint on the spin-independent cross section (see 
\citeeqp{eq:approxsigma}), comparing full samples (top row panels) to the same 
ones but after applying the cut $SP_{\rm tot}>1$ (middle row panels). In each 
top panel, the region where the relic density is set by $\chi\,\chi\to h\,h$,
P-wave case irrelevant to indirect detection, covers almost the full area but 
the extreme right-hand side, where $\chi\,\chi\to a\,h$ becomes dominant. The 
(anti)correlation between 
$\sigv_0\propto [\tilde{c}_{\chi\chi a}/\mchi]^2$ and $\sigma^{\rm SI}\propto 
[c_{\chi h}/(m_h^2\,\cbe)]^2$ appears precisely there, which is more explicit
in the middle-row panels, where the cut $SP_{\rm tot}>1$ was applied. From these
plots, one can extract the following approximate upper limit on the direct
detection phase space:
\ben
\label{eq:uplimsi}
\left[\frac{c_{\chi h}}{\cbe\, m_h^2}\right]^2
\approx \left[ \frac{c_{\chi h}\,\tbe }{m_h^2} \right]^2
\lesssim 10^{-4} \, {\rm GeV^{-4}}\,
\left[ \frac{10^{-4}\,{\rm GeV^{-2}}}{\left[c_{\chi a}/\mchi \right]^2}\right] 
\;.
\een
In the bottom row panels of~\citefig{fig:am}, we report a more explicit 
translation of our antiproton constraints in terms of spin-independent
cross section, which can be directly compared with~\citefig{fig:sigsi}.
We have applied the cut $SP_{\rm tot}>1$. We see that antiprotons provide
quite strong constraints for $\mchi\lesssim 10$ GeV in this regime, 
independently of $\tbe$ and $\tqcd$. This nicely illustrates why the antiproton
signal must be considered carefully in the singlino-like phenomenology.

\begin{figure}[!t]
\includegraphics[width=\columnwidth]{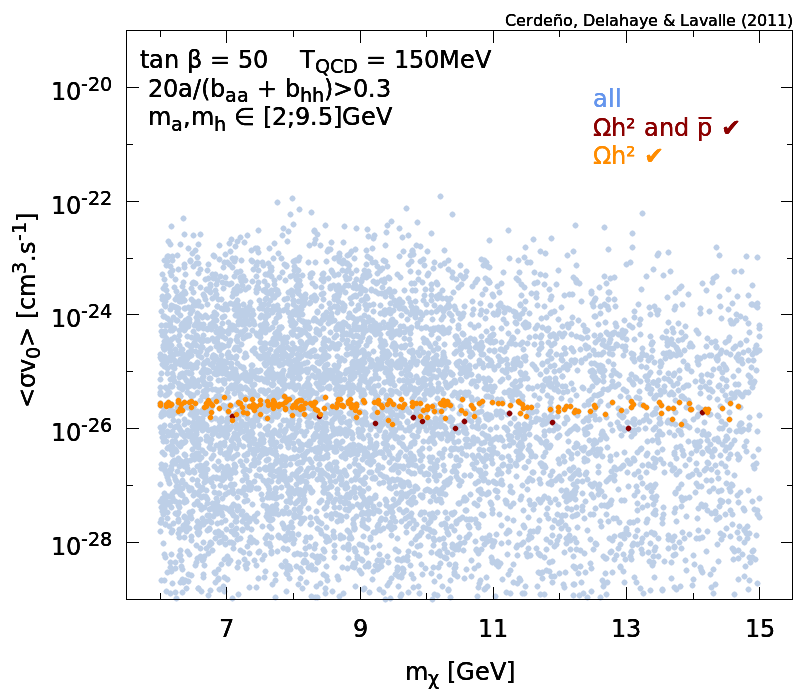}
\caption{Example of generic antiproton constraints on singlino models. 
  See~\citeeq{eq:uplim}.}
\label{fig:optcut}
\end{figure}

\begin{figure*}[!t]
\includegraphics[width=0.5\columnwidth]{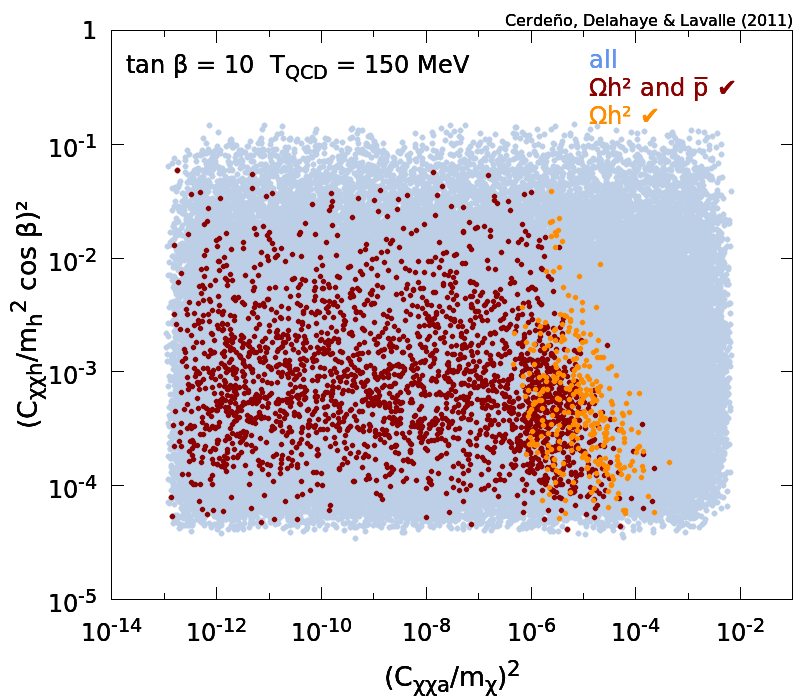}
\includegraphics[width=0.5\columnwidth]{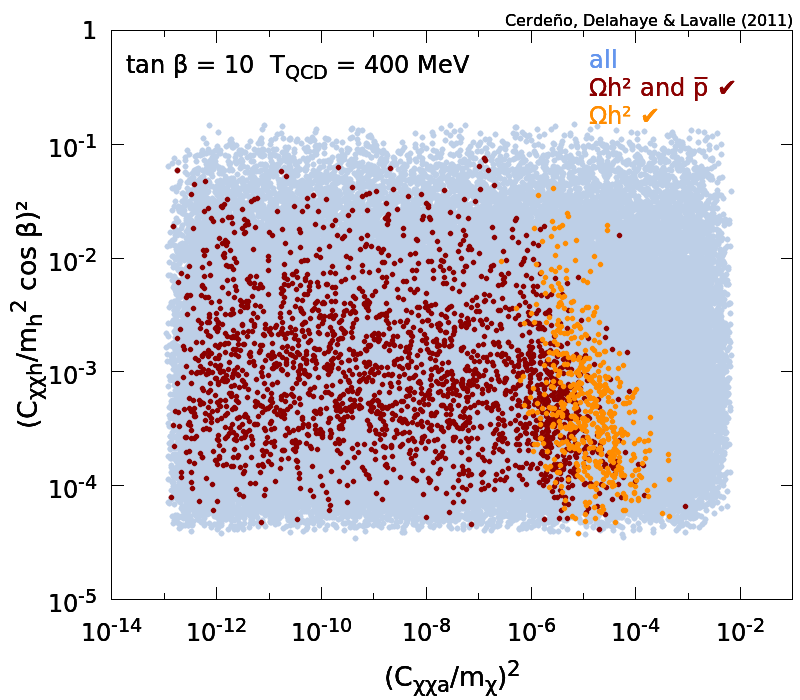}
\includegraphics[width=0.5\columnwidth]{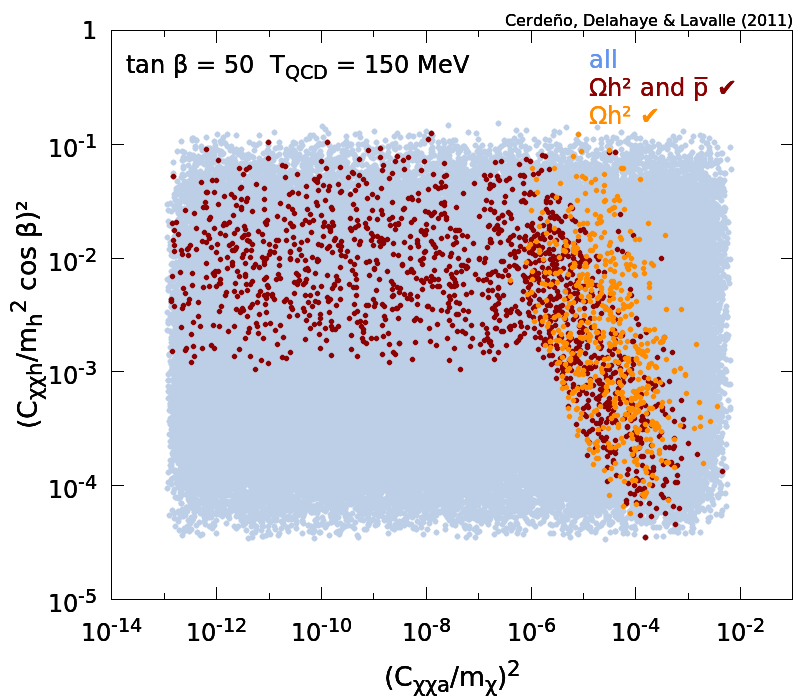}
\includegraphics[width=0.5\columnwidth]{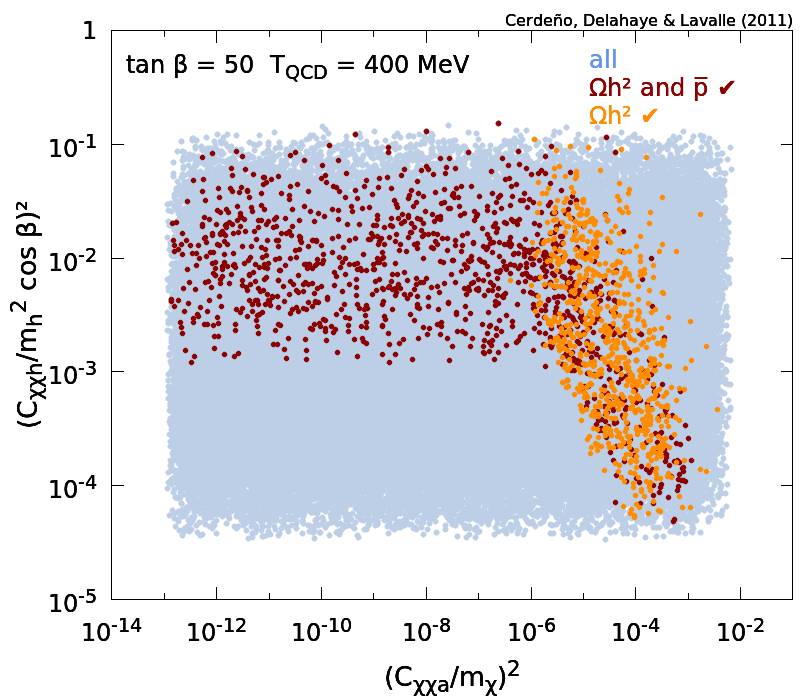}\\
\includegraphics[width=0.5\columnwidth]{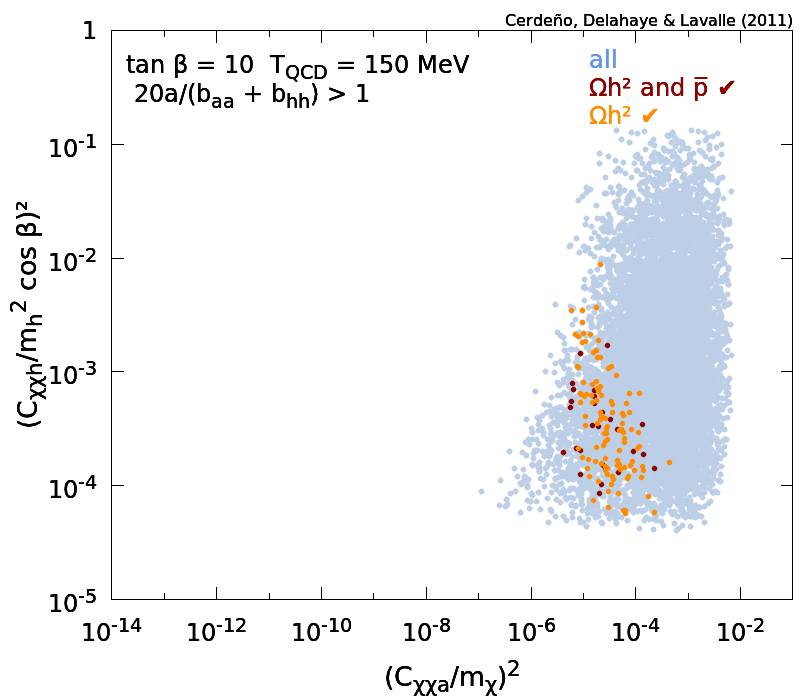}
\includegraphics[width=0.5\columnwidth]{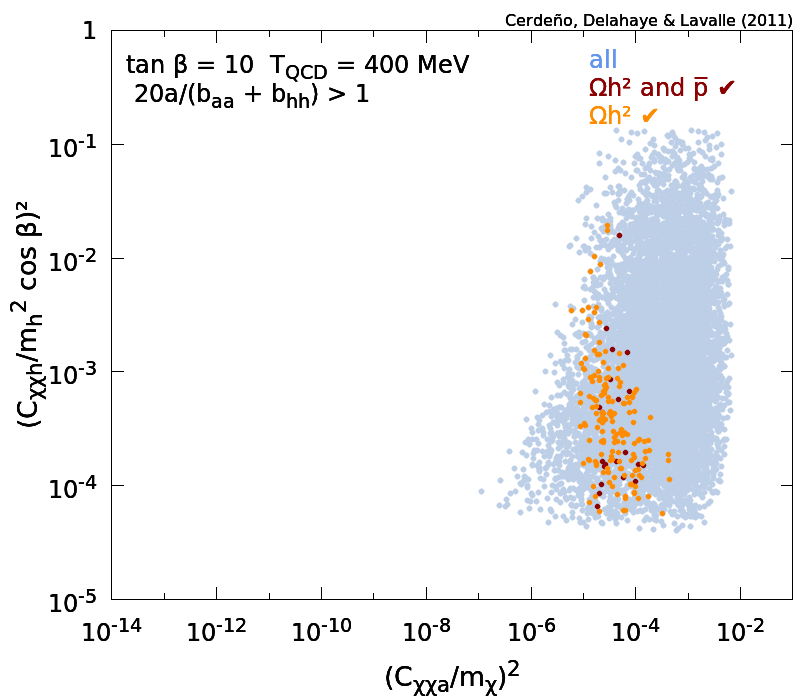}
\includegraphics[width=0.5\columnwidth]{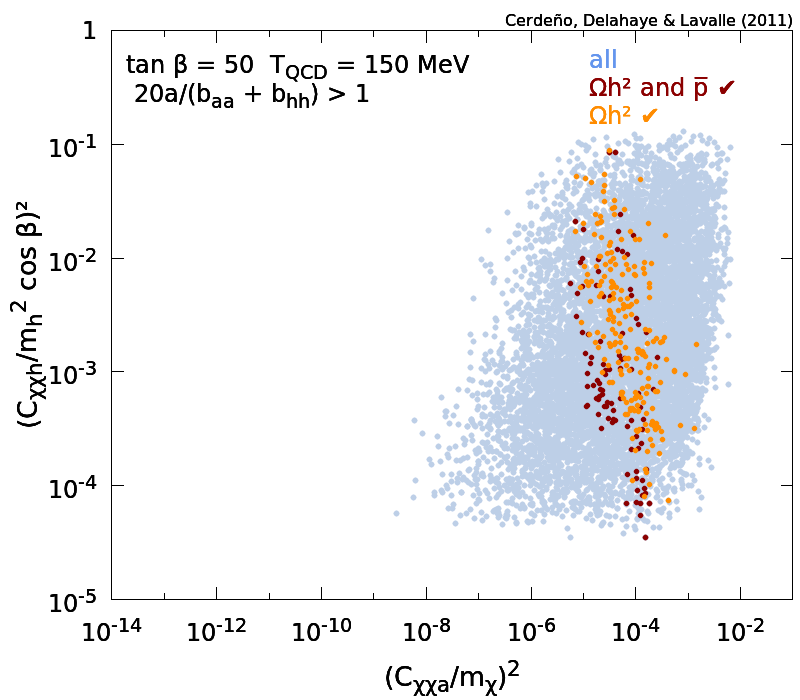}
\includegraphics[width=0.5\columnwidth]{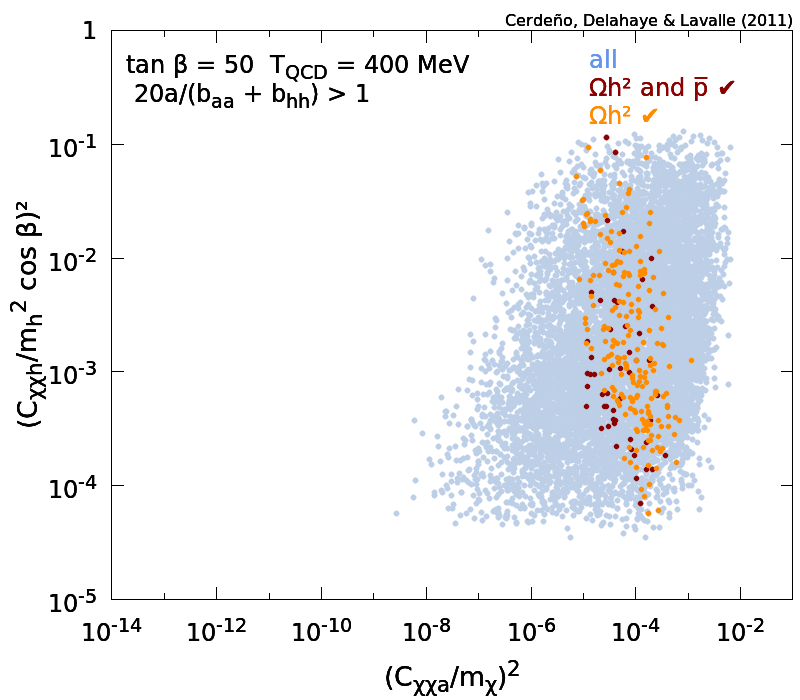}\\
\includegraphics[width=0.5\columnwidth]{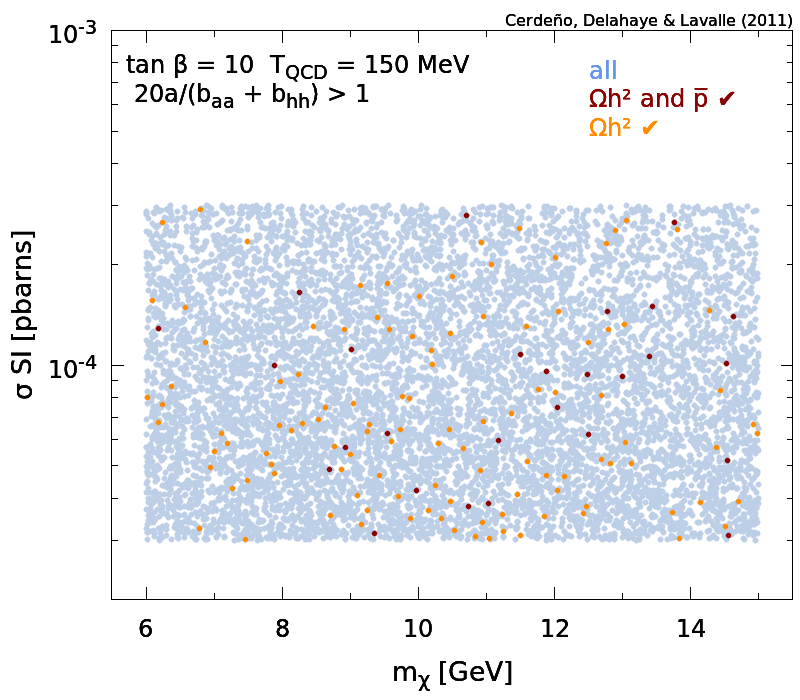}
\includegraphics[width=0.5\columnwidth]{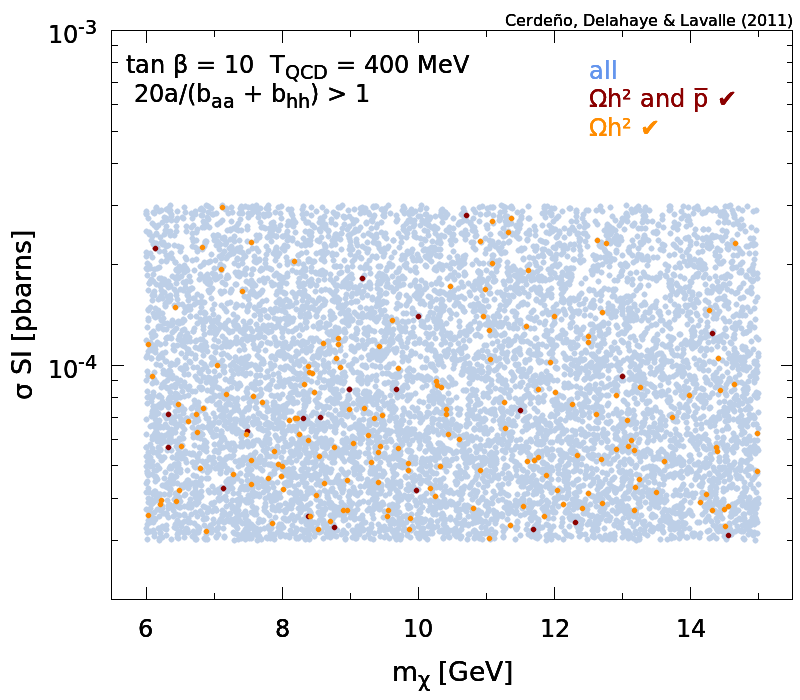}
\includegraphics[width=0.5\columnwidth]{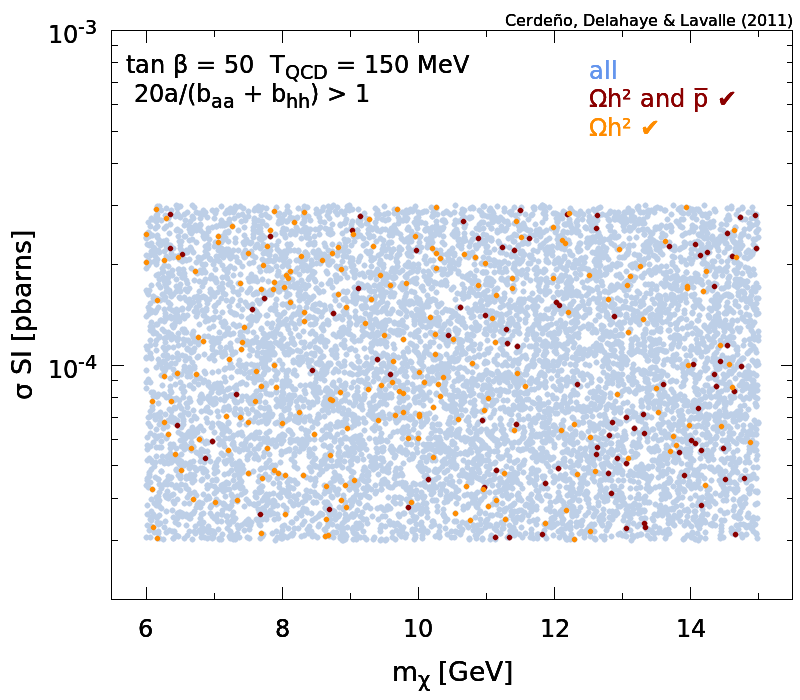}
\includegraphics[width=0.5\columnwidth]{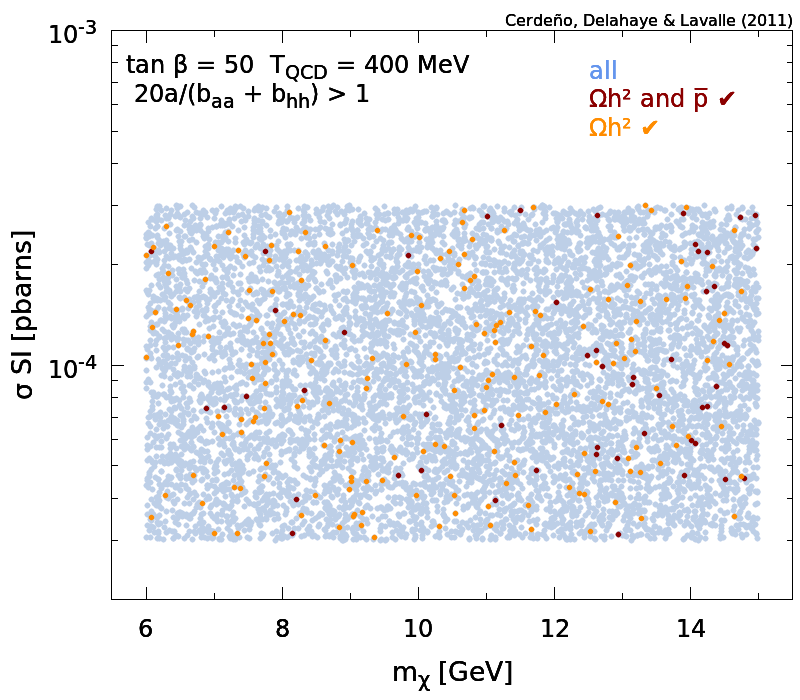}
\caption{Potential correlation between 
$\sigv_0\propto [\tilde{c}_{\chi\chi a}/\mchi]^2$ and $\sigma^{\rm SI}\propto 
[c_{\chi h}/(m_h^2\,\cbe)]^2$. Top/bottom panels: $\tbe =$10/50. Left/right 
panels: $\tqcd = 150/400$ MeV. A cut $SP_{\rm tot}>1$ was applied to
all panels.}
\label{fig:am}
\end{figure*}

\subsubsection{Uncertainties and potential detection prospects for PAMELA and 
  AMS-02}
\label{subsubsec:unc}

\begin{figure*}[!t]
\includegraphics[width=\columnwidth]{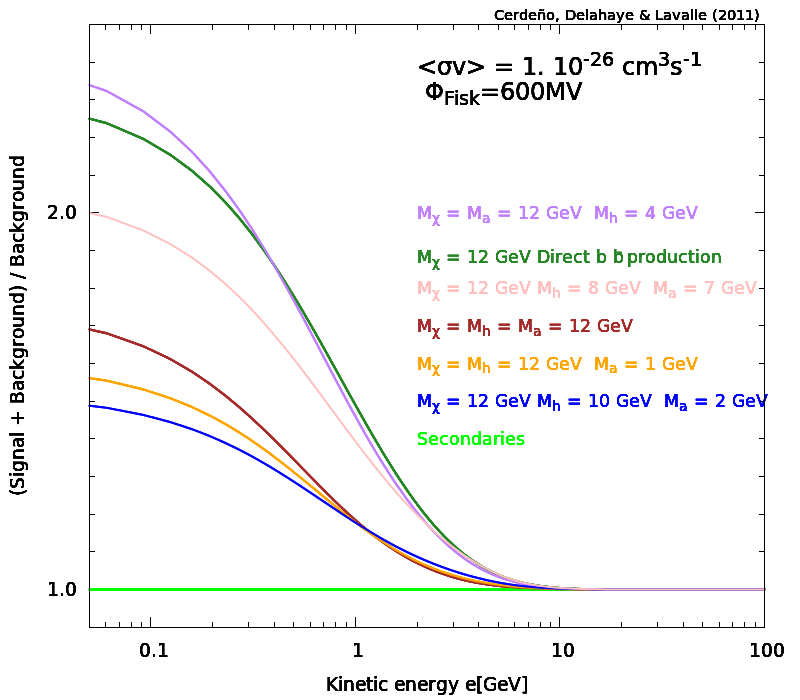}
\includegraphics[width=\columnwidth]{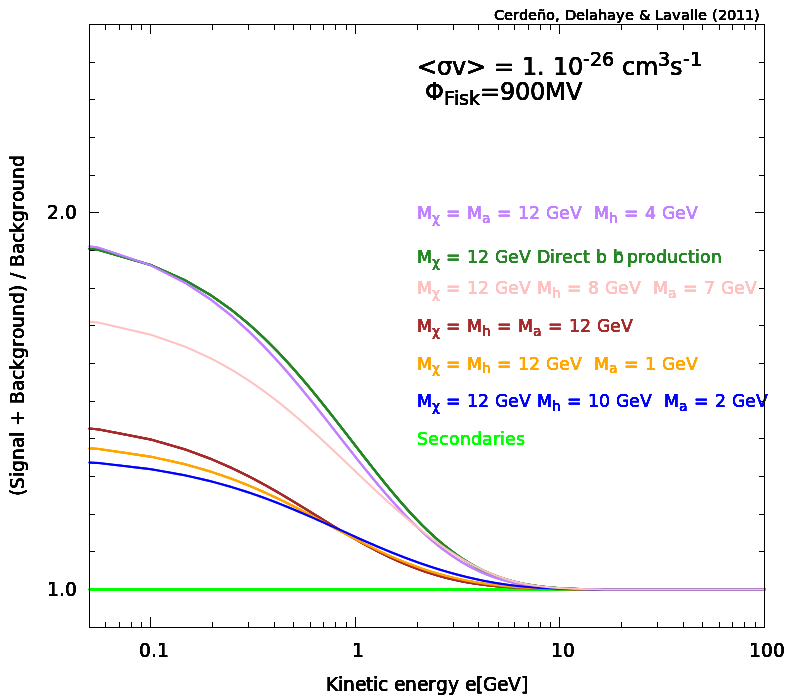}\\
\includegraphics[width=\columnwidth]{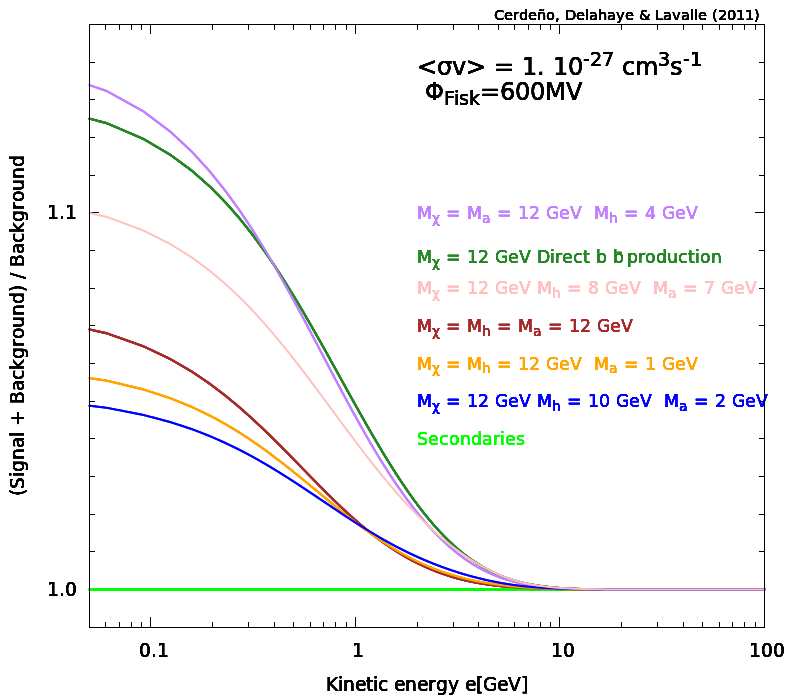}
\includegraphics[width=\columnwidth]{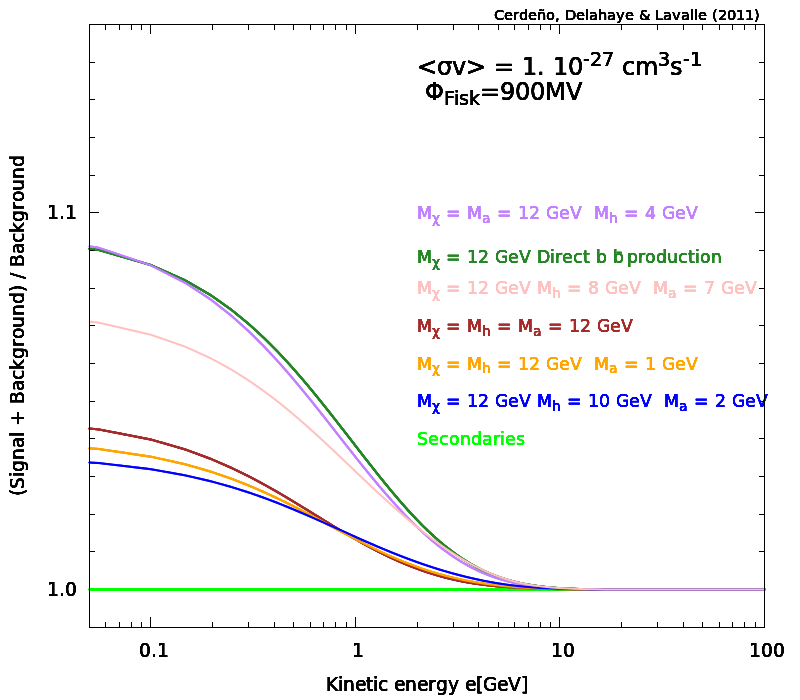}
\caption{Signal to secondary ratio. Top: signals are computed assuming an 
annihilation cross section of $\sigv = 10^{-26}\,{\rm cm^3/s}$. Bottom:
$\sigv = 10^{-27}\,{\rm cm^3/s}$. Left: solar modulation with a Fisk potential
of $\phi = 600$ MV. Right: $\phi = 900$ MV.}
\label{fig:sig_to_sec}
\end{figure*}

There are different sources of theoretical uncertainties that might affect our
predictions. Among important effects which are related to completely 
different domains, some uncertainties may be connected to the QCD 
phase transition in the early universe (relic density), the strangeness content 
of nucleons (spin-independent cross section), the solar modulation modeling and 
the cosmic-ray transport parameters (antiproton flux).

Uncertainties in the QCD phase transition have been shown to lead to a
factor of 2 uncertainty in the annihilation cross section at freeze out in
the WIMP mass range $\sim 4$-12 GeV, with larger transition temperatures 
inducing an annihilation cross section for $\mchi / x_f \lesssim \tqcd$ larger
by a factor of 2 than the one obtained for $\mchi / x_f \gtrsim \tqcd$, due
to current constraints on the dark matter abundance (see~\citefig{fig:rd}).
We tried to account for this uncertainty by using both a sharp
QCD phase transition and two different temperatures.

Regarding uncertainties in the hadronic matrix elements, we refer
the reader to Refs.~\cite{2000APh....13..215B,2002APh....18..205B,2005PhRvD..71i5007E,2008PhRvD..77f5026E,2011arXiv1106.4667B} for more details. The 
consequences on our analysis are twofold, since these uncertainties can 
lead to select either higher (larger strangeness fraction) or smaller 
(smaller strangeness fraction) values of ${\cal C}_{\chi h}/
(\cbe \, m_h^2)$
--- see~\citeeq{eq:ddcondition}. In the former case, then annihilation 
into $h\,h$ through the t-channel would be increased, which would likely 
make the population obeying $\sphh <\spaa < 1$ more prominent, thereby leading 
to a P-wave dominated annihilation cross section. This would further squeeze the
configuration range available to indirect detection. Nevertheless, 
we point out that requiring a correct relic density poses significant 
constraints then, since the annihilation rate may increase too much and lead to 
a too small dark matter abundance. On that account, it may become difficult to 
find cosmologically viable configurations, especially at small \tbe 
--- this is illustrated in the top panels of~\citefig{fig:am}, where the upper 
regions are shown less favored by the relic density constraint. In contrast, the
latter case, namely selecting smaller values of ${\cal C}_{\chi h}/
(\cbe \, m_h^2)$, is more favorable to indirect detection, since a correct 
relic density can be achieved by means of a larger coupling ${\cal C}_{\chi a}$,
increasing the annihilation into $a\,h$; so that the antiproton 
constraints get stronger.

Coming back to astrophysics, the low energy cosmic-ray flux below a 
few GeV is sensitive to the solar 
activity. In our calculation, the solar modulation is taken into account by
using the force-field approximation
\cite{1968ApJ...154.1011G,1971JGR....76..221F}, which is known to provide
a rather satisfying understanding and fit, though missing some details, of the 
impact of solar activity on cosmic rays~\cite{1987A&A...184..119P,2007APh....28..154S,2010AdSpR..46..402P}. We refer the reader to 
Ref.~\cite{2010AdSpR..46..402P} for a more complete review about the current 
state-of-the-art. Most of the solar modulation models are based on numerical or 
analytical solution to the Parker transport equation~\cite{1965P&SS...13....9P},
the force-field approximation providing an analytical solution by assuming
perfect spherical symmetry and a static electromagnetic field.
More complete models lead to charge sign effects depending on the solar 
magnetic polarity, which reverts at each 11-year solar cycle, but they are 
expected to be small, at the $\sim$10-20\% level in terms of flux at the Earth. 
This roughly sizes the theoretical uncertainties affecting the force-field 
approximation. This has to be compared
with the error bar associated with the lowest energy PAMELA data points, 
typically $\lesssim 1.5$ GeV, which turn out to be the most constraining
in our analysis (see~\citefig{fig:fluxes}) --- the observational error is 
$\sim 40$\% in this energy range, which is twice more important than the solar 
modulation error. We have tuned our Fisk potential to a low/intermediate value 
of $\phi = 600$ MV, a rather conservative assumption with respect to the solar 
activity monitored during the PAMELA data taking~\cite{2010ApJ...723L...1M}. 
This means that the error boxes that we have considered to derive our 
constraints safely encompass the theoretical uncertainty affecting the solar 
modulation modeling. In~\citefig{fig:sig_to_sec}, we show the ratio of the total
flux to the secondary flux, which aims at illustrating the spectral break due 
to additional primaries for different amplitudes characterized
by the annihilation cross section --- $\sigv = 10^{-26}\,{\rm cm^3/s}$ 
($10^{-27}\,{\rm cm^3/s}$) on top (bottom) panels --- and for different solar
activities --- $\phi = 600$ MV (900 MV) on left (right) panels. We see
that going from weak to strong solar activity can decrease a predicted
excess down to $\sim 40$\%. The PAMELA data, which were taken close to
a solar minimum, are therefore very well suited for to derive low energy 
limits.

As regards the uncertainties affecting the cosmic-ray transport parameters, 
we stress that one of the most sensitive parameter for dark matter-induced 
cosmic rays is the diffusion halo size $L$, which characterizes the volume 
which the annihilation products are integrated over and which turns out to be 
strongly correlated with the diffusion coefficient normalization $K_0$ --- 
$K_0/L\approx {\rm cst}$ from stable secondary-to-primary nuclei ratio 
constraints like boron-to-carbon (\eg~\cite{2001ApJ...547..264J}). Increasing
$L$ leads to integrate more annihilation products and thus increase the
dark matter-induced primary 
flux~\cite{2004PhRvD..69f3501D,2008A&A...479..427L,2008PhRvD..77f3527D}.
In this paper, we have used $L=4$ kpc (see~\citetab{tab:prop}). We emphasize
that the most recent works on fitting the transport parameters tend to favor 
a larger diffusion zone, with $L\sim5$-10 kpc, which makes our propagation
model rather conservative. These studies consider the additional input of the
secondary radioactive cosmic ray species which allow to break the
degeneracy between $L$ and $K_0$~\cite{2010A&A...516A..66P,2011ApJ...729..106T}.
Another unexpected problem in small-$L$ diffusion zone models is that they
systematically lead to a secondary positron flux in significant excess with
respect to the data at low 
energy~\cite{2009A&A...501..821D,2011MNRAS.414..985L}.
Few other arguments in favor of larger values of $L$ are recalled 
in~\cite{2010PhRvD..82h1302L} --- see complementary arguments in a recent
study on the diffuse Galactic radio emission~\cite{2011arXiv1106.4821B}. Our 
values for $K_0$ and $L$ can therefore be considered as quite motivated and 
rather conservative.

Finally, it is interesting to make some prospects for the end of the PAMELA
campaign or the new operating space experiment AMS-02\footnotemark, which was 
successfully launched in May 2011 and which is supposed to take data for 
$\gtrsim 10$ years. For the latter, it is not yet clear whether the low energy 
systematic errors can be reduced with respect to PAMELA, but statistics will 
for sure be much better. 
\citefig{fig:sig_to_sec} shows how a finer PAMELA/AMS-02 data analysis could 
improve our current limits or lead to the detection of those singlino 
configurations which still induce cosmic antiprotons, but with a flux too weak 
to be detected or constrained. This typically corresponds to an 
annihilation cross section $\sigv\lesssim 5\times 10^{-27}\,{\rm cm^3/s}$. We 
see in the bottom panels that a precision of $\sim 5$\% makes the data quite 
relevant to such a low annihilation cross section, even in the case of strong 
solar activity. Therefore, if the PAMELA or AMS-02 collaborations manage to 
reach such a low precision, we expect much tighter constraints on light dark 
matter models annihilating into antiprotons in the future, even if the solar
activity is also expected to increase in the coming years. In particular, 
much deeper phase-space regions with $SP_{\rm tot}\lesssim 0.1$ 
(see~\citeeqp{eq:spratio}) could be probed.

\footnotetext{\url{http://www.ams02.org/}}

\subsubsection{Connecting this effective approach to more constrained
particle phenomenologies}
\label{subsubsec:nmssm}

The question may arise of how to connect this effective approach to 
existing and more constrained particle phenomenologies beyond the SM.

First, an interesting and very general link can be made by focusing on 
the kinematics of the final states. Indeed, we have considered that annihilation
proceeds into two bodies that may further decay into quarks, leading to the 
production of antiprotons. We have shown that each time the annihilation
cross section is S-wave dominated, then this kinematical configuration may lead 
to an antiproton excess quite generically. Our effective approach is therefore
fully relevant to cases for which such two-exotic-body production is
predicted, as in~\eg~Ref.~\cite{2011arXiv1108.1391C}, where the so-called 
DAMA/CoGeNT signal is explained from a model incorporating two additional light 
gauge bosons that dark matter can annihilate into. Since one of these gauge
fields has a mass larger than 2 GeV, it can in principle decay into an 
antiproton.
This model can therefore suffer the antiproton constraints. Our constraints
may also be important for completely different models, like in the recent study 
performed in~\cite{2011arXiv1108.0978C} of a light right-handed sneutrino dark 
matter that can arise in the frame of the NMSSM; annihilation can then indeed 
go into light pseudo-scalars without being velocity suppressed, while this 
case is not necessarily adapted to fit the CoGeNT region.

Given our effective Lagrangian of~\citeeq{eq:lag}, a natural link
can be made with singlet extensions of the MSSM, which was the original
motivation for this work. This includes general extensions as well as more
constrained ones, like for instance the NMSSM, in which the dark matter 
candidate is a singlino-dominated neutralino. An example of a model of singlino
dark matter can be found in Ref.~\cite{2011PhLB..695..169K}, where the specific
annihilation into light (pseudo)scalar Higgs bosons is considered. Their
case (3), which corresponds to $2\mchi>m_a+m_h$, can be directly compared with
our analysis. In this specific configuration, they find $|{\cal C}_{\chi h}| 
\approx |{\cal C}_{\chi a}| \lesssim |\lambda_{ijk}| $, with $|\lambda_{hhh}| 
\gtrsim |\lambda_{aah}|$. Having similar couplings between $\chi\chi a$ 
and $\chi\chi h$ implies that they both have to be large to explain the 
DAMA/CoGeNT signal, which also implies that the S-wave contribution of the 
$t$-channel annihilation diagram into $a\, h$ is always significant. The 
fact that $|\lambda_{hhh}| \gtrsim |\lambda_{aah}|$ induces a competition with 
the $s$-channel annihilation into $h\,h$ (P-wave), but more suppressed than the
$t$-channel annihilation into $a\, h$ (velocity suppression, and a squared 
propagator factor of $\sim 1/(4 \mchisq-m_h^2)^2$ as compared with 
$\sim 1/m_\chi^4 $ for the $t$-channel). The antiproton limit must therefore
be regarded in this case.

For comparisons with the NMSSM, we may use 
Refs.~\cite{2011PhRvL.106l1805D,2011arXiv1104.1754C,2011arXiv1107.1614A,2011arXiv1107.1604C}. Although the reason is not clear from the mass matrix 
elements only~\cite{2010PhR...496....1E}, it appears to be difficult to 
generically find viable NMSSM configurations obeying $2\mchi>m_a+m_h$. Bosons 
$a$ and $h$ can be found very light, but not simultaneously 
\cite{2011arXiv1104.1754C,2011arXiv1107.1614A,2011arXiv1107.1604C} --- an 
exception can be found in Ref.~\cite{2011PhRvL.106l1805D}, but the authors get 
an annihilation going dominantly into $f\,\bar{f}$. The above 
cited studies still find complementary viable regions in the NMSSM parameter
space. When the relic density is set from the $s$-channel resonant exchange of 
a light $h$, then the annihilation is P-wave dominated and indirect detection 
fails to yield interesting bounds. If, in contrast, the main annihilation final 
state consists in an S-wave production of quarks (even if not with a branching 
ratio of 1), then an antiproton excess very likely follows in most of cases
--- this may also happen for a right-handed sneutrino dark matter within the 
NMSSM~\cite{2011arXiv1108.0978C}. 
This situation was already discussed in Ref.~\cite{2010PhRvD..82h1302L}, where
the antiproton limit was shown to be very sharp.

%% Section 5: conclusion.

\section{Conclusion}
\label{sec:concl}

In this paper, we have designed a generic effective Lagrangian to survey the
light singlino dark matter phenomenology, as soon as annihilation mostly 
proceeds into light singlet-dominated scalar $h$ and/or pseudo-scalar $a$ Higgs 
bosons (see~\citesec{sec:efflag}). Within this general framework, we have 
computed the annihilation cross section (\citesec{subsec:ann}), the relic 
density (\citesec{subsec:relic}), the spin-independent nucleon-singlino cross 
section (\citesec{sec:direct}), the antiproton spectra arising from the 
decays of these light (pseudo-)scalar bosons (\citesec{subsec:pbar_prod}), 
and the subsequent antiproton flux at the Earth (\citesec{subsubsec:prop}). We 
have focused on a specific mass configuration that guarantees the existence of 
an S-wave contribution to the annihilation cross section, ensuring the relevance
of indirect detection, \ie~$2\,\mchi\geq m_a+m_h$. 
We then have selected the 
mass and coupling parameter space to get a spin-independent cross section 
range encompassing the so-called CoGeNT region, and drawn several samples of 
$\sim 10^{5}$ models for two different values of $\tbe$, 10 and 50, for which
we have calculated the relic density. Then, we 
have computed the antiproton fluxes associated with these configurations and 
compared them to the PAMELA data (\citesec{subsec:results}).

We emphasize that there are still theoretical uncertainties affecting the 
estimate of the annihilation cross section at freeze out for $\sim 10$ GeV mass 
WIMPs, up to a factor of $\sim 2$ depending on whether WIMPs decouple after or 
before the QCD phase transition. This is due to the rapid change in the 
relativistic degrees of freedom that occurs at that time and acts as a 
weight on the annihilation cross section. We have taken this uncertainty into
account by considering extreme cases, as illustrated in \citefig{fig:rd}.

Our results demonstrate (\citesec{subsec:results}) that the antiproton 
constraint is generically relevant to the whole CoGeNT region 
(see~\citefig{fig:sigsi}), though there are some simple ways to escape it.
The most trivial way to suppress the S-wave contribution to the annihilation
cross section is to suppress the coupling between the singlino and 
the light pseudo-scalar $c_{\chi a}$, as shown by~\citeeq{eq:swave}. Another
way is to have the light scalar and pseudo-scalar Higgs bosons so light that
they cannot decay into antiprotons, or cannot generate antiprotons energetic
enough to be observed. Anyway, we have shown that as soon as the S-wave is 
significant at freeze out, the antiproton signal is 
in most of the cases in excess with respect to the PAMELA data 
(see~\citefig{fig:s0m}, which translates into \citefig{fig:s0ma} after imposing 
the S-to-P wave ratio to be larger than 1). Depending on the singlino/scalars
mass configuration, we have derived a quite generic upper limit on the
annihilation cross section lying in the range 
$\sigv_0 \lesssim 10^{-26}\,{\rm cm^3/s}$, valid for the 
singlino mass range $\sim$6-15 GeV. This limit is very competitive with
respect to those obtained from gamma-rays~\cite{2011JCAP...01..011A}, and
complementary to limits coming from high-energy neutrinos from the Sun
\cite{2011NuPhB.850..505K}. We recall 
that the antiproton bounds are even more stringent for direct annihilation 
into quarks (see~\citefig{fig:fluxes} and~\cite{2010PhRvD..82h1302L}).

Uncertainties affecting our analysis have been discussed 
in~\citesec{subsubsec:unc}, where we have shown that our procedure is to be
considered as rather conservative. In addition to the points raised there,
we remind that our antiproton flux calculation is poorly sensitive to the 
details of the dark matter distribution far away from the Earth, in contrast 
with the calculation of the gamma-ray flux. The main parameter which sets the
flux amplitude is the local density $\rho_\odot$, so one can easily rescale
our results and limits by a mere factor of $(\rho_\odot^{\rm new}/\rho_\odot)^2$.
Still, we recall that we have considered a dynamically constrained profile, and
that we did not include any substructure. Although their survival against
tidal effects and encounters with baryonic systems (stars and/or Galactic disk)
still needs to be clarified, the survival fraction is expected to be larger
and larger for smaller and smaller substructures, which are more 
concentrated~\cite{2006PhRvD..73f3504B}. Self-consistency would then require
to include their effect in our calculation~\cite{2009NJPh...11j5027B}, 
which turns out to be a slight increase of the antiproton flux at low 
energy~\cite{2008A&A...479..427L,2011PhRvD..83b3518P}. This would therefore
make our present constraints even more stringent.

Finally, we have shown that provided systematic errors can be reduced down 
to $\sim 5$\%, PAMELA, from an improved analysis, or AMS-02 could have the 
capability of detecting a still unseen signal characterized by a spectral break 
(see \citefig{fig:sig_to_sec}) which would hide just within the current PAMELA 
error boxes we have considered in this paper. A null detection with better
errors would in any case significantly ameliorate the limits derived in this
paper.

\paragraph{{\bf Acknowledgements}:
We are grateful to Gilbert Moultaka for inspiring and early discussions
about effective Lagrangians, and to Pierre Salati for interesting comments 
about the relic density calculation. We also thank Genevi\`eve 
B\'elanger, Fawzi Boudjema and Alexander Pukhov for helping us compare our relic
density calculation results with MicrOMEGAS, and Cyril Hugonie for providing
us with useful technical details about the NMSSM.
D.G.C. is supported by the Ram\'on y Cajal program of the Spanish MICINN. 
This work was supported by the Spanish MICINN's Consolider-Ingenio 2010 
Programme under grants MultiDark CSD2009-00064 and CPAN CSD2007-00042. We also 
thank the support of the MICINN under grant FPA2009-08958, the Community of 
Madrid under grant HEPHACOS S2009/ESP-1473, and the European Union under the 
Marie Curie-ITN program PITN-GA-2009-237920.}

\appendix
%% Appendix 1: annihilation cross section

\section{Annihilation cross section}
\label{app:cross_section}

In this appendix, we summarize our calculations of the annihilation cross
section and of the relic density for pedagogical purposes (mostly dedicated
to those who wish to revisit the case without any formal calculation software). 
The scattering amplitudes associated with the process $\chi(p_1)\,\chi(p_2)
\longrightarrow \phi_i(k_i)\,\phi_j(k_j)$ and consistent with the Lagrangian of 
\citeeq{eq:lag}, the Feynman diagrams of which are shown 
in~\citefig{fig:diagrams}, are the following (for the s, t and u channels, 
respectively):
\besubeq{eq:amplitudes}
\ben
{\cal M}_s &=& 
\frac{\lambda_{ijk}}{s-m_k^2+i\,\Gamma_k\, m_k}\,
\vbar(p_2)\,{\cal C}_{\chi k}\,u(p_1)\label{eq:m_s}\\
{\cal M}_t &=& 
\frac{1}{t-\mchi^2} \,\vbar(p_2)\,{\cal C}_{\chi j}\, 
(\myslash{q_t}+\mchi)\,{\cal C}_{\chi i} \,u(p_1) \label{eq:m_t}\\
{\cal M}_u &=& 
\frac{1}{u-\mchi^2} \,\vbar(p_2)\,{\cal C}_{\chi i}\, 
(\myslash{q_u}+\mchi)\,{\cal C}_{\chi j} \,u(p_1) \label{eq:m_u} \;,
\een
\eesubeq
where the couplings $\lambda$ and ${\cal C}$ are defined in 
\citeeqss{eq:lag}{eq:coupl}\footnotemark, $s$, $t$ and $u$ are the Mandelstam 
variables and $q_t$ and $q_u$ the related 4-momenta:
\besubeq{eq:kin1}
\ben
s &\equiv& (p_1+p_2)^2 = (k_i+k_j)^2\;; \;\;
t \equiv q_t^2\;; \;\;
u \equiv q_u^2\;; \\ 
q_t &\equiv& p_1-k_i = k_j-p_2\;; \;\; 
q_u \equiv p_1-k_j = k_i-p_2\;.
\een
\eesubeq
\footnotetext{We stress that this general writing introduces non-physical 
degrees of freedom when going to the total squared amplitude, 
notably for the interference terms (mixing of two different final states, 
\eg~scalar--scalar with pseudo-scalar--scalar). One has to remember that
interaction terms involving $\cchii\times\cchiit$ are forbidden --- this is 
clear from the Lagrangian of~\citeeq{eq:lag} --- and can therefore be dropped.
As a more general rule, one can replace 
$\cchii\times\cchijt\longrightarrow \cchii\times\cchijt\,(1-\delta_{ij})$. 
Despite this caution, which is relevant when trying to give physical 
interpretations to the general-case result, such a general approach proves 
powerful since a single expression can be used for different annihilation final 
states; one has just to switch the appropriate couplings on/off to get the 
selected one, which automatically removes the unphysical terms.}

We also introduce the functions $g(s)$ and $h(s)$ such that in the center of 
mass system:
\besubeq{eq:kin2}
\ben
t &=& \mchi^2 - g(s) + h(s)\,\cos(\theta)\;;\\
u &=& \mchi^2 - g(s) - h(s)\,\cos(\theta)\;;\\
{\rm with}\;\; 
g(s) &\equiv& (s-m_i^2-m_j^2)/2\;;\\ 
h(s) &\equiv& \frac{f_{ij}}{2}\,\sqrt{s(s-4 \mchi^2)}\;,
\een
\eesubeq
where $f_{ij}(s)$ is further defined in \citeeq{eq:dlips}, and $\theta$ is the 
angle between $\vec{p_1}$ and $\vec{k_i}$ in the center of mass frame.

The differential annihilation cross section is given by:
\ben
\label{eq:dsigma}
d\sigma_{12\rightarrow ij} &=& \frac{d{\rm LIPS}}{4\,E_1\,E_2\,\beta_{12}}
\left[ \frac{1}{g_\chi^2\,S_{ij}}\sum_{\rm d.o.f}|{\cal M}|^2 \right]\\
{\rm with}\;\; \beta_{kl} &\equiv& \sqrt{(p_k.p_l)^2 -m_k^2m_l^2}/(E_kE_l)\;,\nn
\een
where $g_\chi$ is the number of degrees of freedom of particle $\chi$ 
($g_\chi=2$ here), $S_{ij}\equiv 1 + \delta_{ij}$ is a symmetry factor, and where
the integrated Lorentz invariant phase space factor reads:
\besubeq{eq:dlips}
\ben
\int d{\rm LIPS} &=& \int 
\frac{d^3\vec{k}_i}{(2\pi)^3 2\,E_i}\,\frac{d^3\vec{k}_j}{(2\pi)^3 2\,E_j}\\
&& \times \, (2 \pi)^4 \, \delta^4(p_1+p_2-k_i-k_j)\nn \\
&=& \frac{f_{ij}(s)}{16\,\pi}\,
\int_{-1}^{1} dz\nn\\
{\rm with}\;\;
f_{ij}(s) &\equiv& \left\{ \left[ 1 - \frac{(m_i-m_j)^2}{s}\right] 
\left[ 1 - \frac{(m_i+m_j)^2}{s}\right]\right\}^{1/2}\;.\nn\\
\een
\eesubeq
Note that $z\equiv\cos(\theta)$.

To compute the temperature-averaged annihilation cross section $\sigv$,
we need to express the velocity $v$, which is not the relative velocity
$v_r/c \equiv c\,|\vec{p}_1/E_1 - \vec{p}_2/E_2| = 
\sqrt{\beta_{12}^2+c^4|(\vec{p}_1/E_1)\times(\vec{p}_2/E_2)|^2}$, but the 
M{\o}ller 
velocity $v/c \equiv \beta_{12}$~\cite{1988NuPhB.310..693S,1991NuPhB.360..145G}.
Incidentally, it is useful to note that $(E_1E_2\sigma v)$ is Lorentz invariant.
Now, let us express the cross section in terms of a sum of the different 
contributions (s-, t-, u-channels and interferences), for a deeper understanding:
\ben
\sigma_{12}\,\beta_{12} &=& 
\sigma_{12}\,\beta_{12}|_{\rm s} +
\sigma_{12}\,\beta_{12}|_{\rm t} +
\sigma_{12}\,\beta_{12}|_{\rm u}+\\ 
&&
\sigma_{12}\,\beta_{12}|_{\rm st} + 
\sigma_{12}\,\beta_{12}|_{\rm su} + 
\sigma_{12}\,\beta_{12}|_{\rm tu}\;.\nn
\een
%
%\subsection{General case}
We are armed enough to derive the cross section in the most general case, 
\ie~we can take particles $\phi_{i,j,k}$ having all a scalar and a pseudo-scalar
component; we will take care of removing the non-physical degrees of freedom
at the end\footnotemark[\value{footnote}].
Denoting $\sigma\,\beta = \sigma_{12}\,\beta_{12}$ and using $g=g(s)$ and 
$h=h(s)$, we find:
\besubeq{eq:sigv_full}
\begin{align}
%
% s-channel
\sigma\,\beta|_{\rm s} &=
\frac{ K\,\lambda_{ijk}^2 \,
\left[  c_{\chi k}^2 (s - 4 \mchi^2) + \tilde{c}_{\chi k}^2 s \right]}
{(s - m_k^2)^2 + \Gamma_k^2 m_k^2}\\
%
% t-channel
\sigma\,\beta|_{\rm t} &=
\frac{K\,\left[g^2-h^2\right]^{-1}}{2} \times \\
& \Bigg\{  \alpha_{0t} +
g \left[\alpha_{1t} + 2 \alpha_{2t} g\right] - \alpha_{2t} h^2 -\nn\\ 
&\frac{1}{2 h} \Big[\alpha_{1t} + 2 \alpha_{2t} g \Big] 
\left[ g^2 - h^2\right]
  \ln\left[ \frac{g + h}{g - h} \right] \Bigg\} \nn\\
%
% u-channel
\sigma\,\beta|_{\rm u} &= \sigma\,\beta|_{\rm t} \\
{\rm with}\;\; & \alpha_{mt}\to (-1)^m \alpha_{mt} \;;\;\; 
\text{and index}\;i \leftrightarrow {\rm index}\;j\;\text{for odd}\;m \nn\\
%
% interference st
\sigma\,\beta|_{\rm st} &= 
\frac{1}{4}\frac{K\,\lambda_{ijk} (s-m_k^2)}{(s-m_k^2)^2 + \Gamma_k^2 m_k^2}
\times \\
& \left\{ 2 \alpha_{1st} - \frac{1}{h}
(\alpha_{0st} + \alpha_{1st} g)\, \ln\left[\frac{g + h}{g - h}\right] \right\} 
\nn\\
%
% interference su
\sigma\,\beta|_{\rm su} &=  \sigma\,\beta|_{\rm st}\\
{\rm with}\;\;& \alpha_{mst}\to (-1)^m \alpha_{mst}\;;\;\;
\text{and index}\;i \leftrightarrow {\rm index}\;j \nn\\
%
% interference tu
\sigma\,\beta|_{\rm tu} &= -\frac{K}{4} \times \\
& \left\{ 2 \alpha_{2tu} - 
\frac{1}{g\,h}  (\alpha_{0tu} + \alpha_{2tu} g^2) 
\ln\left[ \frac{g + h}{g - h}\right ] \right\}  \;. \nn
\end{align}
\eesubeq
Parameter $K$, which carries dimensions of inverse squared mass, is defined as
\ben
K \equiv K(s) = \frac{f_{ij}(s)}{64\,\pi\,E_1\,E_2\,S_{ij}}\;.
\een
The coefficients $\alpha$ are associated with the numerators of terms involving 
the t- or u-channel, which all exhibit an angular dependence. This angular 
dependence can be expressed as a polynomial function of $z$ reading 
$P_x(z)=\sum_n \alpha_{nx}[h(s)\times z]^n$. These coefficients are 
explicitly given below:
\besubeq{eq:alphat}
\ben
\alpha_{0t} &\equiv &
2 \Bigg\{ \cip\,\cjp \left(g^2 - m_i^2\,m_j^2\right) + \\
&& 4 \mchi^2 \Big[
\cchiitsq \cchijsq m_i^2 + \nn\\
&& \cchiisq \left( \cchijsq (s- 4 \, \mchisq) + \cchijtsq m_j^2 \right) 
\Big] \Bigg\} \nn \\
\alpha_{1t} &\equiv & - 16 \, \mchisq \, \cchiisq \, \cchijsq \\
\alpha_{2t} &\equiv & - 2 \, \cip \, \cjp\;,
\een
\eesubeq
\besubeq{eq:alphast}
\ben
\alpha_{0st} &\equiv &
8 \mchi \Bigg\{ \cchii \left( \cchij \, \cchik (s - 4 \mchi^2) + 
\cchijt \,\cchikt \, s \right) - \\
&& \cchikt \ccijm (g + m_i^2) \Bigg\}\nn\\
\alpha_{1st} &\equiv & - 8 \, \cchik \, \cijp  \mchi \;,
\een
\eesubeq
\besubeq{eq:alphatu}
\ben
\alpha_{0tu} &\equiv & 
16\,\mchisq \left[ \cchiisq 
\left( \cchijsq (s-4\, \mchisq) + \cchijtsq m_j^2 \right) + 
\cchiitsq \cchijsq m_i^2 \right]
\nn\\
\alpha_{1tu} &\equiv & 0 \\
\alpha_{2tu} &\equiv & 
4 \left( \left[\cijp\right]^2 - \left[\ccijm\right]^2 \right) \;,
\een
\eesubeq
where we have used:
\besubeq{eq:cijk}
\ben
c_{l\pm}^2 &\equiv& c_{\chi l}^2 \pm \tilde{c}_{\chi l}^2 \\
c_{lk\pm}^2 &\equiv& \cchil \cchik \pm \cchilt \cchikt \\
c_{l/k\pm}^2 &\equiv& \cchil \cchikt \pm \cchilt \cchik \;.
\een
\eesubeq
The null-velocity limit of the annihilation cross section corresponds 
to requiring $s\to 4\,\mchi^2$. This implies that function $h(s)\to 0$, which
{\it a posteriori} justifies the form of \citeeq{eq:sigv_full}, from which
the full limit can be readily calculated. In terms of the above $\alpha$ 
coefficients (we now denote them $\bar{\alpha}$ to make explicit that they also 
have to be taken in the limit $s\to 4\,\mchi^2$), we obtain the following 
suitable expressions:
% Again, I apply a factor of 1/4 (which is therefore not in K) JL.
\besubeq{eq:sigv_limit}
\ben
\sigma\,\beta|_{\rm s,0} & \to & 
\frac{ 4\,\mchisq K\,\lambda_{ijk}^2 \,\tilde{c}_{\chi k}^2}
{(4\,\mchisq - m_k^2)^2 + \Gamma_k^2 m_k^2}\\
\sigma\,\beta|_{\rm t+u+tu,0} & \to & 
\frac{K\,\left[\bar{\alpha}_{0t}+ \bar{\alpha}_{0u} + \bar{\alpha}_{0tu}\right]}
     {2\,g^2(4\,m_\chi^2)}
\\
%\sigma\,\beta|_{\rm u,0} & \to & \frac{K\,\bar{\alpha}_{0u}}{2\,g^2(4\,m_\chi^2)}
%\\
\sigma\,\beta|_{\rm st+su,0} & \to & -\frac{K}{2}
\frac{\lambda_{ijk} \left[  \bar{\alpha}_{0st} + \bar{\alpha}_{0su} \right] 
    (4\,m_\chi^2-m_k^2)}
     {g(4\,m_\chi^2)\left( (4\,m_\chi^2-m_k^2)^2 + \Gamma_k^2m_k^2 \right) }
\;. \nn \\
%\sigma\,\beta|_{\rm su,0} & \to & -\frac{K}{2}
%\frac{\bar{\alpha}_{0su} \lambda_{ijk}(4\,\mchin{2}-m_k^2)}
%     {g(4\,m_\chi^2)\left(  \Gamma_k^2m_k^2 + (4\,m_\chi^2-m_k^2)^2 \right) }
%\\
%\sigma\,\beta|_{\rm tu,0} & \to & \frac{K\,\bar{\alpha}_{0tu}}{2\, 
%g^2(4\,\mchin{2})}
\een
\eesubeq
The above expressions are the ones which are relevant to indirect detection.
We note that for the s-channel, only the pseudo-scalar exchange gives a
non-zero contribution, as expected. It is also easy to check from these
expressions that only annihilation into a scalar and a pseudo-scalar is
non-zero in the null velocity limit.

To get the Taylor expansion terms of the thermally averaged cross section 
of~\citeeq{eq:sigv}, we refer the reader to Ref.~\cite{1988NuPhB.310..693S}.
Defining the Lorentz invariant function
\ben
W(s) = 4\,E_1\,E_2\,\sigma\,\beta
\een
\eg~from~\citeeq{eq:dsigma}, one can demonstrate that:
\ben
a &=& \frac{W(s)}{4\,\mchisq}|_{s\to 4\,\mchisq} = \sigma\,\beta|_0\\
b &=& -\frac{3}{8\,\mchisq}
\left[ 2 \, W(s) - 4 \, \mchisq \frac{dW(s)}{ds} \right]_{s\to 4\,\mchisq}\nn\;.
\een
The zeroth order term $a$ (often called S-wave, though it is only a part of it) 
is therefore fully determined by~\citeeq{eq:sigv_limit}, while the first order 
term $b$ requires a bit more of algebra. The latter case is left to the patience
of the reader, though explicit results are given for the cases of 
$\chi\,\chi\to h\,h$ and $\chi\,\chi\to a\,a$ in \citeeqss{eq:baa}{eq:bhh}, 
respectively. For the indirect detection signal, we have used the analytical 
expression obtained for the S-wave, while for the relic density calculation
we have computed the thermal average numerically following 
Refs.~\cite{1991NuPhB.360..145G,1997PhRvD..56.1879E}. 
We recall the full form
of the thermally averaged annihilation cross section below:
\ben
\sigv &=& 
\frac{ \int \frac{d^3\vec{p}_1}{2\,E_1\,(2\,\pi)^3} \, 
\frac{d^3\vec{p}_2}{2\,E_2\,(2\,\pi)^3} \, 
f_\chi(\vec{p}_1)\,f_\chi(\vec{p}_2) \, W(s)}   
{\int \frac{d^3\vec{p}_1}{(2\,\pi)^3} \frac{d^3\vec{p}_2}{(2\,\pi)^3} \,
f_\chi(\vec{p}_1)\,f_\chi(\vec{p}_2)}\,,
\label{eq:def_sigv}
\een
where function $f$ is the WIMP phase-space distribution, 
usually taken as a Maxwell-Boltzmann distribution. The above expression can be 
further developed and simplified, and is finally found to be a 
temperature-dependent integral over simply two variables: $s$, the squared 
center-of-mass energy, and $\mu$, the cosine of the angle between the 
annihilating particles 
--- see~\cite{1991NuPhB.360..145G,1997PhRvD..56.1879E} for more details. As
shown above, the integral over $\mu$ is performed analytically in our case.
In practice, we compute \sigv\ numerically from~\citeeq{eq:def_sigv}, 
where function $W(s)$ is tabulated up to $\sqrt{s-4\,\mchi^2}=10\,\mchi$, 
including all potential resonance and thresold effects. The Taylor expansion 
provided in~\citeeq{eq:sigv} is only meant to clarify the discussion. It is 
indeed well known that such an expansion is a bad approximation in some 
cases~\cite{1991PhRvD..43.3191G}. Still, the S-wave expression given 
in~\citeeq{eq:swave} turns out to be quite accurate to compute the indirect 
detection signals, which is not surprising given that the WIMP velocity 
distribution in the Galaxy peaks around $x=\mchi/T\approx 10^{7}$.

%% Appendix 2: Relic density

\section{Relic density}
\label{app:rd}
We refer the reader to Refs.~\cite{1991NuPhB.360..145G,1997PhRvD..56.1879E} for
more details on the relic density calculation, which we have followed in this 
study. We still provide a few pieces of information below, mostly related to 
the effective degrees of freedom, which is an independent part, except for the
last paragraph which gives some information about the numerical algorithm.

Assuming entropy conservation and a Hubble rate set by the energy density 
$\rho$ as $H^2 = 8\pi\rho/3$, the Boltzmann equation that drives the 
evolution of the dark matter fluid of density $n=s\,Y$ in the early universe 
reads:
\ben
\label{eq:rd}
\frac{dY}{dx} &=& - \frac{1}{3\,H}\frac{ds}{dx}\frac{\mchi}{x^2}\,\sigv \,
\left(Y^2 - Y_{\rm eq}^2 \right) \\
&=& - \sqrt{\frac{\pi}{45}} \, \frac{M_p\,\mchi}{x^2} \, 
\sqrgstar \, \sigv \,\left( Y^2 - Y_{\rm eq}^2 \right)\;, \nn
\een
where $M_p$ is the Planck mass, $s=(\rho+p)/3$ is the entropy density ($p$ being
the pressure). The dynamical variable is $x=\mchi/T$, increasing when the 
universe expands. We further define the effective relativistic degrees of 
freedom $\geff$ and $\heff$ relevant to the energy density and the entropy 
density, respectively, such that:
\ben
\label{eq:dof}
\rho &=& \frac{\pi^2}{30}\, \geff(T)\,T^4\\
s &=& \frac{2\,\pi^2}{45}\,\heff(T)\,T^3\nn\\
\sqrgstar &=& \frac{\heff}{\sqrt{\geff}}
\left[1 + \frac{T}{3\,\heff} \, \frac{d\heff}{dT} \right] \nn \;.
\een
It is useful to recall the expressions of the energy density and the entropy
density, for a fermion or a boson species labelled $i$:
\ben
\rho_{i, {\rm b/f}}(T) &=& g_i\, \int\frac{d^3\vec{p}}{(2\,\pi)^3}\,
\frac{E}{\exp\left\{\frac{E}{T}\right\} \mp 1  }\\ 
s_{i,b/f} &=& \frac{g_i}{T} \int\frac{d^3\vec{p}}{(2\,\pi)^3}\,
\frac{E+\vec{p}\cdot\vec{v}/3}{\exp\left\{\frac{E}{T}\right\} \mp 1  } \nn
\een
These integrals can actually be evaluated quite easily from series involving
modified Bessel functions of order $n$, $K_n$, which quickly converge to the 
exact results such that each species $i$ carries~\cite{1989PhRvD..40.3168B}:
\ben
g_{{\rm eff},i,{\rm b/f}} &=& 
\frac{15}{\pi^4} \, g_i \, \sum_{n=1}^{\infty}(\pm 1)^{n-1}\times\\
&&\left\{ \frac{x_i^3}{n}\,K_1(n\,x_i)+
\frac{3\,x_i^2}{n^2}\,K_2(n\,x_i) \right\}\nn\\
h_{{\rm eff},i,{\rm b/f}} &=& 
\frac{45}{4\,\pi^4} \, g_i \, \sum_{n=1}^{\infty}(\pm 1)^{n-1}\times\nn\\
&&\left\{ \frac{x_i^3}{n}\,K_1(n\,x_i)+
\frac{4\,x_i^2}{n^2}\,K_2(n\,x_i) \right\}\nn\;,
\een
where $x_i = m_i/T$. Note that in the ultra-relativistic limit $x_i\to 0$, these
expressions tend to $g_i$ and $(7/8)\,g_i$, respectively for bosons and 
fermions, as expected. We have used the following property of Bessel 
functions~\cite{1972hmfw.book.....A}:
\ben
\int_y^{+\infty}\left(x^2 - y^2\right)^{\alpha-1}  e^{-\mu\,x}  dx = 
\frac{1}{\sqrt{\pi}}\left[\frac{2\,y}{\mu}\right]^{\frac{2\,\alpha-1}{2}}
\Gamma(\alpha)\, K_{\frac{2\,\alpha-1}{2}}(\mu\,y)\;.\nn
\een

Given these expressions, one can calculate the complete set of effective degrees
of freedom by summing over all physical degrees of freedom at a given 
temperature. Before the QCD phase transition in the very early universe down
to $T\sim M_W$, this amounts to all degrees of freedom of the standard model, 
including quarks and gluons, except the $W^{\pm}$, $Z$ and Higgs bosons (the 
electroweak phase transition occurs earlier when $T\gtrsim M_W$). After the
QCD phase transition, quarks and gluons are no longer free and remain confined
into hadrons; then resonances have also to be included in addition to nucleons,
an can provide extra-contribution depending on whether they are relativistic 
or not~\cite{1996RPPh...59.1493S}.

Note that in principle, one should also add those additional
degrees of freedom coming from non-standard particles related to the particular
dark matter model under scrutiny. In this paper, the light Higgs bosons and
the singlino should therefore be added. Nevertheless, we can safely neglect 
them since they are also non-relativistic when singlinos decouple, around
$\mchi/T\sim 20$ (we used $m_a\gtrsim 1$ GeV and $m_h\gtrsim 4$ GeV, and
focused on $5\,{\rm GeV} \lesssim \mchi \lesssim 15\,{\rm GeV}$). Our
results are illustrated in~\citefig{fig:rd}, where a sharp QCD phase transition
has been considered. Such a transition implies that $\gstar$ goes to infinity
because of the derivative of $\heff$ in~\citeeq{eq:dof}. We have suppressed
this divergence based on that the actual transition is expected to be 
smoother~\cite{2005PhRvD..71h7302H}. We have checked that considering this 
effect or not has no significant impact on our results, at the percent level.

As regards the numerical method, we have implemented a Runge-Kutta-6 scheme to
track the evolution of the comoving density $Y$ featuring~\citeeq{eq:rd}.
To avoid starting the calculation too early (too small $x$), we first estimate 
the decoupling temperature by implicitly solving \citeeq{eq:rd}, demanding
$Y(x_\epsilon) = (1+\epsilon)Y_{\rm eq}(x_\epsilon)$ (with $\epsilon<1/2$, 
typically) --- before decoupling, $Y$ tracks its equilibrium value. We then 
start the calculation at $x_{\rm in}\lesssim x_\epsilon$, up to after the
decoupling. Indeed, to achieve an accuracy at the percent level, we need
to solve~\citeeq{eq:rd} up to $x_{\rm end}$ such that 
$Y(x_{\rm end})>10\,Y_{\rm eq}(x_{\rm end})$, because of the term 
$(Y^2-Y_{\rm eq}^2)$. In practice, we use 
$Y(x_{\rm end})=30\,Y_{\rm eq}(x_{\rm end})$. Then, we neglect $Y_{\rm eq}$
and can therefore use the solution of~\citeeq{eq:y0}, taking 
$x_{>f} = x_{\rm end}$ and performing a Simpson integral over $x$ in logarithmic
steps.

%%\input{app1_annihilation_cross_section.tex}
%%\input{app2_relic_density.tex}

%%\input{app3_answers_to_referee.tex}

%% References


\bibliographystyle{elsarticle-num}
\bibliography{lavalle_bib}

% \end{thebibliography}

\end{document}